\documentclass{article}

\usepackage{amsmath}
\usepackage{amsfonts}
\usepackage{multirow}
\usepackage{graphicx}
\usepackage{caption}
\usepackage{subcaption}
\usepackage{longtable}
\usepackage{pdflscape}
\usepackage[a4paper, margin=0.7in]{geometry}
\usepackage{bm}
\usepackage{afterpage}
\usepackage{multicol}
\usepackage{float}
\usepackage{longtable}
\usepackage[toc,page]{appendix}
\usepackage{breqn}
\usepackage{lscape}
\usepackage{amssymb,mathtools}
\usepackage{rotating}
\usepackage{authblk}\usepackage{siunitx}
\usepackage[backend=bibtex, minbibnames=3,style=numeric,citestyle=numeric-comp,sorting=none, giveninits ]{biblatex}
\DeclareNameAlias{author}{last-first}

\usepackage{verbatim}
\usepackage{xcolor}
\usepackage{hyperref}

\usepackage{titlesec}
\usepackage{enumitem}

\titleformat{\paragraph}[runin]
{\normalfont\normalsize\bfseries}{\theparagraph}{1em}{}[:]
\titlespacing*{\paragraph}
{0pt}{3.25ex plus 1ex minus .2ex}{1.5ex plus .2ex}

\providecommand{\keywords}[1]{\textbf{\textit{Keywords---}} #1}

\DeclareMathAlphabet\mathbfcal{OMS}{cmsy}{b}{n}

\title{A physiologically realistic virtual patient database for the study of arterial haemodynamics}
\author[1]{Gareth Jones}
\author[2]{Jim Parr}
\author[1]{Perumal Nithiarasu}
\author[1, $\dagger$]{Sanjay Pant}
\affil[1]{College of Engineering, Swansea University, Swansea, United Kingdom.}
\affil[2]{McLaren Technology Centre, Woking, United Kingdom.}

\affil[$\dagger$]{Corresponding author: Sanjay.Pant@swansea.ac.uk}

\date{}      
\bibliography{Large_network_VPD_arxiv.bib}

\begin{document}

\maketitle

\begin{abstract}
\noindent This study creates a physiologically realistic virtual patient database (VPD), representing the human arterial system, for the primary purpose of studying the affects of arterial disease on haemodynamics. A low dimensional representation of an anatomically detailed arterial network is outlined, and a physiologically realistic posterior distribution for its parameters constructed through the use of a Bayesian approach. This approach combines both physiological/geometrical constraints and the available measurements reported in the literature. A key contribution of this work is to present a framework for including all such available information for the creation of virtual patients (VPs). The Markov Chain Monte Carlo (MCMC) method is used to sample random VPs from this posterior distribution, and the pressure and flow-rate profiles associated with each VP computed through a physics based model of pulse wave propagation. This combination of the arterial network parameters (representing a virtual patient) and the haemodynamics waveforms of pressure and flow-rates at various locations (representing functional response and potential measurements that can be acquired in the virtual patient) makes up the VPD. While 75,000 VPs are sampled from the posterior distribution, 10,000 are discarded as the initial burn-in period of the MCMC sampler. A further 12,857 VPs are subsequently removed due to the presence of negative average flow-rate, reducing the VPD to 52,143.  Due to undesirable behaviour observed in some VPs---asymmetric under- and over-damped pressure and flow-rate profiles in left and right sides of the arterial system---a filter is proposed to remove VPs showing such behaviour. Post application of the filter, the VPD has 28,868 subjects. It is shown that the methodology is appropriate by comparing the VPD statistics to those reported in literature across real populations. Generally, a good agreement between the two is found while respecting physiological/geometrical constraints. The pre-filter database is made available at \url{https://doi.org/10.5281/zenodo.4549764}. \\[3pt]

\noindent \keywords{virtual patients, stenosis, aneurysm, pulse wave haemodynamics, MCMC, screening, virtual patient database}

\end{abstract}


\section{Introduction}

Arterial disease refers to any disease affecting the arterial system. Two of the most common forms of arterial disease are stenosis and aneurysm. These two forms of disease are estimated to affect between 1\% and 20\% of the population \cite{fowkes2013comparison, shadman2004subclavian, mathiesen2001prevalence, li2013prevalence}. Ruptured abdominal aortic aneurysms alone are estimated to be the cause of between 6,000 and 8,000 deaths a year within the United Kingdom \cite{darwood2012twenty}. 
Current methods for the detection of arterial disease are primarily based on imaging techniques which can be expensive and thus unsuitable for mass screening. A potential solution is offered by the recent emergence of data-driven Machine Learning (ML) algorithms which can be trained on easily acquirable pressure and flow-rate measurements. The key idea is that the presence of disease---both stenosis (a reduction in vessel area) and aneurysm (an increase in vessel area)---will create changes in such measurements, which may be detectable by the ML algorithms. To explore this possibility, however, a large database of both healthy and diseased subjects along with their respective measurements is necessary. This study presents a computational method for the creation of such a physiologically realistic database.

As opposed to acquiring measurements from a real population, which may be impractical, a virtual patient database (VPD) based on a validated physics-based model is proposed. This idea of exploiting virtual patients (VPs) to gain meaningful information has been previously explored in the context of human arterial networks \cite{willemet2015database, chakshu2020towards, jones2021proof}, and  in the context of path clamp physiology \cite{celik2020deep}. In \cite{willemet2015database}, a VPD of healthy subjects is created by varying seven parameters of a large 55-artery network to assess the relation between foot-to-foot pulse wave velocities and aortic stiffness. In \cite{chakshu2020towards}, 1D arterial network parameters are scaled through empirical relations based on weight, height, and age to create a virtual patient database. In \cite{jones2021proof},  nine parameters of a small network of abdominal aortic bifurcation are varied to create the VPD and ML algorithms subsequently applied to detect the presence of stenosis. The limitations of these studies are that either not all the parameters are varied in both healthy and diseased states, or \emph{a priori} distributions are assumed for the parameters. Both of these are important considerations for the creation of a general purpose realistic VPD which captures the variability as close as possible to the real population. Furthermore, the VPD must account for known variation of the measurements reported in the literature along with the physical constraints, while making minimal assumptions. This study presents a probabilistic framework for creation of such a VPD based on a previously proposed, anatomically correct, 56-vessel and 77-segment arterial network  \cite{blanco2014blood, blanco2015anatomically}. The proposed framework may also be useful for the novel and active research area of in-silico clinical trials \cite{pappalardo2019silico}, for which generation of realistic virtual patients is a key consideration.

This article is organised as follows. First, a brief description of the physics-based pulse wave propagation model is presented. This is followed by a description of the arterial network topology---vessels, locations of disease and potential measurements---and its reduction. Next, parameterisation of the arterial network---geometry, vessel properties, and boundary conditions---is presented, followed by the Bayesian framework for the VPD creation. Finally, the solution to the Bayesian problem through Markov Chain Monte Carlo method is presented, before discussing the results and statistics of the VPD.

\section{Pulse wave propagation model}
\label{section_1D_model}
A validated one dimensional model of pulse wave propagation \cite{alastruey2012arterial,boileau2015benchmark} is adopted in this study. The model describes the arterial vessels as one-dimensional compliant tubes, i.e. all the properties are described by a single coordinate $x$ along the length of the vessel. Blood flow is assumed to be Newtonian, laminar, incompressible, and axisymmetric. With the arterial cross-sectional area represented as $A$, and considering the average lumen pressure $P$ and the average velocity $U$ to be the primary variables, the mass and momentum conservation equations at any coordinate $x$ can be, respectively, written as
%
 %
\begin{equation}
\frac{\partial{A}}{\partial{t}}+\frac{\partial{(UA)}}{\partial{x}}=0,
\label{eq_mass_conservation}
\end{equation}
\begin{equation}
\frac{\partial{U}}{\partial{t}}+U\frac{\partial{U}}{\partial{x}}+\frac{1}{\rho}\frac{\partial{P}}{\partial{x}}=\frac{f}{\rho{A}},
\label{eq_momentum_conservation}
\end{equation}
where $\rho$ represents blood density and $f$ represents the frictional force per unit length.  Depending on the assumption of velocity profile across the arterial cross section, $f$ can be written as
\begin{equation}
f(x, t)=-2({\zeta}+2){\mu}{\pi}U,
\label{eq_f}
\end{equation}
where $\zeta$ is a velocity-profile dependant constant, and $\mu$ represents dynamic viscosity of blood. The system of equations is closed by specification of the mechanical behaviour of the wall, commonly referred to as the ``tube law''. A commonly used form \cite{boileau2015benchmark} is
%
\begin{equation}
P-P_{ext}=P_d+\frac{\beta}{A_d}{(\sqrt{A}-\sqrt{A_d})},
\label{eq_pressure_area}
\end{equation}
with 
\begin{equation}
\beta=Eh\frac{4}{3}\sqrt{\pi},
\label{eq_beta}
\end{equation}
where $P_{ext}$ represents the external pressure, $P_{d}$ represents the diastolic blood pressure, $A_d$ represents the diastolic area of the vessel, $E$ represents the vessel wall's Young's modules, and $h$ represents the vessel wall's thickness. This system of equations has been widely employed, tested extensively, and validated against more complex three-dimensional models \cite{boileau2015benchmark, formaggia2003one, alastruey2012arterial, olufsen2000numerical, reymond2009validation, matthys2007pulse}.

\begin{figure}[tb]
\centering 
\includegraphics[width=2.5in, trim={0 0 0 0.5cm}, clip] {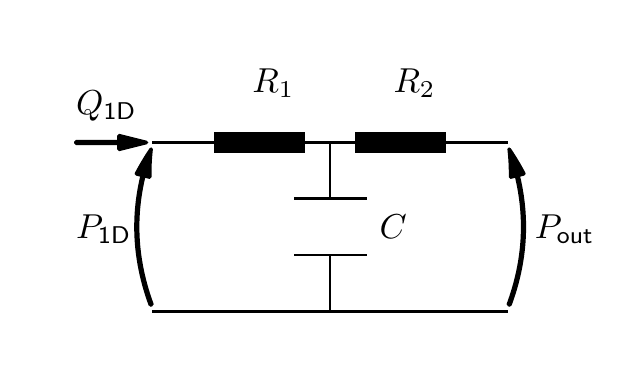}
\caption{Configuration of a three-element Windkessel model. $Q_{\text{1D}}$ and $P_{\text{1D}}$ represent the volumetric flow-rate and pressure at the terminal boundary of the 1D system respectively.}
\label{figure_windkessel}
\end{figure}

Since it is computationally impractical to model every single artery in the circulatory network, it is common to truncate the 1D network by only considering relatively large arteries which are of interest. The effect of all the vessels that come before and after the network are then incorporated as appropriate boundary conditions. Typically, at the inlet of the network a volumetric flow-rate is specified, and at the outlets Windkessel models \cite{westerhof2009arterial} are employed. A three-element Windkessel model, shown in Figure \ref{figure_windkessel}, comprises of two resistors, $R_1$ and $R_2$, which replicate the viscous resistance of large arteries and the micro-vascular system, respectively, and a capacitor $C$, which replicate the compliance of large arteries. 



\noindent The 1D model is solved numerically by a subdomain collocation (SDC) method \cite{carson2019development}. This scheme is chosen for its computational efficiency. The SDC implementation is validated against a discontinuous Galerkin (DCG) scheme \cite{alastruey2012arterial}, which in turn has been successfully validated against a 3D model of blood-flow through stenosed arterial vessels \cite{boileau2018estimating}.

%

\section{Arterial network topology}
\label{section_toplogy}

\subsection{Initial network}

The arterial network used as a basis for the creation of VPs is based on a version of the ADAN network \cite{boileau2015benchmark}. This network contains 56 vessels within the human arterial system, described by 77 arterial vessel segments. The topology of this network is shown in Figure \ref{fig_full_network}, and each vessel within the network is identified in Appendix \ref{appendix_list_of_vessels}. 
This network is described by:
\begin{itemize}[leftmargin=*]
\item \textbf{38 vessel segments} with constant reference radii ($r_0$) and vessel wall mechanical properties ($\beta$) along their lengths. The description of each of these segments requires specification of three parameters: the length $L$, reference radius $r_0$, and mechanical property $\beta$ of the segment.
\item \textbf{39 vessel segments} with linearly varying $r_0$ and $\beta$  along their lengths. The description of each of these segments requires specification of five parameters: the length $L$; the reference radii $r_0$ at the proximal and the distal ends of the segment; and  $\beta$ at the proximal and  distal ends.
\item \textbf{31 Windkessel models.} Every Windkessel model at the outlets requires specification of 2 resistances and a compliance. 
\item \textbf{One inlet flow-rate profile.} In this study, the inlet flow-rate is parameterised by a $5^{\text{th}}$ order Fourier Series (see Section \ref{sec_inlet_boundary} and \cite{jones2021proof}). This requires specification of 11 coefficients.
\end{itemize}

\noindent Thus, to describe a VP through direct specification of all the above parameters results in the total dimensionality of 413 per VP. This high dimensionality is problematic leading to increased complexity in creating VPs \cite{kuo2005lifting}. In what follows, methods to reduce the dimensionality of VP description are presented. These are primarily related to either reducing the network or employing a parsimonious parameterisation.
Since the primary purpose of the VPD is deployment of ML methods for screening of aneurysm and stenosis through easily acquirable measurements, it is important that the reduction in dimensionality does not compromise i) the locations where disease and measurements are possible, and ii) the precision and variability of measurements at such locations.


 \begin{figure*}
\centering
\includegraphics[width=5.5in]{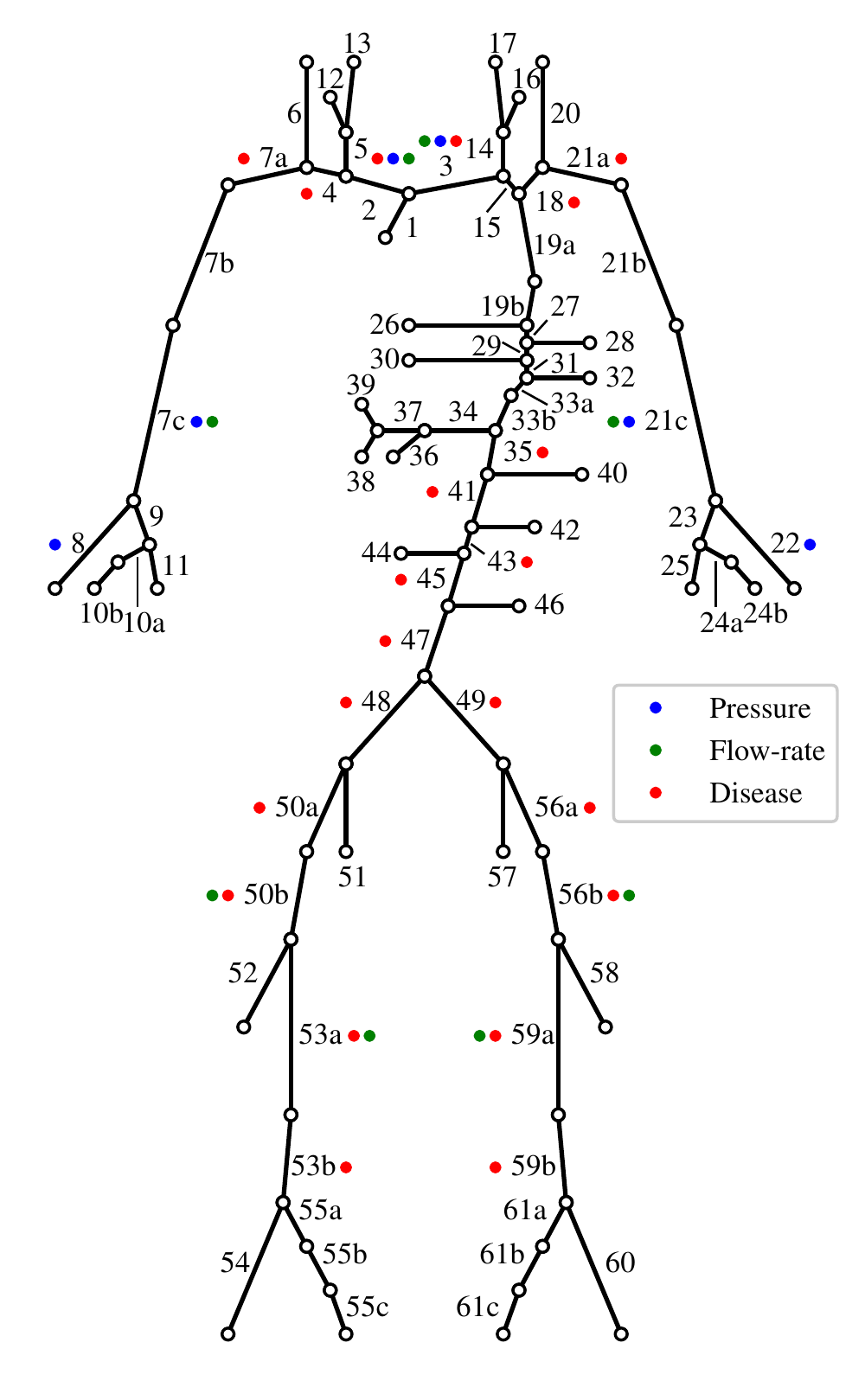}
\caption{The connectivity of the reference arterial network, taken from \cite{boileau2015benchmark}. At the inlet (free end) of vessel 1, a volumetric flow-rate is specified and at all the outlets (free ends of the terminal vessels), a Windkessel model is specified. Locations at which pressure and flow-rates can be measured; and disease is likely to occur are also highlighted, see Section \ref{section_locations}.}
\label{fig_full_network}
\end{figure*}

\subsection{Important locations}
\label{section_locations}

A review of literature is carried out to understand both where disease is likely to occur and where measurements can be obtained. The latter are restricted to locations at which continuous profiles can be recorded non-invasively. While arterial disease can occur in a large number of vessels, for this study vessels with only high prevalence of stenosis or aneurysm are considered.

\subsubsection{Locations of measurements}
Based on literature, the following locations where measurements can be taken are identified:\\

\noindent \textbf{Pressure in the radial and common carotid arteries:} Continuous non-invasive arterial blood pressure profiles can be obtained in the radial and common carotid arteries using applanation tonometry \cite{adji2006clinical, o2015carotid}. Applanation tonometry is already clinically used \cite{langwieser2015radial}. The right and left radial and common carotid arteries are identified in Figure \ref{fig_full_network} as vessels 8 and 22; and 5 and 14, respectively.\\

\noindent \textbf{Pressure in the brachial arteries:} It is possible to estimate continuous blood pressure at the brachial arteries through reconstruction of finger arterial pressure \cite{guelen2008validation}. Although the use of a model to estimate brachial pressure will introduce errors, these recreated brachial pressure profiles have been shown to meet the requirements set by the Association for the Advancement of Medical Instrumentation \cite{guelen2008validation, guelen2003finometer}. The right and left brachial arteries are identified in Figure \ref{fig_full_network} as vessels 7c and 21c respectively. \\
 
\noindent \textbf{Flow-rate in brachial, carotid, and femoral arteries:} Using Doppler ultrasound techniques it is possible to measure blood velocity in both the upper and lower extremities. Doppler ultrasound has been shown to work on the brachial \cite{bystrom1998ultrasound}, common carotid \cite{oglat2018review}, and femoral \cite{radegran1997ultrasound} arteries. The first and second, right and left femoral arterial segments are identified in Figure \ref{fig_full_network} as vessels 50b and 53a; and 56b and 59a respectively.

\subsubsection{Locations of disease} 

 Four of the most common forms of arterial disease are subclavian artery stenosis (SAS), carotid artery stenosis (CAS), peripheral arterial disease (PAD), and abdominal aortic aneurysm (AAA).\\

\noindent \textbf{SAS locations:} The ADAN network splits the subclavian arteries into two segments. The right and left instances of the first and second subclavian artery segments are identified in Figure \ref{fig_full_network} as vessels 4 and 7a; and 18 and 21a respectively.\\

\noindent \textbf{CAS locations:} The carotid arteries consist of the common carotid, external carotid, and internal carotid segments. While the narrowing of an arterial vessel can occur within any of the three carotid segments \cite{dyken1974complete}, it is chosen to limit its occurrence to the common carotid arteries. The right and left instances of the common carotid arteries are identified in Figure \ref{fig_full_network} as vessels 5, and 14 respectively.\\
  
\noindent \textbf{PAD locations:} The third frequent form of stenosis is PAD, referring to the stenosis of any peripheral vessel. Isolating arterial vessels that are likely to experience stenosis at a high prevalence in patients with PAD is more difficult relative to SAS or CAS. Both SS and carotid artery stenosis have short and relatively easy to define lists of possible vessels they can affect. PAD, on the other hand, can cover a large range of different vessels depending on the definition prescribed. Studies into PAD primarily focus on the lower extremities \cite{kullo2016peripheral, aboyans2010general, chen2013disease} and thus it is assumed that  VPs with PAD experience change in these vessels. Based on  \cite{chen2013disease} and \cite{aboyans2010general}, PAD in VPs is restricted to the common iliacs, external iliacs, first and second femoral segments, and the first popliteal segments. The right and left instances of each of these vessels are identified in Figure \ref{fig_full_network} as vessels 48 and 49; 50a and 56a; 50b and 56b; 53a and 59a; and 53b and 59b, respectively.\\

\noindent \textbf{AAA locations:} The ADAN network splits the abdominal aorta into five segments. These five segments are identified in Figure \ref{fig_full_network} as vessels 35, 41, 43, 45, and 47 respectively.

\subsection{Network reduction}

\label{section_removal_vessels}
\begin{figure}[tb]
 \center
\includegraphics[width=6in]{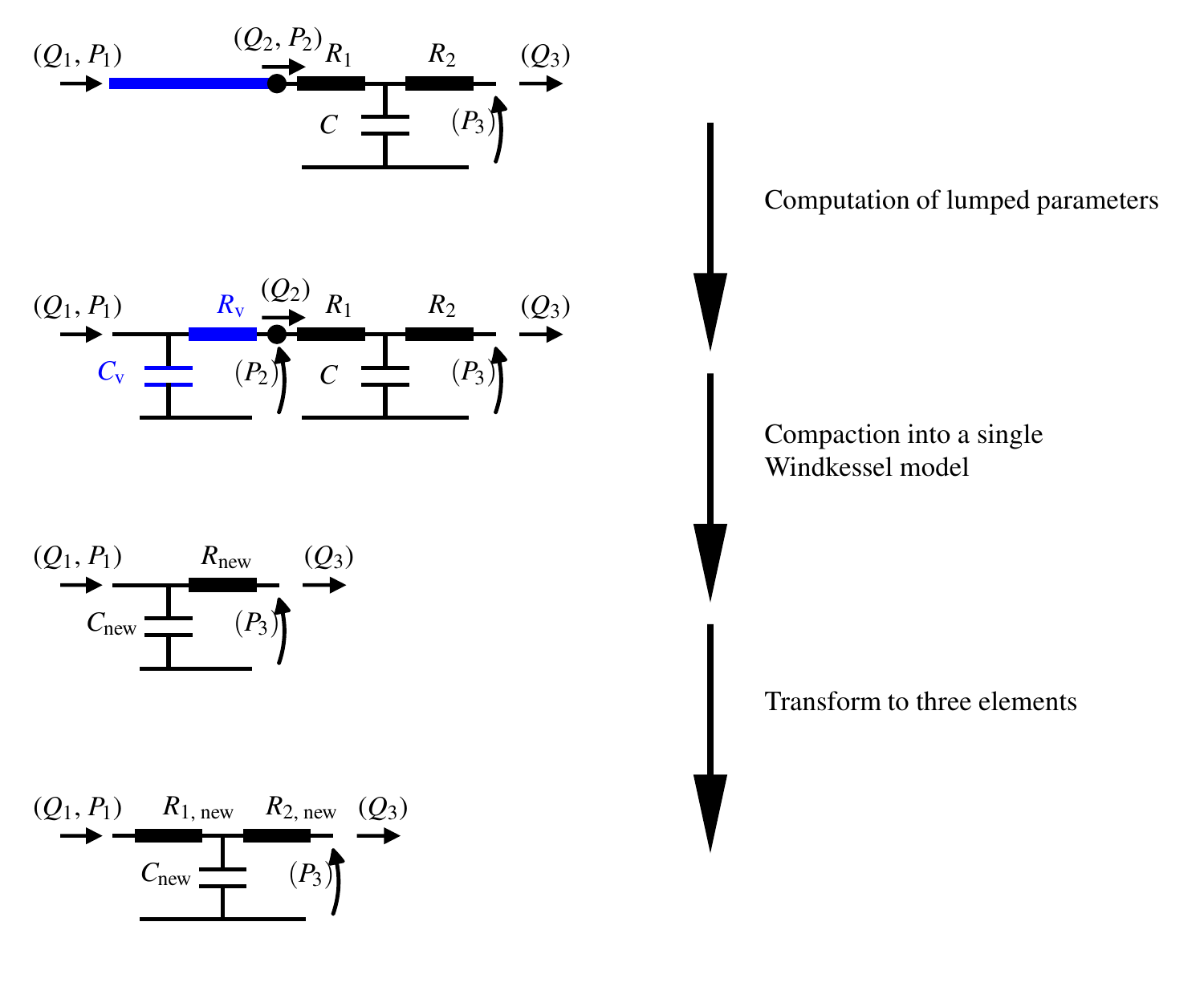}
\caption{The process of incorporating peripheral 1D vessels into the 0D terminal boundary Windkessel model parameters  \cite{epstein2015reducing}. $Q_1$ and $P_1$ represent the flow-rate and pressure at the proximal end of the vessel that is being removed; $Q_2$ and $P_2$ represent the flow-rate and pressure at the distal end of the vessel that is being removed; $Q_3$ and $P_3$ represent the the flow-rate and pressure at the outlet of the system; $R_1$, $R_2$ and $C$ represent the resistances and compliance of the original terminal Windkessel model; $R_v$, and $C_v$ represent the viscous resistance and compliance of the vessel being removed; $R_{\text{new}}$ and $C_{\text{new}}$ represent the resistance and compliance of the new 2 element Windkessel model after the incorporation of the 1D vessel; and $R_{\text{1,new}}$, $R_{\text{2,new}}$, and $C_{\text{new}}$ represent the resistances and compliance of the final 3 element Windkessel model. Details on computation of $R_v$ and $C_v$ and then  $R_{\text{new}}$ and $C_{\text{new}}$ can be found in \cite{epstein2015reducing}. $R_{\text{1, new}}$ is assumed to be equal to the characteristic impedance of the connecting vessel (see \cite{epstein2015reducing}), and $R_{\text{2,new}} = R_{\text{new}}-R_{\text{1,new}}$.}
\label{fig_network_reduction}
\end{figure}

In this section, the reduction of the network by removing vessels while preserving the disease and measurement locations is presented.
It is important, however, to ensure that the removal of any vessels does not have a significant impact on the upstream pressure and flow-rate profiles. Network reduction can be performed by removing terminal vessels and merging them into the lumped parameter (Windkessel) boundary conditions \cite{epstein2015reducing}. 
This process is adopted in this study and is summarised in Figure \ref{fig_network_reduction}.
 %
 The following three  groups of vessels eligible for removal are identified:
 \begin{itemize}[leftmargin=*]
 \item \textbf{Group 1:} The first and second splenic segments; the left gastric; and the common hepatic (identified in Figure \ref{fig_full_network} as vessels 37 and 39; 38; and 36).
 \item \textbf{Group 2:} The common interosseous, the posterior interosseous, and the second ulnar segment (identified within Figure \ref{fig_full_network} as vessels 10a and 24a; 10b and 24b; and 11 and 25).
 \item \textbf{Group 3:} The second popliteal segments; the anterior tibial; the tibiofibular trunk; and the posterior tibial (identified within Figure \ref{fig_full_network} as vessels 55a; 54; 55b; and 55c on the right side and 61a; 60; 61b; and 61c on the left side). 
 \end{itemize}

\noindent Three possible reduced networks are proposed, each with the removal of a single group of vessels (group 1, 2, or 3 in isolation). The pressure and flow-rate profiles produced by each reduced network are compared against the full ADAN network at all the measurement locations (see Section \ref{section_locations}). The error metrics proposed in  \cite{epstein2015reducing} are used. 
If the maximum error is less than 2\%, the full group of vessels are omitted from the arterial network. Otherwise, the vessel segment at the proximal end of the group is re-introduced into the  reduced network. 
This process is iteratively repeated until the maximum number of vessels that can be removed from each group individually are found. 
Finally, these vessels from the three groups are removed simultaneously.

\begin{figure*}
\centering
\includegraphics[width=\textwidth]{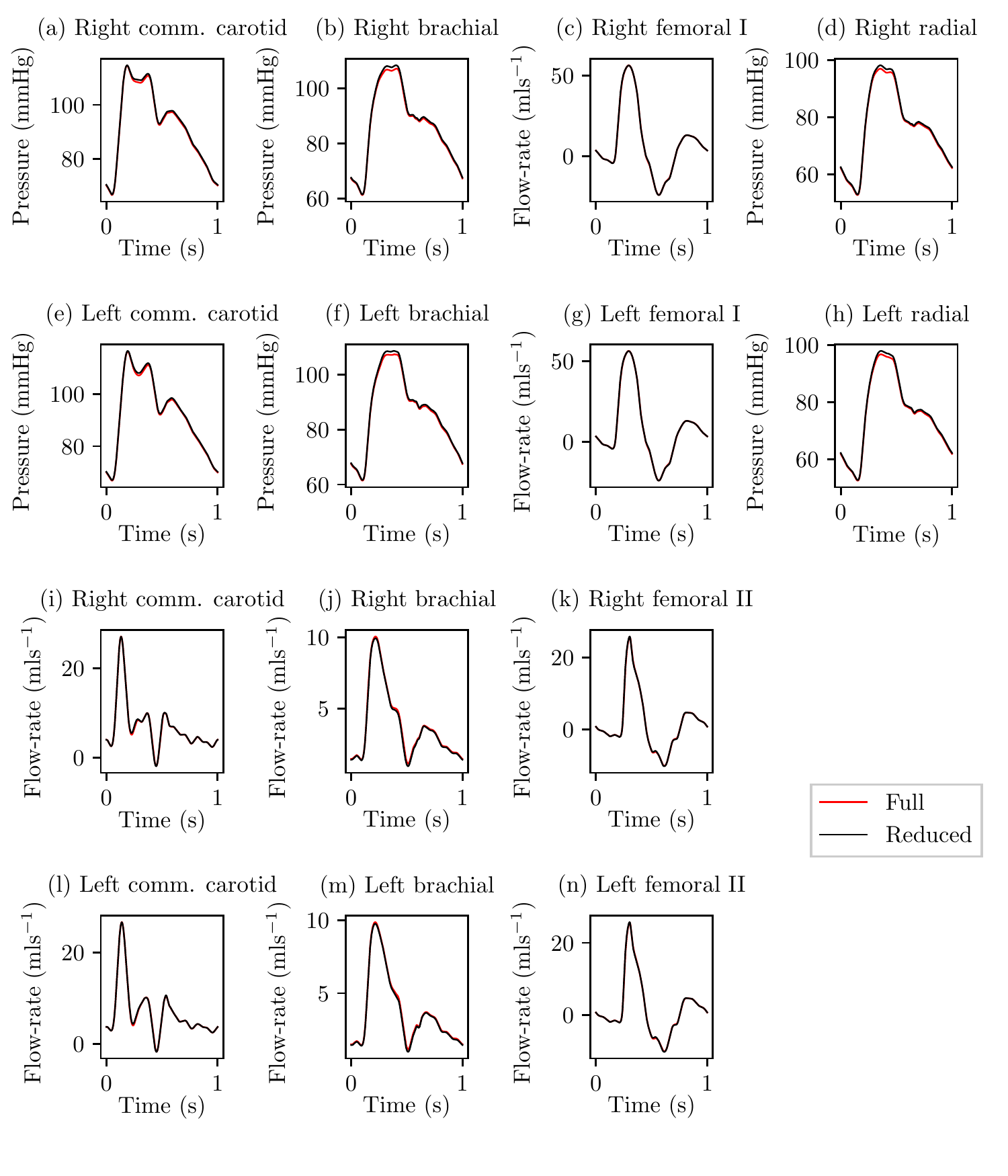}
\caption{Comparison of pressure or flow-rate profiles at all the measurement locations taken from the full ADAN network and the reduced network produced by the removal of the first and second splenic segments, left gastric, common hepatic, and posterior interosseous.}
\label{fig_reduction_6}
\end{figure*}

\begin{figure*}
\centering
\includegraphics[width=5in]{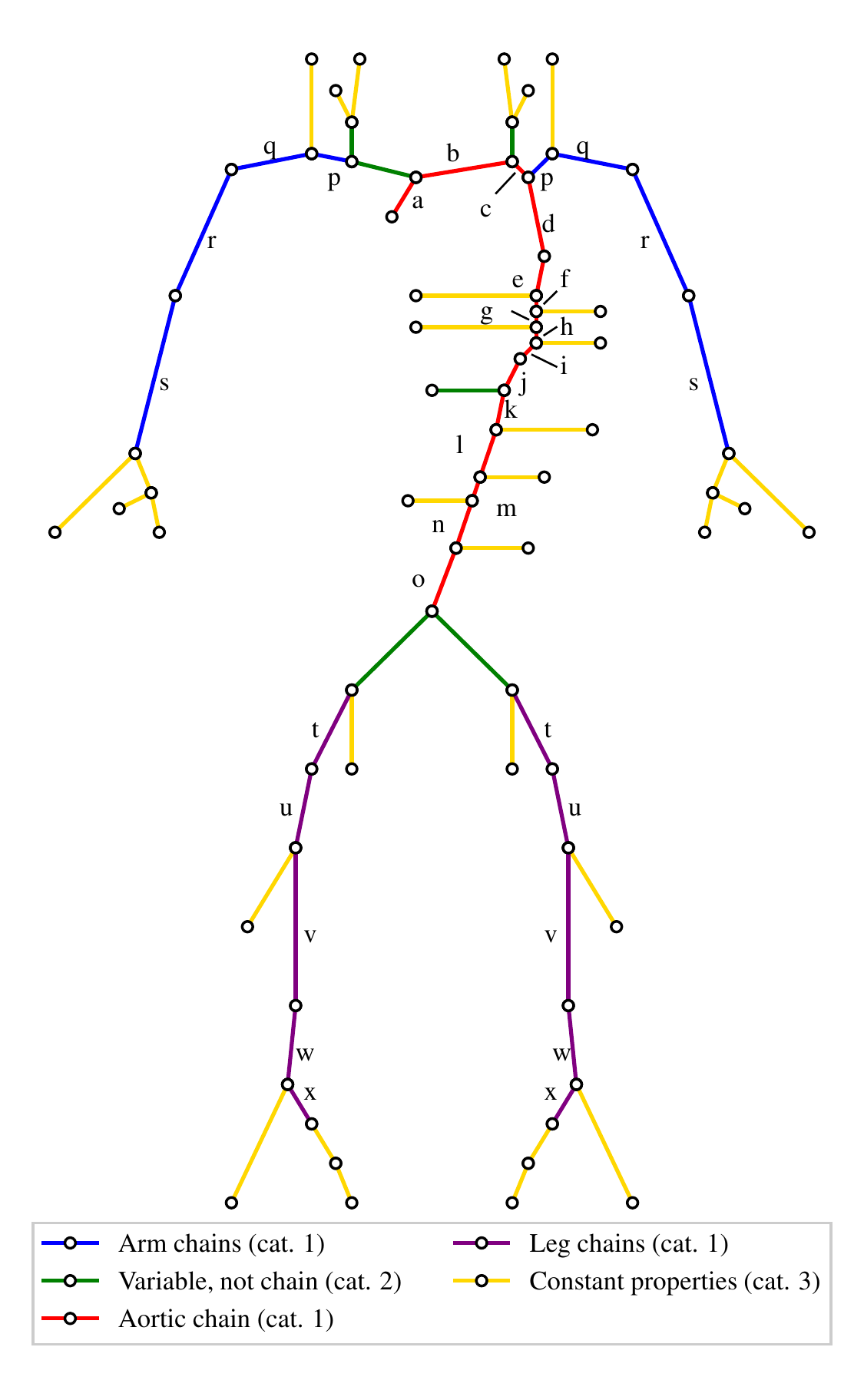}
\caption{The reduced network and grouping of vessels. The aortic chain is shown in red and consists of: the first to forth aortic arch segments denoted by `a' through to `d', the first to sixth thoracic aorta segments denoted by `e' through to  `j', and the first to fifth abdominal aorta segments denoted by `k' through to `o'. The right and left arm chains are shown in blue and consist of: the first and second subclavian segments denoted by `p' and `q' respectively, the axillary artery denoted by `r', and the brachial artery denoted by `s'. The right and left leg chains are shown in purple and consist of: the external iliac denoted by `t', the first and second femoral segments denoted by `u' and `v' respectively, and the first and second popliteal segments denoted by `w' and `x' respectively. Vessels which have variable properties but are not part of a chain are shown in green. Finally, vessels with constant properties are shown in yellow.}
\label{fig_vessel_chains}
\end{figure*}

It is found that the first and second splenic segments; left gastric; common hepatic; and the left and right posterior interosseous vessels (identified within Figure \ref{fig_full_network} as vessels vessels 37 and 39; 38; 36; and 10b and 24b) can all be removed from the arterial network without introducing errors larger than 2\% relative to the full ADAN network. 
The pressure and flow-rate profiles produced by the full and reduced networks are shown in Figure \ref{fig_reduction_6} and the errors are reported in Table \ref{table_reduction_6}. The reduced network is shown in Figure \ref{fig_vessel_chains} and has a dimensionality of 389, a reduction of 24 from the full network. This network and its associated parameters---either directly taken from or computed based off of the ADAN network---forms the reference  network for this study.

\begin{table*}
\centering
\def\arraystretch{1.2}
\begin{tabular}{|c| c| c c c|}
\hline
 \textbf{Location} & \textbf{Measurement}  & \textbf{Average error [\%] } &  \textbf{Systolic  error [\%] } &  \textbf{Diastolic error [\%] } \\
\hline
\multirow{ 2}{*}{Right common carotid} & Pressure & 0.4380 & 0.114 & 0.316\\
& Flow-rate & 0.250 & -0.085 & -0.271\\
\multirow{ 2}{*}{Left common carotid} & Pressure & 0.438 & 0.157 & 0.318\\
& Flow-rate & 0.246 & -0.111 & -0.186\\
\multirow{ 2}{*}{Right brachial} & Pressure & 0.643 & 1.108 & 0.378\\
& Flow-rate & 0.875 & -1.223& -1.833\\
\multirow{ 2}{*}{Left brachial} & Pressure & 0.644 & 1.179 & 0.380\\
& Flow-rate & 0.889 & -1.131 & -1.788\\
Right radial & Pressure & 0.654 & 1.236 & 0.383 \\
Left radial & Pressure & 0.657 & 1.231 & 0.386\\
Right femoral I & Flow-rate & 0.256 & -0.174 & -0.114\\
Right femoral II & Flow-rate & 0.365 & 0.515 & -0.463\\
Left femoral I & Flow-rate & 0.254 & -0.166 & -0.115\\
Left femoral II & Flow-rate & 0.363 & 0.514 & -0.463\\
\hline
\end{tabular}
\caption{Percentage discrepancy between the measurements taken from the full arterial network and the reduced network produced by the removal of the first and second splenic segments; left gastric; common hepatic; and posterior interosseous. Percentage errors have been computed using the error metrics presented in \cite{epstein2015reducing}.}
\label{table_reduction_6}
\end{table*}

\section{Network parameterisation}
\label{section_parameterisation}
A parsimonious representation is sought for the aforementioned reference (reduced) network so that its dimensionality can be further reduced.
%
The reference network has 389 parameters.
The vessels in this network can be split into the following three categories, which are also indicated in Figure \ref{fig_vessel_chains}: 
\begin{itemize}[leftmargin=*]
\item \textbf{Category 1:} 33 vessel segments with varying $\beta$ and $r_0$, and continuous variation with respect to the prior and subsequent vessels.
\item \textbf{Category 2:} six vessel segments with varying $\beta$ and $r_0$ , and discontinuous variation with respect to the prior and subsequent vessels.
\item \textbf{Category 3:} 32 vessel where $\beta$ and $r_0$ are constant along their lengths.
\end{itemize}
For Categories 1 and 2 with varying intra-vessel property profile, the reference network assumes a linear variation  \cite{boileau2015benchmark}. Thus any property $g(x)$ (reference radius $r_0$ or $\beta$) varies along length $x$ as: 
\begin{equation}
g(x) = g_{0}-\left(g_{0}-g_{L}\right)\frac{x}{L},
\label{eq_linear_profile}
\end{equation}
where $g_{0}$ and $g_{L}$ represent the properties at the proximal end $x=0$ and the distal end $x=L$, respectively. A re-parameterisation of these properties to reduce the total dimensionality is presented next.
For all the aforementioned three categories of vessels, a unified parameterisation is proposed, which also results in reduction in the dimensionality associated with the description of Category 1 vessels. 
This parameterisation includes description of the mechanical properties (Section \ref{sec_material_properties}), geometric properties (Section \ref{sec_geometric_properties}), and boundary conditions (Section \ref{sec_boundary_properties}).

\subsection{Mechanical properties}
\label{sec_material_properties}

As opposed to describing the property variations linearly and individually for every vessel segment, see Equation \eqref{eq_linear_profile}, it proposed to use exponential variations, which allows for a simultaneous description of many successive vessels, thus reducing the total dimensionality.
Such re-parameterisation of the mechanical property $\beta$ for the three categories of vessels is as follows: \\

\noindent \textbf{Category 1:} These vessels have continuous variation in properties between successive vessels. Thus, successive vessels can be lumped together into chains and a single parameterisation be adopted. 
The 33 vessel segments (Category 1 vessels described in Section \ref{section_parameterisation} and labelled in Figure \ref{fig_vessel_chains} as cat.~1),  that meet this description can be lumped into the following five chains:
\begin{itemize}[leftmargin=*]
\item \textbf{The aortic chain.} This chain of vessel segments includes the first to forth aortic arch segments; the first to sixth thoracic aorta segments; and the first to fifth abdominal aorta segments. 
\item \textbf{The right and left arm chains.} These chains of vessels include the first and second subclavian segments; the axillary artery; and the brachial artery. 
\item \textbf{The right and left leg chains.} These chains of vessels include the external iliac; the first and second femoral segments; and the first and second popliteal segments.
\end{itemize}

\noindent The $\beta$ profiles for an entire vessel chain can now be described using a single function as opposed to separate functions (and hence separate parameters) for each individual vessel. An appropriate choice
for such intra-vessel variation is
\begin{equation}
\beta(x)=\beta_{0} \exp(- \Omega_d x),
\label{eq_property_profile}
\end{equation}
where $\beta(x)$ and $\beta_{0}$ represent the material properties of the chain at the spatial coordinate $x$ and the proximal end of the chain respectively; and $\Omega_d\geq 0$ represents a decay parameter. \\

\noindent \textbf{Category 2:}
 These vessels do not form part of any chain but show an intra-vessel property variation. For consistency and uniformity, their variation is also described by the exponential re-parameterisation of Equation \eqref{eq_property_profile}. These vessels are  labelled in Figure \ref{fig_vessel_chains} as cat.~2.\\
 
\noindent  \textbf{Category 3:} 
These vessels are also not part of a chain and have constant intra-vessel property variation. Again, for uniformity, the re-parameterisation of Equation \eqref{eq_property_profile} is used, with the added explicit condition that $\Omega_d=0$. These vessels are  labelled in Figure \ref{fig_vessel_chains} as cat.~3.\\

\noindent With the above unified description for the three vessel categories, $\beta$ properties can be hierarchically assigned.

\subsubsection{Hierarchical property assignment}
\label{sec_hierarchial_assignment}
Proximal-distal coherence dictates that the $\beta$ properties at the distal end of a vessel segment must be greater than or equal to the corresponding property at the proximal end of any subsequent daughter segments \cite{blanco2015anatomically}. To ensure this a hierarchical procedure is adopted.
The $\beta$ properties of the arterial network are initialised by explicitly assigning a value at the inlet of the first aortic arch segment. The property profile of the aortic chain, and so consequently the vessel wall mechanical property at the proximal and distal ends of each segment within the chain, is computed using Equation \eqref{eq_property_profile}. The properties at the proximal end of any vessel segments branching from the aorta are computed by applying a scaling term to the corresponding property at the distal end of the parent aortic segment: 
\begin{equation}
\beta_{0}^{d}=\beta_{L_p}^{p}\Omega_s,
\label{eq_parent_daughter_scaling}
\end{equation}
where $\beta_{0}^{d}$ represents the property at the proximal end of the daughter segment, $\beta_{L_p}^{p}$ represents the property at the distal end of the parent segment, and $\Omega_s$ represents a daughter-parent scaling term, where:
\begin{equation}
0<\Omega_s \leq 1.
\label{eq_daughter_parent_bounds}
\end{equation}
The intra-vessel or intra-chain property profiles of all daughter branches bifurcating from the aorta are computed using Equation \eqref{eq_property_profile}. This process is sequentially repeated through each generation of the arterial network until all terminal boundaries are reached.

\subsection{Geometric properties}
\label{sec_geometric_properties}
Here, the re-parameterisation of geometric properties---reference radius and length---is presented.
\subsubsection{Arterial vessel reference radius}

Since the variation of reference radius needs the same requirements as those for the variation of $\beta$, the variation of $r_0$ 
is described exactly in the same manner as $\beta$, as presented in Section \ref{sec_material_properties}.  
Lumping arterial vessels into chains, where appropriate, and applying exponential functions to describe the properties $\beta$ and $r_0$ of all the vessels reduces the number of dimensions from 389 in the reference network to 269.

\subsubsection{Arterial vessel length}
\label{section_vessel_length}

Each arterial vessel requires specification of its length. To individually assign a length to every vessel requires 71 parameters, accounting for a large proportion of the 269 remaining dimensions. It is thus proposed to reduce the dimensionality of the network by applying a singular scaling term to the lengths of all the vessels. 
It is empirically found that behaviours, patterns, and overall variability observed in the pressure and flow-rate profiles  when allowing for maximum freedom to the length of vessels, \textit{i.e.} assigning independent lengths to each arterial vessel, are not lost when applying a singular vessel length scaling term to all the vessels. This analysis justifies the use of a single scaling term and can be found in Appendix \ref{Appendix_vessel_length}. With this assumption, the dimensionality reduces from 269 to 199.

\subsection{Boundary conditions}
The boundary conditions consist of the inlet flow rate and the terminal lumped parameters. These are parameterised as follows:
\label{sec_boundary_properties}

\subsubsection{Inlet flow-rate}
\label{sec_inlet_boundary}

The volumetric inlet flow-rate to the network, vessel `a' in Figure \ref{fig_vessel_chains}, is described using a Fourier series (FS) \cite{jones2021proof}:
\begin{equation}
Q_{\text{inflow}}(t)=\sum_{n=0}^N a_n \sin (n \omega t) + b_n \cos(n \omega t),
\label{eq_FS_rep_1}
\end{equation}
where and $a_n$ and $b_n$ represent the $n^{\text{th}}$ sine and cosine FS coefficients, respectively; $N=5$ represents the truncation order; and $\omega={2 \pi}/{T}$, with $T$ as the time period of the cardiac cycle. 
Thus, the following 11 parameters are required to specify the inlet flow-rate: $\{a_0=0, b_0, a_1, b_1,...,a_5, b_5\}$.

%
%

\subsubsection{Terminal lumped models}
\label{section_terminal_boundaries}

There are 29 terminal boundaries in the network. Each of these terminal boundaries is coupled to a Windkessel model. Each Windkessel model requires three parameters: two resistances, and a compliance. 
These Windkessel models therefore require specification of a total of 87 parameters.\\

 The above description (presented in Sections \ref{sec_material_properties}, \ref{sec_geometric_properties}, and  \ref{sec_boundary_properties})
  has 199 parameters. These represent 22 decay parameters $\Omega_d$, 78 daughter-parent scaling parameters $\Omega_s$, 87 Windkessel parameters at the outlets, 11 FS coefficients describing the inlet flow-rate, and 1 length scaling term. These 199 parameters represent a reduction of approximately 48\% with respect to the original description with 389 parameters. These  parameters are denoted individually by $\Omega_i$, for the $i^{\mathrm{th}}$ parameter, and collectively with the vector $\bm\Omega = [\Omega_1, \Omega_2, \cdots, \Omega_{199}]$. The final step in parameterisation is to describe all these 199 parameters relative to the reference network, and is described next.

\subsection{Scaling with respect to the reference network}

Instead of describing the networks by directly stating values of the 199 parameters, they are specified relative to the reference network.
\begin{equation}
\bm{\Omega} = \bm{\theta}^{\mathrm{T}}\;\bm{\Omega}_{\text{ref}},
\label{eq_scaling_space}
\end{equation}
where $\bm{\theta} = [\theta_1, \theta_2, \cdots, \theta_{199}]^{\mathrm{T}}$ represents the scaling vector in relation to the reference network  and $\bm{\Omega}_{\text{ref}}$ represents the vector of reference network parameters. 
Random sampling from the distribution of $\bm{\theta}$ will result in the VPD. This is described in the next section.


\section{Statistical modelling}
\label{section_stat_model}
The pulse-wave propagation model, denoted as $\mathcal{M}$, takes the parameters $\bm\theta$ as inputs and outputs the pressure and flow-rates, collectively denoted as the vector $\bm{y}$, at all the locations:
\begin{equation}
\bm{y} = \mathcal{M}(\bm{\theta}),
\label{eq_forward_model_1}
\end{equation}
The specific measurements are denoted by $\bm\tau$, which are essentially transformations of $\bm{y}$ corrupted by measurement noise $\bm{\mathcal{E}}$:
\begin{equation}
\bm{\tau} = \mathcal{H}(\bm{y})+ \bm{\mathcal{E}},
\label{eq_forward_model_2}
\end{equation}
where $\mathcal{H}$ represents the observation operator. Commonly, the observation operator $\mathcal{H}$ represents an identity transform of selected components of $\bm{y}$. 
The measurement error $\bm{\mathcal{E}}$ is typically assumed to be a zero-mean multivariate normal:
\begin{equation}
\bm{\mathcal{E}} \sim \mathcal{N}(\bm{0}, \bm{\Sigma}_{\text{error}}),
\label{equation_error}
\end{equation}
where $\bm{0}$ is a zero-vector and $\bm{\Sigma}_{\text{error}}$ represents the error covariance matrix (diagonal when errors are independent).\\[-5pt]

To create the VPD, random samples of $\bm\theta$ from their joint probability distribution $\mathrm{p}(\bm\theta)$ are required.
This distribution $\mathrm{p}(\bm\theta)$ should satisfy two broad pieces of information: first,  it should satisfy known physiological and geometric constraints, and second, it should be consistent with what is known about the distributions of measurements $\bm{\tau}$ from literature. The latter ensures that the when the physics-based model is run with the parameters distributed as $\mathrm{p}(\bm\theta)$, the resulting distributions of the measurements are close to those reported in real populations.

\subsection{Bayesian formulation}
\label{section_bayes_theorm}
A Bayesian formulation is adopted to find the posterior distribution of the parameters $\bm\theta$. This is suitable because the geometric and physiological constraints, in the form a prior, can be combined with the literature reported measurements, in the form of a likelihood, to result in a posterior distribution that considers both the pieces of available information. The Bayes' theorem allows such a combination naturally:
\begin{equation}
\underbrace{\text{p}\left(\bm{\theta} \mid  \bm{\tau} \right) }_{\text{posterior}}= \frac{
\overbrace{\text{p}\left( \bm{\tau}  |\bm{\theta}  \right) }^{\text{likeihood}} \;
\overbrace{\text{p}\left(\bm{\theta}\right) }^{\text{prior}}
}
{
\underbrace{\text{p}\left(\bm{\tau} \right)}_{\text{evidence}}
}.
\label{equation_Bayes}
\end{equation}
The prior distributions $\mathrm{p}(\bm\theta)$ are described in Section \ref{sec_bayes_prior_evidence} and are chosen to be either uninformative or weakly informative. This avoids imposition of strong prior beliefs which may influence the posterior. Their primary purpose is to impose appropriate bounds on the supports of the distributions based on geometric and physiological constraints. For the computation of the likelihood, the literature reported measurements are first represented as multivariate Normal distribution with mean $\bm{\mu}_{\text{lit}}$ and covariance $\bm{\Sigma}_{\text{lit}}$. Then, in the context of the statistical model, Equations \eqref{eq_forward_model_1}--\eqref{equation_Bayes}, it is assumed that $\bm\tau$ is measured to be equal to $\bm\tau = \bm{\mu}_{\text{lit}}$ and the error covariance is $\bm{\Sigma}_{\text{error}} = \bm{\Sigma}_{\text{lit}}$. The likelihood for individual measurements is described in Section \ref{sec_bayes_likelihood}. Finally, in Equation \eqref{equation_Bayes}, the evidence term is a normalising constant, which is not of interest and need not be evaluated.

\subsection{Prior distributions}
\label{sec_bayes_prior_evidence}
 Note that the network parameterisation is relative to the reference network, see Equation \eqref{eq_scaling_space}, and hence the distributions are for the scaling terms $\bm\theta$.
Since scarce information is available about the variations of the parameters, the prior distributions are primarily constructed based on the bounds that should be observed in the parameters. In future studies, another approach which may be employed to assign tighter prior distributions could be the use of empirical relationships between the parameters and weight, age, height, etc., see for example \cite{chakshu2020towards}. The following three types of uninformative or weakly informative priors \cite{bernardo2009bayesian} are employed:\\[-5pt]

\begin{figure*}
\centering
\includegraphics[width=6in]{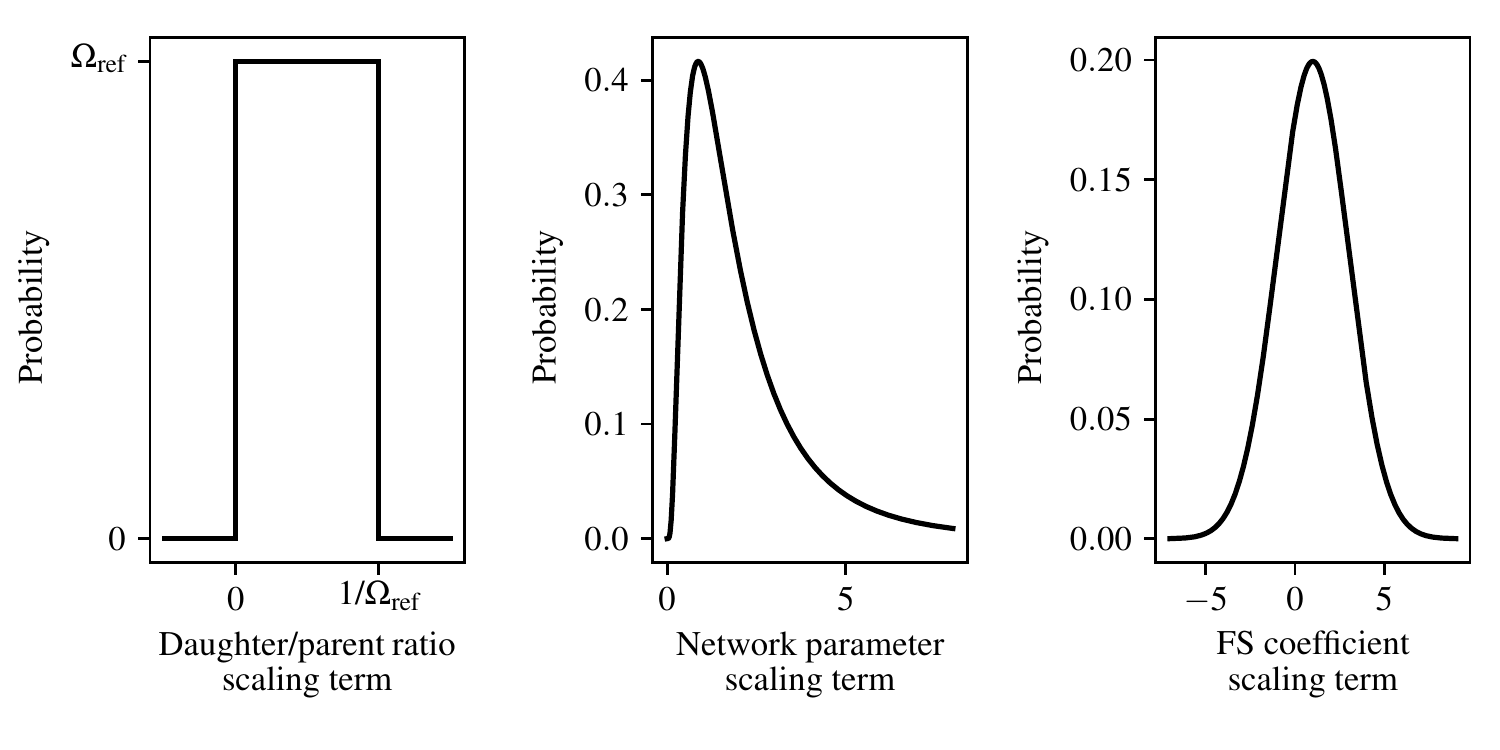}
\caption{The three type of prior distributions used for all arterial network scaling parameters. 
}
\label{fig_prior_probabilities}
\end{figure*}

\noindent \textbf{1. Bounded parameters:} A uniform prior distribution is chosen for all the scaling parameters which are supported on a bounded interval. The daughter-parent ratio, see Equations \eqref{eq_parent_daughter_scaling}, is the only example of such a parameter. For the  $i^{\mathrm{th}}$ pair of vessels with reference parent-daughter ratio $\Omega_{i,\text{ref}}$, it can be seen from Equations  \eqref{eq_daughter_parent_bounds} and \eqref{eq_scaling_space} that  the corresponding scaling term $\theta_i$ is bounded between 0 and  $1/\Omega_{i,\text{ref}}$.
The prior probability density function (PDF) is thus:
 \begin{equation}
  \text{p}(\theta_i) =
  \begin{cases}
    \Omega_{i,\text{ref}} & \text{if} \quad\theta_i \in (0, 1/\Omega_{i,\text{ref}}]\\
    0 & \text{otherwise}\\
  \end{cases},
 \label{eq_classify_OVA}
\end{equation}
and shown in the left plot of Figure \ref{fig_prior_probabilities}.   \\[-5pt]

\noindent \textbf{2. Semi-bounded parameters:}  A log-normal prior distribution is chosen for all the scaling parameters with semi-infinite support.  These parameters with a lower bound of zero and no upper bound are: 
\begin{itemize}[leftmargin=*]
\item initial values of radii $r_0$ and vessel wall mechanical properties $\beta$ at the inlet of the aorta (see Sections  \ref{sec_material_properties}, \ref{sec_hierarchial_assignment}, and  \ref{sec_geometric_properties});
\item  the decay terms used within all exponential property profiles (see Sections \ref{sec_material_properties}, and  \ref{sec_geometric_properties}); 
\item the length of arterial vessels (see Section \ref{section_vessel_length}); and
\item the terminal boundary Windkessel model parameters (see Section \ref{section_terminal_boundaries})
\end{itemize}
The prior PDF of the $i^{\mathrm{th}}$ such parameter is thus described by the following log-normal distribution:
\begin{equation}
\text{p}(\theta_i) = \frac{1}{\theta_i\sigma_i\sqrt{2\pi}}\exp\left(-\frac{\left(\ln{(\theta_i)}-\mu_i \right)^2}{2\sigma_i^2}\right),
\end{equation}
where $\mu_i$ and $\sigma_i$ represent the mean and standard deviation of the underlying Normal distribution, i.e. $\ln{(\theta_i)} \sim  \mathcal{N}(\mu_i, \sigma_i^2)$.
%
This log-normal distribution is created with large variance, resulting in a weakly informative prior. For all the scaling parameters $\mu_i$ is set to 0.5 and and $\sigma_i^2$ is set to 0.8, respectively. This PDF is shown in the middle plot of Figure \ref{fig_prior_probabilities}. \\

\noindent \textbf{3. Unbounded parameters:} A normal prior distribution is chosen  for all the scaling parameters with infinite support. 
The only such parameters are the inlet flow-rate FS coefficients (see Section \ref{sec_inlet_boundary}).
The prior PDF of the $i^{\mathrm{th}}$ FS coefficient with mean $\mu_i$ and standard deviation $\sigma_i$ is thus described as:
\begin{equation}
\text{p}(\theta_i) = \frac{1}{\sigma_i \sqrt{2 \pi}}\exp \left(-\frac{1}{2}\left(\frac{\theta_{i}-\mu_i}{\sigma_i}\right)^2\right).
\end{equation}
While the FS scaling parameters do have infinite support, they are expected to be within physiological range and hence within several multiplications of the reference FS coefficients. Thus, $\mu_i=1$ and a standard deviation $\sigma_i = 2$ is set. This PDF is shown in the right plot of Figure \ref{fig_prior_probabilities}.
%
It is assumed that all the priors are independent of each other, thus the joint prior PDF is:
\begin{equation}
\text{p}(\bm{\theta})= \prod_{i=1}^{N} \text{p}(\theta_i),
\label{eq_prior}
\end{equation}
where $N=199$ is the total number of scaling parameters.

\subsection{Likelihood}
\label{sec_bayes_likelihood}
The likelihood corresponds to the literature reported measurements. These measurements can be categorised as follows

\subsubsection{Scalar pressure and flow-rate measurements}
\label{sec_discrete_pressure_and_flowrate}
Pressure (diastolic and systolic) and flow-rate (average) measurements at  locations (radial artery, ascending aorta, common carotid, and femoral artery) reported in literature are incorporated into the likelihood. Their statistics and sources are shown in Table \ref{table_target_distributions}. When multiple measurements are available for the same quantity, they are pooled together into a single mean and variance \cite{rudmin2010calculating}. For such a measurement $\tau_i$ that was measured (or pooled together) to be $\mu_{i, \text{dis}}$ with standard deviation  $\sigma_{i, \text{dis}}$, the likelihood is:
\begin{equation}
\begin{split}
\text{p}\left( \tau_i \!=\! \mu_{i,\text{dis}}  |\bm{\theta}  \right)
 = \frac{1}{\sigma_{i, \text{dis}} \sqrt{2 \pi}}
\exp \left(-\frac{1}{2}\left(\frac{\mu_{i,\text{dis}}-\mathcal{H}_{i}(\mathcal{M}(\bm{\theta}))}{\sigma_{i, \text{dis}}}\right)^2\right),
\label{eq_likelihood_pressure_flow_mag_ind}
\end{split}
\end{equation}
where $\mathcal{H}_{i}$ represents the observation operator (see Section \ref{section_stat_model}) that extracts the $\tau_i$ component of the model output. Denoting all such discrete measurements collectively with the vector $\bm{\tau}_{\text{dis}}$, the combined likelihood with the assumption of independence is
\begin{equation}
\text{p}\left( \bm{\tau}_{\text{dis}} =\bm{\mu}_{\text{dis}}  |\bm{\theta}  \right) = \prod_{i} \text{p}\left( \tau_i \!=\! \mu_{i,\text{dis}}  |\bm{\theta}  \right),
\label{eq_likelihood_pressure_flow_mag}
\end{equation}

\begin{table*}
\begin{center}
\def\arraystretch{1.2}
\begin{tabular}{|c | c | c c c c |}
\hline
\textbf{Measurement} & \textbf{Measurement } & \textbf{No. of } & \textbf{Measured} & \textbf{Measured } & \textbf{Sources}\\
\textbf{location} & \textbf{type} & \textbf{patients} & \textbf{mean} $(\mu_{\text{dis}})$ & \textbf{std.}  $(\sigma_{\text{dis}})$ & \\
\hline
\multirow{ 2}{*}{Radial artery} & Diastolic pressure (mmHg)& \multirow{2}{*}{71} & 65 & 12.84 & \multirow{ 2}{*}{\cite{pauca1992does}, \cite{chen1997estimation}}\\
& Systolic pressure (mmHg)& & 123 & 22.75 &\\
\hline
\multirow{ 2}{*}{Ascending aorta} & Diastolic pressure (mmHg)& \multirow{2}{*}{69} & 65 & 7.14 &  \multirow{ 2}{*}{\cite{pauca1992does}, \cite{murgo1980aortic}}\\
& Systolic pressure (mmHg)& & 103 & 8.24 & \\
\hline
Common carotid & Diastolic pressure (mmHg)& \multirow{2}{*}{134} & 75.58 & 6.01 &  \cite{sugawara2000relationship}, \cite{arndt1968diameter},\\
artery & Systolic pressure (mmHg)& & 122.78 & 12.01 & \cite{hansen1995diameter} \\
\hline
Femoral artery & \multirow{ 2}{*}{Average flow-rate ($ml/$sec)}& \multirow{ 2}{*}{63} & \multirow{ 2}{*}{5.84} &  \multirow{ 2}{*}{2.11} &\multirow{ 2}{*}{\cite{maroto1993brachial} \cite{lewis1986measurement}}\\
(left and right) & &  &  &   &\\
\hline

\end{tabular}
\caption{The measurement of the discrete pressure and flow-rate taken from literature. When multiple source measurements are available for the same quantity, they are pooled together into a single mean $\mu_{\text{dis}}$ and variance $\sigma_{\text{dis}}$ \cite{rudmin2010calculating}.} 
\label{table_target_distributions}
\end{center}
\end{table*}

\subsubsection{Time varying inlet flow-rate and cardiac output measurement}
\label{sec_timedomain_inflow}
\begin{figure*}
\centering
\includegraphics[width=4.5in]{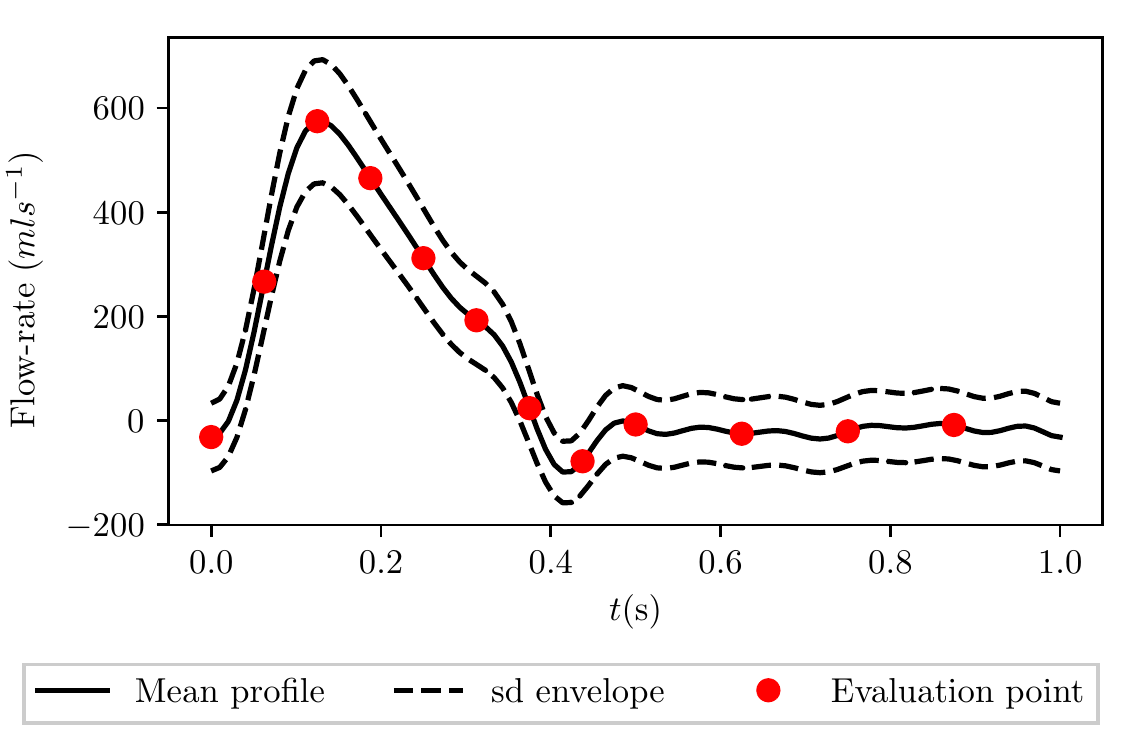}
\caption{The measured time domain inlet flow-rate profile taken from the ADAN network is shown in solid black. The standard deviation (sd) envelope computed using Equation \eqref{equation_flow_sd} are shown with dashed lines and the discrete time points at which the likelihood is evaluated is shown in red circles.}
\label{fig_eval_points}
\end{figure*}

Some time-varying information about the behaviour of pressure and flow-rates over the cardiac cycle is necessary to obtain a physiologically realistic posterior.  Statistics on any such time-varying behaviour is not reported in literature to the authors' knowledge. Thus, a time-varying pseudo-measurement is constructed by combining the reference (ADAN network \cite{boileau2015benchmark}) time varying inlet flow-rate $Q_{\text{ref}}(t)$ and a measurement of average cardiac output \cite{ihlen1984determination, chandraratna1984determination}. This pseudo-measurement $Q_{\text{meas}} (t) $  is described by a Gaussian Random Field (GRF, also referred to as a Gaussian Process) \cite{rasmussen2004gaussian, ibragimov2012gaussian} with the following mean $\mu_{\text{pseudo}} (t)$ and standard deviation $\sigma_{\text{meas}} (t)$
%
\begin{equation}
\label{eq_grf_mean}
\mu_{\text{meas}} (t) = Q_{\text{ref}}(t),
\end{equation}
\begin{equation}
 \sigma_{\text{meas}}\left(t\right)=
\bigg(1+\frac{Q_{\text{ref}}(t)-\text{min}({Q}_{\text{ref}})}{\text{max}({Q}_{\text{ref}})-\text{min}({Q}_{\text{ref}})}\bigg)\delta,
 \label{equation_flow_sd}
 \end{equation}
where $\delta$ is a positive scaling parameter. The scaling of $\sigma_{\text{meas}} (t)$ is to ensure that i) the standard deviation is positive, and ii) the variance is proportional to the magnitude of the flow rate (i.e. the percentage variance with respect to the magnitude remains fixed). The covariance between the pseudo-measurements at any two times $t_i$ and $t_j$ is described through a periodic kernel \cite{rasmussen2004gaussian}
%
 \begin{equation}
 \begin{split}
 \text{cov}\left( Q_{\text{meas}}(t_i), Q_{\text{meas}}(t_j) \right)=
 \sigma_{\text{meas}}\left(t_1\right)\sigma_{\text{meas}}\left(t_2\right)
\exp\bigg(\frac{-2\;\text{sin}^2\left(\pi |t_i-t_j|/T\right)}{(\upsilon T)^2}\bigg),
\end{split}
\label{eq_kernel_used}
\end{equation}
where $T$ represents the cardiac period, and $\upsilon$ represents the ratio of the correlation length to the cardiac period. Through a trial-and-error analysis, this ratio is set to $\upsilon = 0.5$ to produce a physiologically realistic variation in the inlet flow-rate profile. 
%
The scaling parameter $\delta$ is tuned to the average cardiac output (CO) measurement: a mean of 98 ml/sec and standard deviation of 35.55 ml/sec  \cite{ihlen1984determination, chandraratna1984determination}. Through a sweep of $\delta$, an optimal value of $\delta=59.09$ is found, when the GRF produces a mean CO of 97.8 ml/sec and a standard deviation of 34.84 ml/sec. This GRF for $Q_{\text{meas}} (t) $ is shown in Figure  \ref{fig_eval_points} with the  mean and $\pm$ one standard deviation profiles.


The GRF is evaluated at $k=12$ time points, shown in Figure \ref{fig_eval_points}, for the computation of the likelihood. Denoting these evaluation time points as $\{t_1, t_2, \cdots, t_k\}$,
the vector of measurements at these times with $\bm{\tau}_{\text{inflow}}$,  the mean GRF vector as $ \bm{\mu}_{\text{inflow}} = [\mu_{\text{meas}} (t_1), \mu_{\text{meas}} (t_2), \cdots, \mu_{\text{meas}} (t_k)]$ computed through Equation \eqref{eq_grf_mean},  and the vector of inlet flow rates produced by the network parameters through Equation \eqref{eq_FS_rep_1} as $\bm{Q}_{\text{inflow}} = [Q_{\text{inflow}}(t_1), Q_{\text{inflow}}(t_2), \cdots, Q_{\text{inflow}}(t_k)]$, the likelihood for flow-rate can be written as:
\begin{equation}
\begin{split}
\text{p} \left( \bm{\tau}_{\text{inflow}}  = \bm{\mu}_{\text{inflow}}  |\bm{\theta}  \right) = 
 (2 \pi)^{-\frac{k}{2}}\text{det}(\bm{\Sigma})^{-\frac{1}{2}}
 \exp \bigg(-\frac{1}{2} \bm{\xi}^{\text{T}}
\bm{\Sigma}^{-1} \bm{\xi} \bigg),
\end{split}
\label{eq_inflow_likelihood}
\end{equation}
where
\begin{equation}
 \bm{\xi} = \bm{\mu}_{\text{inflow}}- \bm{Q}_{\text{inflow}},
\end{equation}
and $\bm{\Sigma}$ is the GRF covariance matrix whose $i^{\text{th}}$ row and $j^{\text{th}}$ column element $\Sigma_{i,j} = \text{cov}\left( Q_{\text{meas}}(t_i), Q_{\text{meas}}(t_j) \right)
$ can be computed from Equation \eqref{eq_kernel_used}.

\subsubsection{Vessel length measurement}
\label{sec_vessel_length_measurement}
As mentioned in Section \ref{section_vessel_length}, the length of the vessels are parameterised by a single scaling term relative to the reference network. Since statistics of any direct vessel length measurement are not available, it is assumed that the lengths of arterial vessels are directly proportional to the height of a subject. A study of 25,945 twins from eight countries reported the mean and standard deviation of the height of the full cohort to be 172.0 cm and 9.308 cm respectively \cite{silventoinen2003heritability}. Since the reference arterial network has a patient of height 170 cm, the measurement data corresponds to a mean of $\mu_{\text{len}} = 1.0118$ and the standard deviation of  $\sigma_{\text{len}} = 0.0548$ for the vessel length scaling term. Denoting this measurement as $\tau_{\text{len}}$, the likelihood for the scaling parameters $\theta_{\text{len}}$ is thus
\begin{equation}
\begin{split}
\text{p}\left(\tau_{\text{len}} = \mu_{\text{len}}  |\bm{\theta}  \right) =  
 \frac{1}{ \sigma_{\text{len}} \sqrt{2 \pi}}
\exp \left(-\frac{1}{2}\left(\frac{ \mu_{\text{len}} -  \theta_{\text{len}}} { \sigma_{\text{len}}}\right)^2\right),
\end{split}
\label{equation_length_measurement}
\end{equation}

Assuming all the measurements to be independent, the combined likelihood in Equation \eqref{equation_Bayes} is:
\begin{equation}
\begin{split}
\text{p}\left( \bm{\tau}|\bm{\theta}  \right)  = \text{p}\left( \bm{\tau}_{\text{dis}}\! = \! \bm{\mu}_{\text{dis}}  |\bm{\theta}  \right) \times 
\phantom{xx}\text{p}\left( \bm{\tau}_{\text{inflow}}\! = \!\bm{\mu}_{\text{inflow}}  |\bm{\theta}  \right) \times \text{p}\left(\tau_{\text{len}}\! = \!\mu_{\text{len}}  |\bm{\theta}  \right),
\end{split}
\label{eq_likelihood}
\end{equation}
where the RHS terms can be computed from Equations \eqref{eq_likelihood_pressure_flow_mag}, \eqref{eq_inflow_likelihood}, and \eqref{equation_length_measurement}. 

In summary, the prior distributions for the 199 network parameters are specified in Section \ref{sec_bayes_prior_evidence}. These priors will be modified through the likelihood---specified by 7 scalar measurements of pressure and flow-rate, 1 time-varying flow-rate profile evaluated at 12 time points in the cardiac cycle, and a measurement of vessel lengths---to yield the posterior distribution of the parameters through Equation \eqref{equation_Bayes}. Random sampling from this posterior will result in the virtual patient database. The sampling procedure is presented next.

\section{Sampling from the posterior distribution}
With the prior and the likelihood specified, the posterior distribution is given by Equation \eqref{equation_Bayes}, and can be evaluated at any given $\bm{\theta}$ up to a normalising constant (the evidence term in the denominator of the equation). Sampling from this analytically intractable posterior is achieved through the Markov-chain Monte Carlo method \cite{geyer1992practical, gilks2005m}, which is a widely used method to sample from unnormalised distributions.
%
%

\subsection{Markov-chain Monte Carlo }
The particular MCMC method employed is the Metropolis-Hastings algorithm \cite{chib1995understanding}. Starting from an initial sample, a chain of samples is built sequentially. Denoting the $k^{\text{th}}$ sample as $\bm{\theta}^{(k)}$, a candidate for the next sample is generated by a proposal distribution which depends only on $\bm{\theta}^{(k)}$. This candidate sample $\bm{\theta}^{*}$  is proposed by sampling from the following multivariate Normal distribution:
\begin{equation}
\label{eq_proposal_sample}
\bm{\theta}^{*} \sim \mathcal{N}(\bm{\theta}^{(k)}, \bm{\Sigma}_{\text{step}}),
\end{equation} 
where $\bm{\Sigma}_{\text{step}}$ is the covariance matrix, whose $i^{\text{th}}$ row and $j^{\text{th}}$ column element $\Sigma_{\text{step}, ij}$ is given by 
\begin{equation}
{\Sigma}_{\text{step}, ij} = 
\begin{cases}
0 & i=j \\
\sigma^2_{\text{step},i} & i=j
\end{cases},
\end{equation}
where $\sigma^2_{\text{step},i}$ represents the marginal variance for the $i^{\text{th}}$ element of $\bm{\theta}^*$.
This proposed candidate $\bm{\theta}^{*}$ is either accepted or rejected based on the following ratio $\alpha$ of posterior probabilities at $\bm{\theta}^{*}$ and $\bm{\theta}^{(k)}$
\begin{equation}
\alpha  = \frac{ \text{p}\left(\bm{\theta} = \bm{\theta}^{*} \mid  \bm{\tau} \right) }   { \text{p}(\bm{\theta}  = \bm{\theta}^{(k)}\mid  \bm{\tau} ) }
\end{equation}
which, through the Bayes' rule of Equation \eqref{equation_Bayes} applied to both the numerator and the denominator, can be written as
\begin{equation}
\alpha =   
\frac{\text{p}\left(\bm{\tau} |  \bm{\theta}=\bm{\theta}^{*} \right) \text{p}\left(\bm{\theta} = \bm{\theta}^{*}\right)}{\text{p}\left( \bm{\tau} | \bm{\theta}=\bm{\theta}^{(k)} \right) \text{p}\left(\bm{\theta} = \bm{\theta}^{(k)}\right)}.
\end{equation}
The first term and the second term in both the numerator and the denominator are the likelihoods and priors at the candidate and current point, respectively, and can be computed by Equations \eqref{eq_likelihood} and \eqref{eq_prior}. To accept or reject the proposed candidate $\alpha$ is compared to a random number $\gamma$ drawn from a uniform distribution between 0 and 1. If $\alpha \geq \gamma$, the proposed candidate, $\bm{\theta}^{*}$ is accepted and the chain progress with $\bm{\theta}^{(k+1)} = \bm{\theta}^{*}$. Otherwise, the proposed candidate is rejected and a new candidate is proposed through Equation \eqref{eq_proposal_sample}. This process is repeated until the chain is sufficiently long. After discarding the first few thousands of samples from this chain, referred as the burn-in period \cite{raftery1996implementing}, the remaining samples  are independent samples from the posterior, and form the virtual patient database.

The acceptance rate, \textit{i.e.} the proportion of candidate sample points that are accepted, is tuned by varying $\sigma^2_{\text{step},i}$. A high variance results in the Markov chain exploring the posterior density quickly, with a high proportion of the candidate points being rejected. A low variance results in slow movement of the chain moving but with a high proportion of candidate points being accepted. An optimum acceptance rate to balance these two apposing behaviours is stated in literature to be 0.234 \cite{roberts1997weak}. It is empirically found that using a standard deviation of 0.0375, equivalent to 3.75\% of the reference values, for the scaling terms applied to the inlet flow-rate FS coefficients and that of 0.025, equivalent to 2.5\% of the reference values, for all other scaling terms results in an acceptance rate of approximately 0.2.

\subsection{Pre-fetching}
\begin{figure*}
\centering
\includegraphics[width=4in]{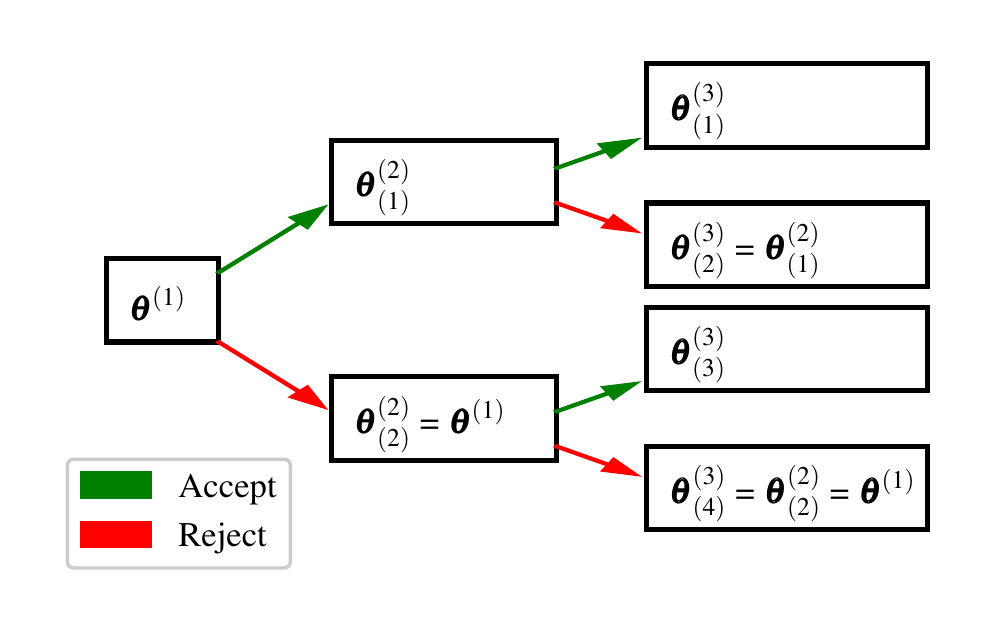}
\caption{An example of an MCMC decision tree across two iterations of a chain.  $\theta_{(j)}^{(i)}$ represents the $j^{\text{th}}$ possible set (candidates) of the arterial network parameter scaling terms at the $i^{\text{th}}$ step of the chain. Green branches denote the path of the chain taken if a candidate is accepted, while red branches show the path of the chain if a candidate is rejected.}
\label{fig_decision_tree}
\end{figure*}

The MCMC algorithm outlined above is inherently sequential---each subsequent sample depends on the previous sample and hence the chain grows only one sample at a time. Generating a long chain this way thus leads to very high computational run times, since computation of the likelihood requires running a pulse-wave propagation simulation at each step. To alleviate this issue and achieve some level of parallelisation, the concept of pre-fetching \cite{brockwell2006parallel, strid2010efficient} is used. The central idea is to think of MCMC accept-reject decisions in the form a of a decision tree. Figure \ref{fig_decision_tree} shows such a decision tree with a depth of $\eta = 2$. This results in $2^\eta$ leaf-nodes at the end of the tree. A careful consideration of the decision tree (see Figure \ref{fig_decision_tree}) shows that the only unique parameters in the entire decision tree correspond to those at the leaf-nodes. Thus, computational simulations for all the leaf-node parameters can be run simultaneously in-parallel. Then, $\eta$ steps of the MCMC algorithm can be taken together without any computational overhead by walking the decision tree. A tree depth of $\eta=4$ is used in this study.


%
%

\section{Results and discussion}
The VPD is created by generating samples through the MCMC algorithm. The results for the MCMC method and the VPD are presented and discussed here.

\subsection{Burn-in analysis}

It is common for the initial iterations of an MCMC chain to be considered as the ``burn-in'' period, during which the chain converges to an equilibrium distribution. This burn-in period is, therefore, discarded from the final sampled posterior distribution. To estimate this  burn-in, trace plots for every parameter at all the iterations of the MCMC  are plotted.  An example of a trace plot is shown for the scaling terms applied to the length of VPs arterial vessels in Figure \ref{fig_TP_length}, and all other trace plots are shown in Appendix \ref{appendix_MCMC_trace_plots}. To aid visualisation, these plots are thinned out by plotting every $100^{\text{th}}$ iteration of the chain. These trace plots are visually assessed to determine burn-in. Essentially, if there is an initial period where a clear migration of the chain is seen away from the sample used to start the chain, then the burn-in is selected to be until this initial migration is complete.  These trace plots also show good movement of the chain around the distribution, referred to as the ``mixing'' of the chain. Figures in Appendix \ref{appendix_MCMC_trace_plots} show that most parameters are initialised in a region of high density and hence no net migration is observed---the samples oscillate around a central mean value. Some interesting behaviours in the trace plots, including features that deviate from such a general/desired behaviour, are: 
 \begin{itemize}[leftmargin=*]
 \item The scaling term applied to $\beta$ at the inlet of the aortic chain are seen to centre around 5. The maximum scaling term applied is approximately 7.5.
 \item The scaling terms applied to the reference decay term of the left leg chain $\beta$ profile initially oscillates around a value of approximately 1 for 25,000 iterations, before migrating and oscillating around a value of approximately 7 (see Figure \ref{fig_TP_beta_chain} in Appendix). This may suggest that this distribution is multimodal. 
 \item The scaling terms applied to the reference decay term of the brachiocephalic trunk $\beta$ profile oscillates around a value of 1 for the first 40,000 iterations, before migrating and oscillating around a value of 9 (see Figure \ref{fig_TP_beta_ind_I}). The scaling terms applied to the reference decay term of the brachiocephalic trunk $r_0$ profile shows no complimentary behaviour and remains within the region of 0--4 throughout. This may again suggest a multimodal distribution.
 \item A spike is seen in the scaling terms applied to the compliance of the right external carotid Windkessel model at approximately 55,000 iterations (see Figure \ref{fig_TP_WK_armR} in Appendix). This behaviour is not seen within the  scaling terms applied to the compliance of the left external carotid Windkessel model which remains within the region of 0--3 throughout.
 \end{itemize}
  
Even though most of the parameters do not show a clear migration, an initial burn-in period of 10,000 iterations is chosen as a precautionary measure to minimise any effect the initial sample position may have. Once this burn-in period has been removed, the VPD contains 65,000 VPs. Along with the burn-in period, all VPs with negative average flow-rate within any arterial vessel are removed from the VPD. These VPs are removed from the VPD as it is physiologically unlikely for a patient to have negative average flow-rate in any vessel. Of the 65,000 post burn-in VPs, 12,857 are removed due to the presence of negative average flow-rate, reducing the VPD to 52,143.

%
 \begin{figure*}
\centering
\includegraphics[width=6in]{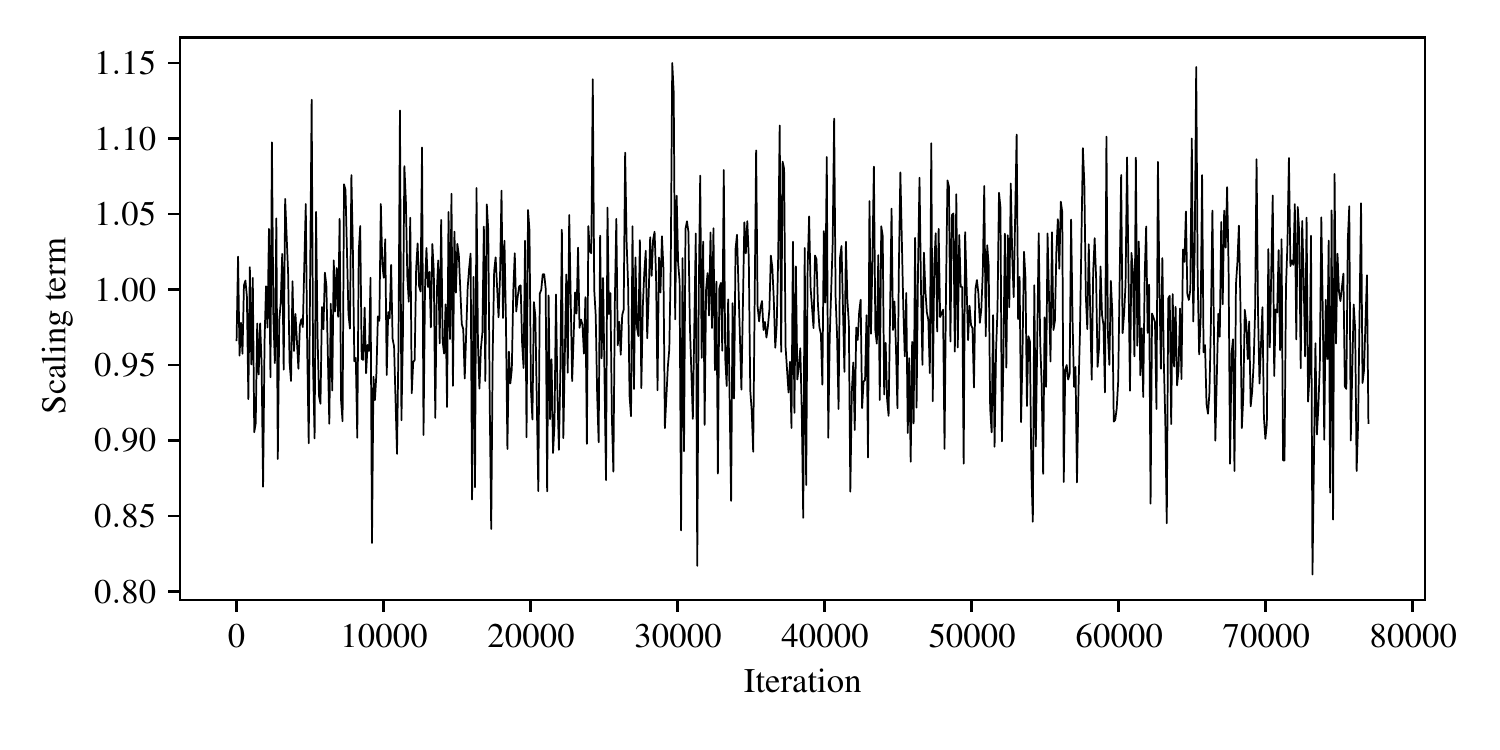}
\caption{MCMC trace plot of the scaling term applied to the length of the arterial vessels at every $100^{th}$ iteration. In this case, not net migration of the chain is observed.}
\label{fig_TP_length}
\end{figure*}

\subsection{Posterior vs. literature reported measurements}

\begin{figure*}
\centering
\includegraphics[width=5.5in]{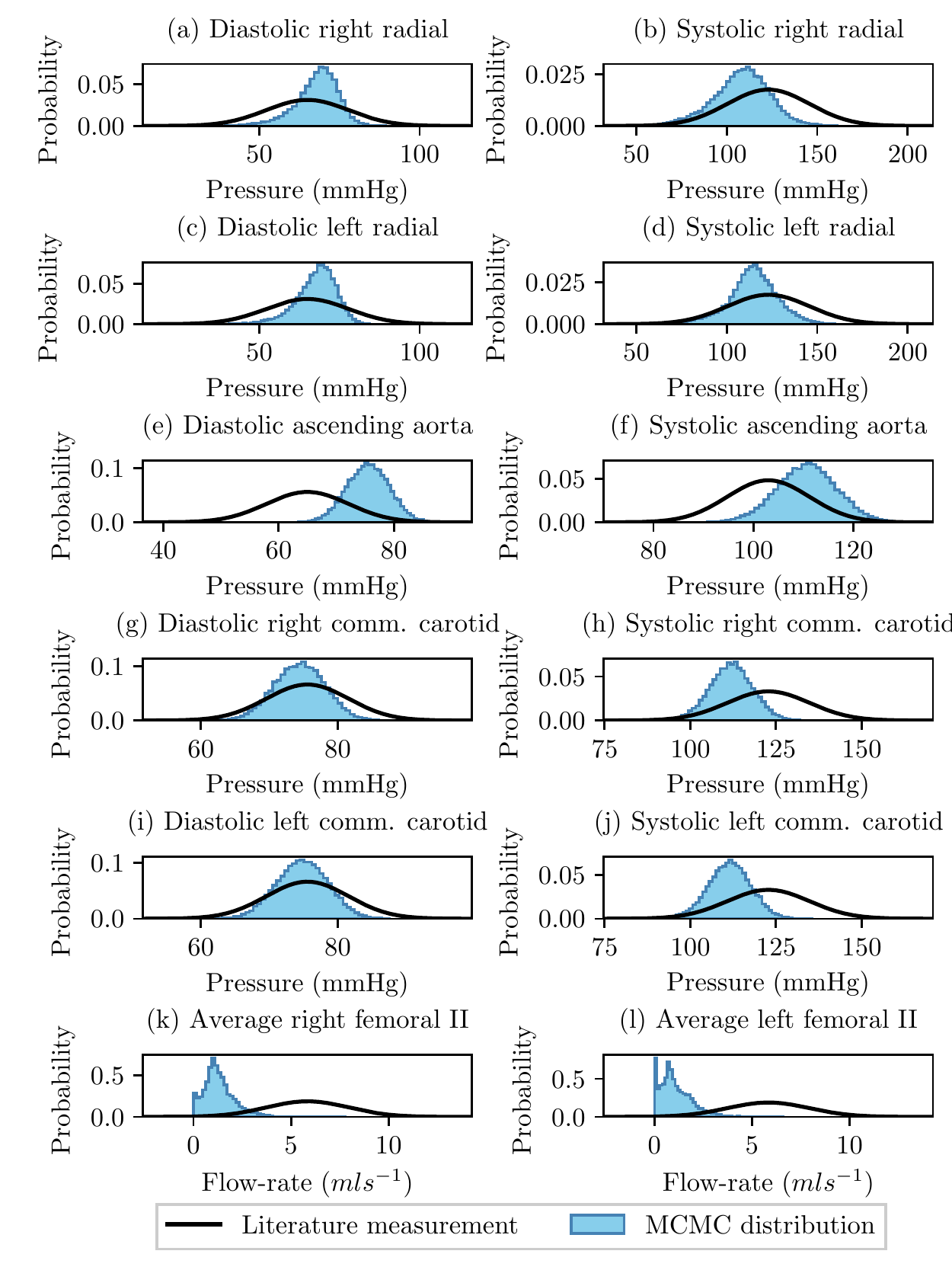}
\caption{Histograms of the MCMC distributions of the pressure and flow-rate measurements at all measurement locations. The  literature based measurements and associated error distribution at each location are overlaid in black. Diastolic and systolic pressure in the right radial artery shown in (a) and (b) respectively; the diastolic and systolic pressure in the left radial artery are shown in (c) and (d) respectively; the diastolic and systolic pressure in the ascending aorta are shown in (e) and (f) respectively; the diastolic and systolic pressure in the right common carotid artery are shown in (g) and (h) respectively; the diastolic and systolic pressure in the left common carotid artery are shown in (i) and (j) respectively; and the average flow-rate in the second segments of the right and left femoral artery are shown in (k) and (l) respectively.}
\label{fig_pressure_flow_target_comparison}
\end{figure*}

\begin{figure}[tb]
\centering
\includegraphics[width=2.5in]{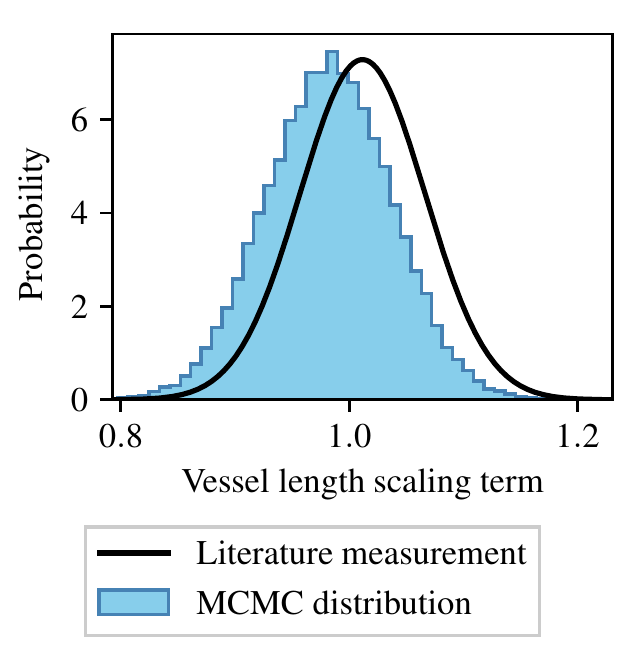}\caption{Histogram of the vessel length scaling terms assigned to VPs across the created VPD . The  literature based measurement of vessel length scaling terms is overlaid in black.}
\label{fig_length_scaling_hist}
\end{figure}

\begin{figure*}
\centering
\includegraphics[width=6in]{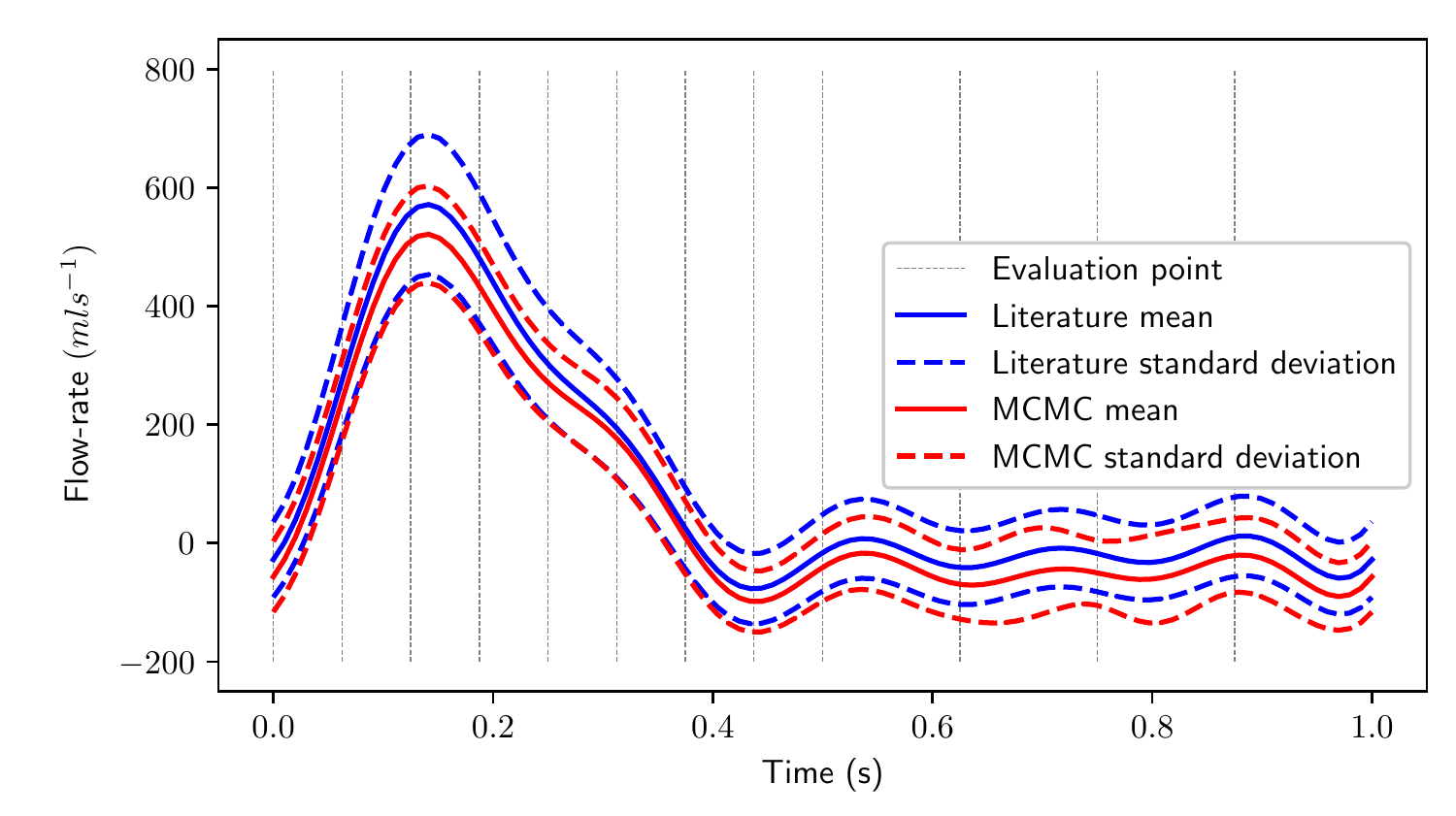}
\caption{Comparison of the empirical distribution of the time varying inlet flow-rate profiles between the MCMC samples and the pseudo-measurement created through a GRF. All standard deviation curves in dashed lines depict mean $\pm$ 1 standard deviation.}
\label{fig_inflow_full_profile}
\end{figure*}

The posterior distribution is a combination of the prior distribution and the likelihood term within Bayes' theorem. As the prior distributions are weakly informative or uninformative, the distributions of measurements produced by the VPD should be close to the literature based measurements incorporated through the likelihood. These are not necessarily identical though, as the prior distribution does correct the posterior to account for geometrical and physiological constraints. Furthermore, since the incorporated measurements are taken from several different sources, the posterior distribution resolves inconsistencies between such measurements, the physics of  pulse-wave propagation, and the constraints of the prior.
%
A comparison of the VPD pressure and flow-rate  distributions to the literature based measurements (see Table \ref{table_target_distributions} and Section \ref{sec_discrete_pressure_and_flowrate}) are shown in Figure \ref{fig_pressure_flow_target_comparison}. A similar comparison for the vessel length scaling term (see Section \ref{sec_vessel_length_measurement}) is shown in Figure \ref{fig_length_scaling_hist}. Finally, the statistics of the time varying inlet flow-rate profiles in the VPD are compared to the GRF (see Section \ref{sec_timedomain_inflow}) in Figure \ref{fig_inflow_full_profile}.





\begin{figure*}
\centering
\includegraphics[width=5in]{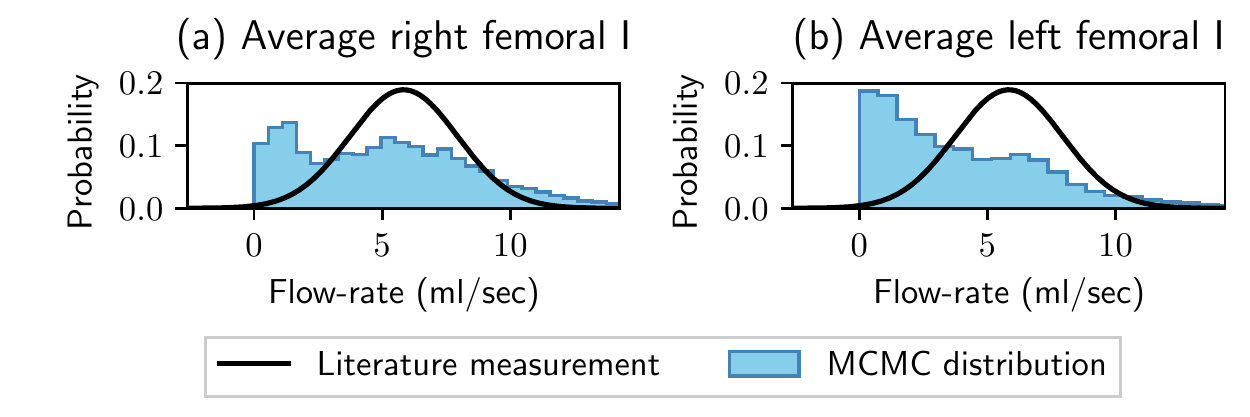}
\caption{Histograms of the MCMC distributions of the average flow-rate in the first segment of the right femoral artery (c), and the first segment of the left femoral artery (d). The literature based measurement of average flow-rate in the femoral  arteries and the associated error distribution is overlaid in black.}
\label{fig_femoral_I_II_comparison}
\end{figure*}

Generally, a good agreement between the scalar pressure and flow-rate measurements are seen in Figure \ref{fig_pressure_flow_target_comparison}. This agreement enforces confidence in the overall approach. However, for the average flow-rates in the left and right femoral arteries, larger than expected differences are observed. This is likely due to the following reasons:

\begin{itemize}[leftmargin=*]
\item A large inconsistency between the femoral flow-rates measurement and the other pressure measurements. Note that they are taken from different sources and hence not from the same population. Furthermore, there may be an inconsistency between the cardiac output measurement used for generating the pseudo-measurement for the time varying inlet flow-rate measurement (see Section \ref{sec_timedomain_inflow}).
%
\item The presence of more measurements for pressure as opposed to flow-rates. Since the likelihoods for all the measurements are weighted equally, it may be possible that the chain is influenced weakly by the few flow-rate measurements. Such an issue can be resolved in future studies by assigning variable weightings to the measurements. 
\item In the network the femoral arteries are split into the first (I) and second segments (II), shown in Figure \ref{fig_vessel_chains} by `u' and `v', respectively. The precise location at which the literature based femoral flow-rate measurement has been taken is unknown. In this study, it is assumed that this measurement was acquired at the centre of the second segment `v'. Since the first segment `u' bifurcates into `v' and another segment \emph{profunda femoris} (Figure \ref{fig_vessel_chains}), the flow-rate in `v' is smaller than that in `u'.  Thus, it is possible that the measured flow-rate was in `u'. A comparison of VPD flow-rate in this vessel `u' against the measurement is presented in Figure \ref{fig_femoral_I_II_comparison}, and shows a significantly better agreement.
\end{itemize}

Figure \ref{fig_length_scaling_hist} shows a good agreement between the vessel length scaling term in the VPD and the literature based measurement. The mean and standard deviation of this term in the VPD are 0.0551 and 0.9819, respectively, which compare well to those reported in literature with values of 0.0548 and 1.0118, respectively. 


Figure \ref{fig_inflow_full_profile} shows a good agreement between the statistics of the time varying inlet flow-rate profile in the VPD and the pseudo-measurement constructed through the GRF (Section \ref{sec_timedomain_inflow}). The mean profile closely follows that of the GRF throughout the cardiac cycle, with maximum difference of approximately at 10\% peak systole. Figure \ref{fig_inflow_full_profile} also shows $\pm$ one standard profiles for both the VPD flow-rate and the GRF pseudo-measurement. Throughout the cardiac cycle, the agreement and the trend is well captured in the VPD, with largest errors of approximately 10\% magnitudes at peak systole.

\subsection{Evaluation of individual VPs}
\begin{figure}[tb]
\centering
\includegraphics[width=6in]{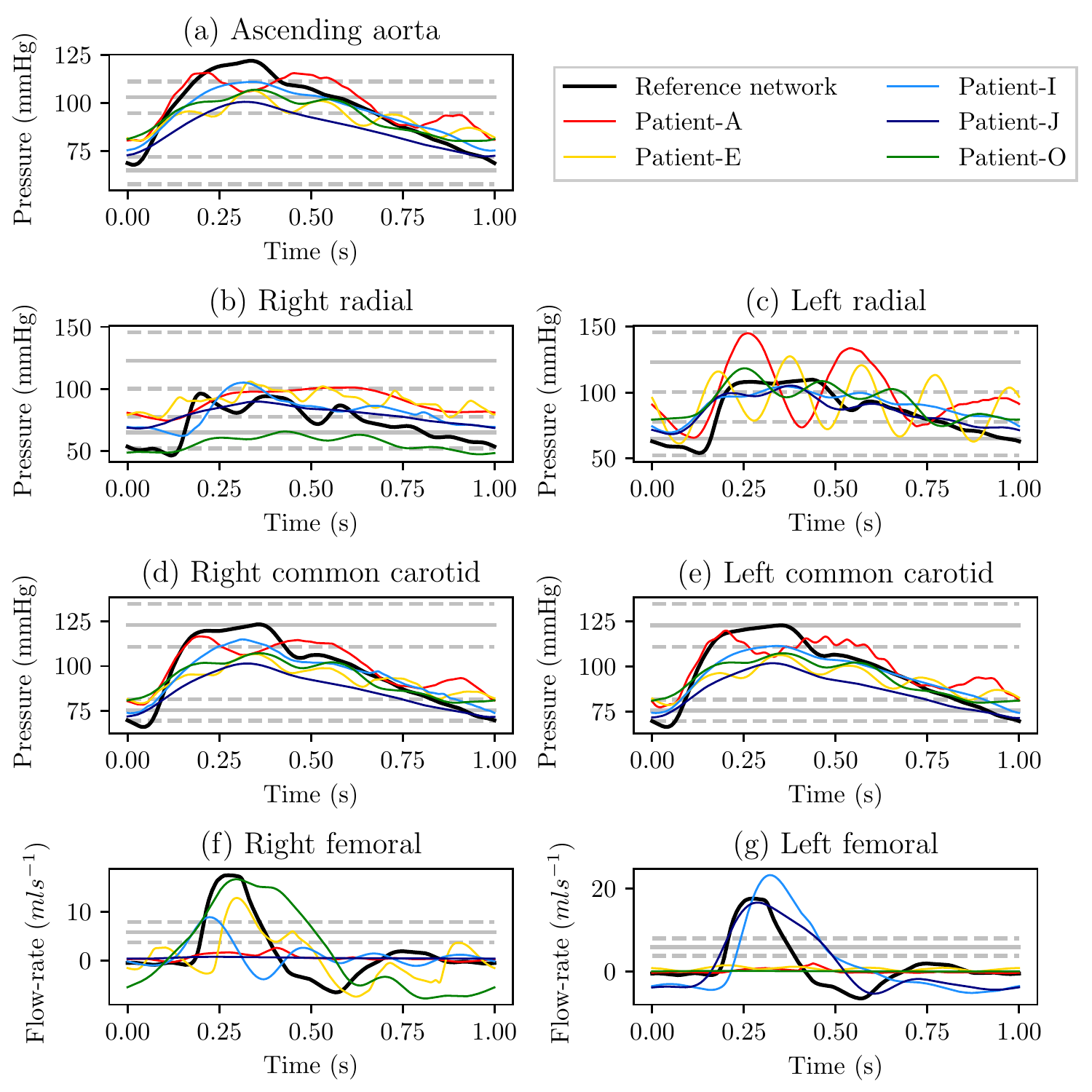}
\caption{Pressure profiles in the ascending aorta (a), pressure profiles in the right radial artery (b), pressure profiles in the left radial artery (c), pressure profiles in the right common carotid artery (d), pressure profiles in the left common carotid artery (e), flow-rate profiles in the right second femoral artery (f), and flow-rate profiles in the left second femoral artery (g). In each figure the profiles taken from the reference network are shown in black; and the literature reported measurements and associated variances are shown by the solid (mean) and dashed grey lines (mean $\pm$ 1 standard deviation) respectively.}
\label{fig_random_profiles_over_view}
\end{figure}

Random samples from the VPD are assessed  to gain further insights on the VPs and the behaviour of pressure and flow-rate profiles. Pressure profiles are examined at the ascending aorta; right and left radial arteries; and right and left common carotid arteries. Flow-rate profiles are examined at the right and left second femoral segments. 15 VPs are randomly drawn from the VPD (excluding the burn-in), and the pressure and flow-rate profiles associated with each are shown at all examination locations in Appendix \ref{appendix_pressure_flow}. Five VPs of interest are extracted from Appendix \ref{appendix_pressure_flow} and shown in Figure \ref{fig_random_profiles_over_view}.

%
%
 
Figures in Appendix \ref{appendix_pressure_flow} show that pressure profiles in proximal vessels, \textit{i.e.} the ascending aorta and two common iliacs, show more consistency and similarity to the reference profiles compared to the pressure and flow-rate profiles in distal vessels, \textit{i.e.} the radial artery pressure and the femoral artery flow-rate. This suggests that the pseudo-measurement of the time varying inlet flow-rate profile is sufficient to impose sufficient control over the shape of pressure and flow-rate profiles in the proximal vessels. As the spatial distance from the inlet of the aorta increases, an increases in the variability of pressure  and flow-rate profiles is observed. This is expected as the measurements that guide the posterior are only available at few locations. Thus, access to more measurements, distributed across the network in both space and time, will result in an even more realistic database as the likelihood at several such locations will guide the posterior.

Figure \ref{fig_random_profiles_over_view} shows a selection of undesirable behaviour in the VPD profiles. In the case of Patient-E, oscillatory behaviour is seen within all  the pressure and flow-rate profiles. This is caused by the shape of the inlet flow-rate profile prescribed for this patient. While this behaviour is not very common in the database, if needed its occurrence can be further reduced by imposing stronger correlations between inlet flow-rates at two time points through the parameter $\upsilon$ in Equation \eqref{eq_kernel_used} in the GRF.

Highly oscillatory pressure profiles are observed in the left radial artery, shown in Figure \ref{fig_random_profiles_over_view}(c), of Patient-A and Patient-E. In the case of VP Patient-A there is no significant oscillatory behaviour in any other profile, but the pressure waveform in the right radial artery is featureless with no clear systolic and diastolic points. 
This apparent over- and under-damping of the pressure profiles in the radial arteries may suggest an imbalance in the compliances or resistances of the right and left arms. This hypothesis of left-right imbalance is further supported by Figures \ref{fig_random_profiles_over_view}(f) and (g) when the femoral flow-rates are observed for Patient-I: the flow-rate in the right femoral artery shows a high-mean high-oscillation behaviour while in the left femoral artery it is low-mean with significantly lower oscillations. Similar behaviour is seen within Patient-J and Patient-O. In future studies, a symmetry metric that balances, while still allowing for some variability, the left and right side parameters for symmetric left and right side vessels may be implemented. In this study, it is chosen to filter out vessels that show a high left-right imbalance with an analysis that is described next.

\subsection{Filter for left-right imbalance}
\label{section_resistance_compliance}
\begin{figure}[tb]
\centering
\includegraphics[width=2.5in]{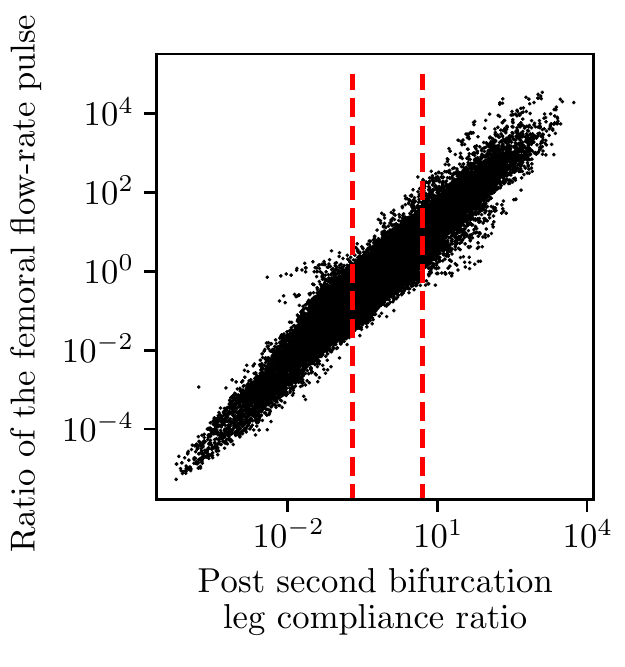}
\caption{Left-to-right femoral pulse ratio against the left-to-right compliance ratio of reduced network (up to vessel `u' in Figure \ref{fig_vessel_chains}). Filter range on the compliance ratio is shown with vertical lines.}
\label{figure_compliance_limits}
\end{figure}

The left-right lower extremity imbalance is assessed by reducing the network (see Section \ref{section_removal_vessels} and Figure \ref{fig_network_reduction}) up to the second bifurcation in the leg vessels, i.e up to the end of vessels labelled `u' in Figure \ref{fig_vessel_chains}. To assess the left-right imbalance, the ratio of femoral pulse (maximum flow-rate minus the minimum flow-rate) on the left and right sides is considered. It is found that this ratio shows highest correlation against the ratio of reduced network compliances on the left and right sides. A plot of the femoral pulse ratio against the compliance ratio is shown in Figure \ref{figure_compliance_limits}. The left-right imbalance is apparent in this figure which shows that the ratio of left-to-right femoral pulses varies from 10$^{-4}$ and 10$^{4}$ across the VPD.  To limit this, a filter on the left-to-right compliance ratio between 0.2--5 is introduced. These limits are shown as vertical lines in Figure \ref{figure_compliance_limits} and constrains the the femoral ratio between 10$^{-2}$ and 10$^{2}$ across the dataset.
With this filter, approximately 45\% (23,275 patients) of the patients are discarded, leaving with 28,868 physiologically realistic patients in the VPD.


This VPD reflects both prior beliefs/constraints and measurements from the population. Thus, it is suitable to study population characteristics and behaviour. More importantly, disease can be virtually created in these patients and subsequent assessment of machine learning be made on the database. The pre-filter VPD is publicly made available at Zenodo repository \footnote{Link to Zenodo repository: \url{https://doi.org/10.5281/zenodo.4549764}}.

\section{Conclusions}
A physiologically realistic virtual patient database is presented for the human arterial network. A methodology to create virtual patients guided by prior beliefs, geometrical/physiological constraints, and literature reported measurements is presented.  Starting from a reference network describing the arterial network, the methodology includes i) network reduction without compromising relevant behaviour; ii) re-parameterisation to reduce dimensionality; iii) incorporation of geometrical and physiological constraints in the form of a prior; iv) incorporation of literature reported clinical measurements in the form of the likelihood; v) combination of the prior and likelihood to generate the posterior; and vi)  sampling from the posterior with MCMC to create the VPD. This generic methodology, given a mathematical description of a biological system, can be adopted to create virtual patients for any biological system while accounting for all available information.

The key conclusion of this study is that despite computational challenges and dimensional complexity, the Bayesian approach to creating virtual patients is viable and yields a realistic database of patients consistent with the expected behaviour in terms of the range of outputs expected in a population. The study also shows that appropriate incorporation of priors and measurements (distributed both spatially and in time) is necessary for the generation of physiologically realistic virtual patients. In the absence of these, the virtual patients may be physiologically unrealistic while still statistically reproducing the literature measurements. In this study this phenomenon was observed as no likelihood or prior term incorporating left-right symmetry was included. While such issues can be fixed by incorporation of a filter, the end result is wasted effort in the generation of virtual patients which are subsequently discarded. Nevertheless, the Bayesian approach and the framework presented is an attractive framework that is capable of incorporating such expectations in the methodology so that such issues are avoided.

A realistic database of 28,868 patients representing human arterial network is presented and made available. This database is created with a view that it is useful in subsequent studies which exploit the advances in machine learning methods for the detection of arterial disease. This forms the primary area of future investigation with this database.

\section*{Funding}
This work is supported by an EPSRC studentship ref. EP/N509553/1 and an EPSRC grant ref. EP/R010811/1.

\clearpage

\tiny

\printbibliography


\clearpage

\appendix
\normalsize

\section{List of vessels within the reference network}
The 77 arterial segments in the reference network are listed in Table \ref{table_full_network}.

\label{appendix_list_of_vessels}
\begin{table*}[htbp]
\footnotesize
\centering
\def\arraystretch{1.2}
\begin{tabular}{|c c|c c|}
\hline
\textbf{Reference number} & \textbf{Vessel name} & \textbf{Reference number} & \textbf{Vessel name} \\
\hline
1 & Aortic arch I & 34 & Celiac trunk \\
2 & Brachiocephalic trunk & 35 & Abdominal aorta I\\
3 & Aortic arch II & 36 & Common hepatic\\
4 & Subclavian R I & 37 & Splenic I\\
5 & Common carotid R & 38 & Left gastric\\
6 & Vertebral R & 39 & Splenic II\\
7a & Subclavian R II & 40 & Superior mesenteric \\
7b & Axillary R & 41 & Abdominal aorta II \\
7c & Brachial R & 42 & Renal L \\
8 & Radial R & 43 & Abdominal aorta III\\
9 & Ulnar R I & 44 & Renal R\\
10a & Common interosseous R & 45 & Abdominal aorta IV\\
10b & Posterior interosseous R & 46 & Inferior mesenteric\\
11 & Ulnar R II & 47 & Abdominal aorta V\\
12 & External carotid R & 48 & Common iliac R\\
13 & Internal carotid R & 49 & Common iliac L\\
14 & Common carotid L & 50a & External iliac R\\
15 & Aortic arch III & 50b & Femoral R I\\
16 & External carotid L & 51 & Internal iliac R\\
17 & Internal carotid L & 52 & Profunda femoris R\\
18 & Subclavian L I & 53a & Femoral R II\\
19a & Aortic arch IV & 53b & Popliteal R I\\
19b & Thoracic aorta I & 54 & Anterior tibial R\\
20 & Vertebral L & 55a & Popliteal R II\\
21a & Subclavian L II & 55b & Tibiofibular trunk R\\
21b & Axillary L21cBrachial L & 55c & Posterior tibial R\\
22 & Radial L & 56a & External iliac L\\
23 & Ulnar L I & 56b & Femoral L I\\
24a & Common interosseous L & 57 & Internal iliac L\\
24b & Posterior interosseous L & 58 &  Profunda femoris L\\
25 & Ulnar L II & 59a & Femoral L II\\
26 & Posterior intercostal R 1 & 59b & Popliteal L I\\
27 & Thoracic aorta II & 60 & Anterior tibial L\\
28 & Posterior intercostal L 1 & 61a & Popliteal L II\\
29 & Thoracic aorta III & 61b & Tibiofibular trunk L\\
30 & Posterior intercostal R 2 & 61c & Posterior tibial L\\
31 & Thoracic aorta IV & & \\
32 & Posterior intercostal L 2 & & \\
33a & Thoracic aorta V & & \\
33b &Thoracic aorta VI & & \\
\hline
\end{tabular}
\caption{The 56 arterial vessels, described by 77 segments, within the reference arterial network, taken from \cite{boileau2015benchmark}, are outlined above. The numbers assigned to each vessel within this table align with those in Figure \ref{fig_full_network}.}
\label{table_full_network}
\end{table*}
\normalsize

\clearpage

\section{Arterial vessel length}
\label{Appendix_vessel_length}

To assess the importance of vessel length on the variability of pressure and flow-rate profiles, two groups of VPs are created using the reference network (see Section \ref{section_removal_vessels}). For both groups, all properties of each VPs arterial network are set to their reference values, except for the length of each vessel. The lengths of VPs arterial vessels are randomised for each group by:
\begin{itemize}
\item \textbf{Group 1:} applying an individual independent scaling term to the reference length of each vessel. These scaling terms are sampled from normal distributions with mean of 1 and a standard deviation of 0.2.
\item \textbf{Group 2:} applying a single scaling term to the reference length of all vessels. As with the first case, this scaling term is sampled from a normal distribution with mean of 1 and standard deviation of 0.2.
\end{itemize}

\begin{figure*}
\centering
\includegraphics[width=5in]{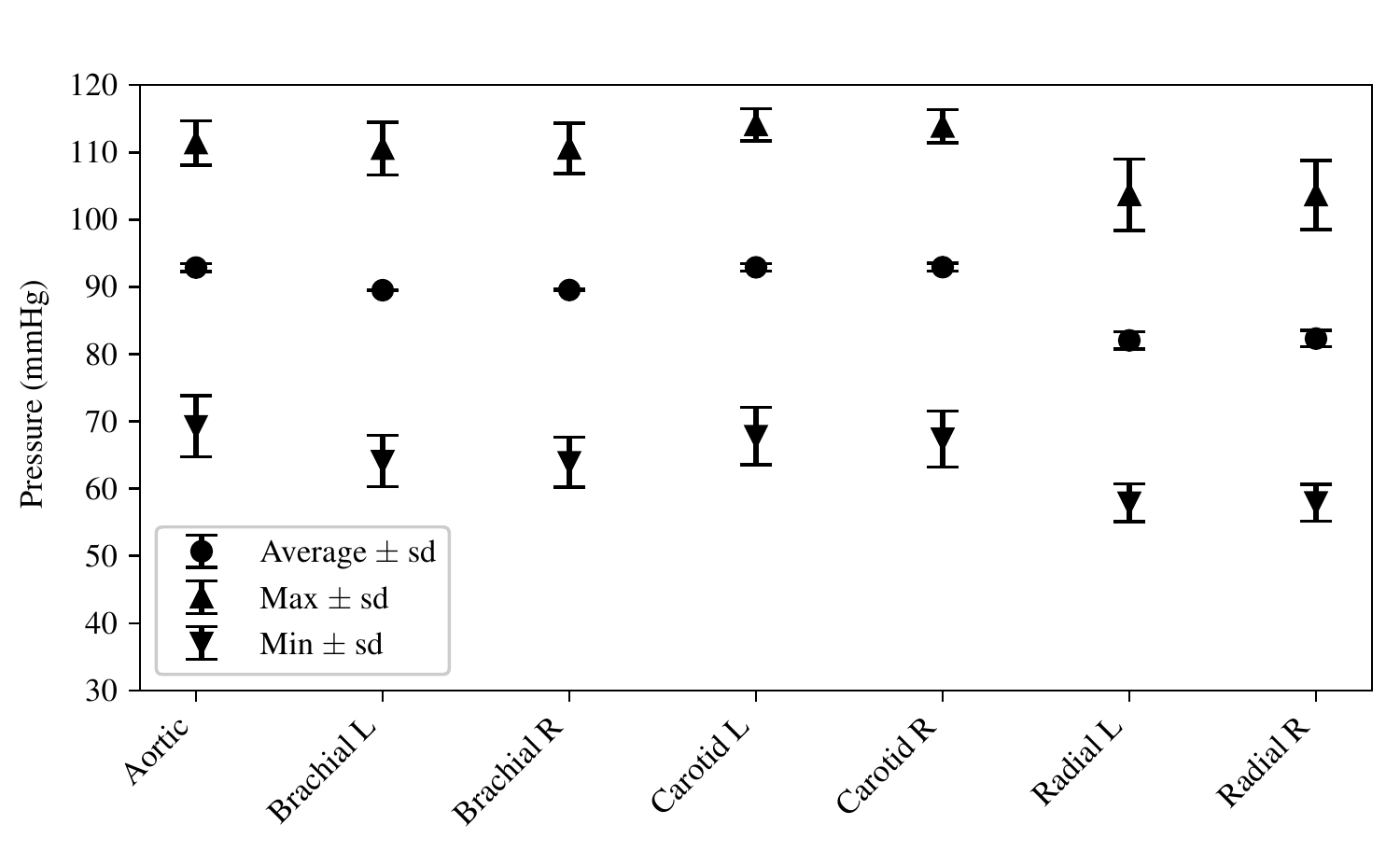}
\caption{The mean and standard deviation of the average, maximum, and minimum pressure at all appropriate examination locations for the 30,000 VPs created using a constant length scaling term.}
\label{fig_len_scaling_pressure_constant}
\end{figure*}

\begin{figure*}
\centering
\includegraphics[width=5in]{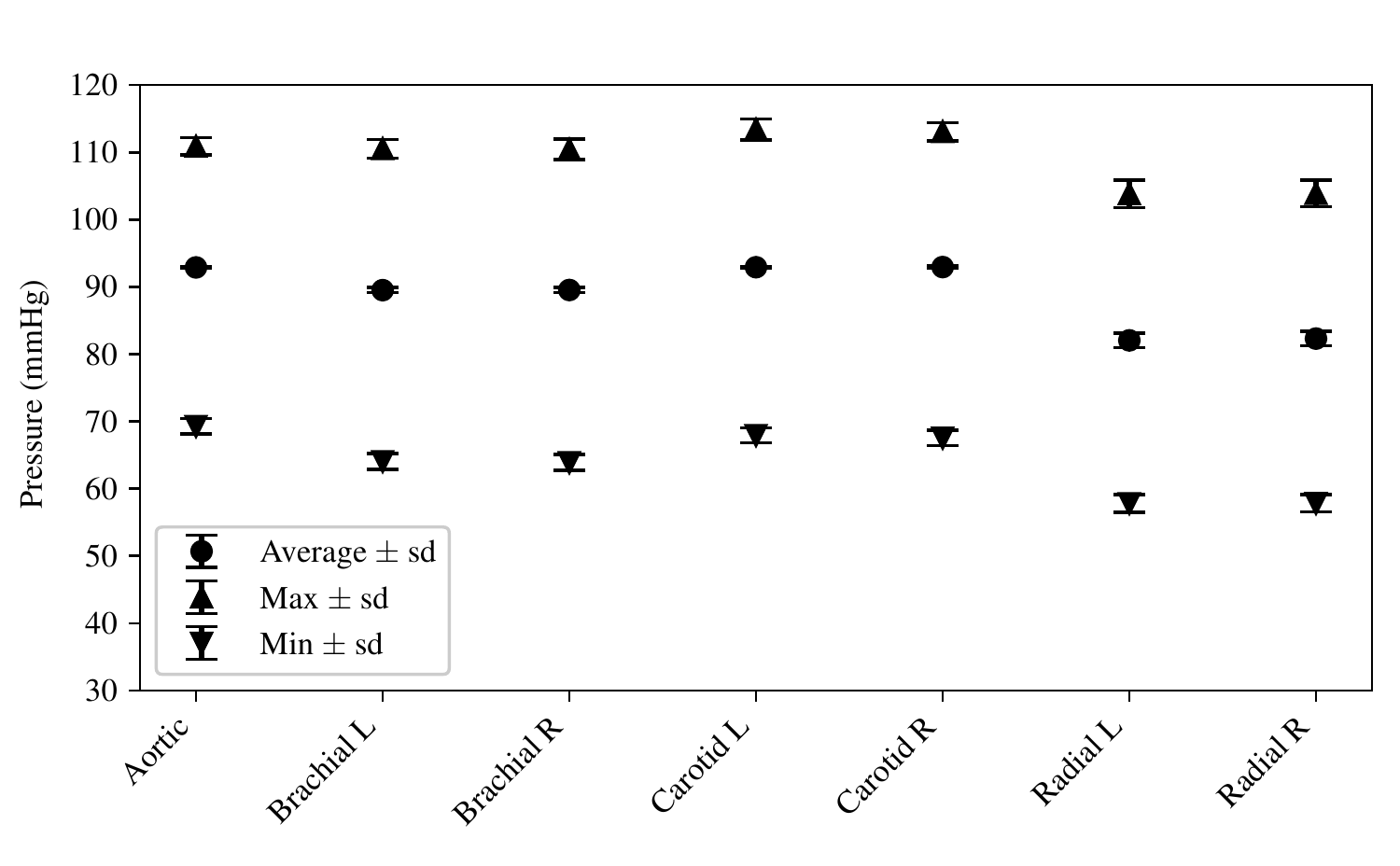}
\caption{The mean and standard deviation of the average, maximum, and minimum pressure at all appropriate examination locations for the 30,000 VPs created using individual length scaling terms applied to each vessel within the network.}
\label{fig_len_scaling_pressure_individual}
\end{figure*}

For each of the two groups outlined above 30,000 VPs are sampled, and the pressure and flow-rate profiles associated with each computed. Pressure or flow rate profiles are taken from each VP at all measurement locations, highlighted within Section \ref{section_locations}, and the average, maximum, and minimum of each profile recorded.  The average, maximum, and minimum pressure is also recorded at the inlet of the aorta, to compare the affect of vessel length on pressure profiles at a location with known flow rate. The mean and standard deviation of the average, maximum, and minimum pressure at each appropriate examination location is shown for patients with a constant length scaling term in Figure \ref{fig_len_scaling_pressure_constant}, and individual scaling terms in Figure \ref{fig_len_scaling_pressure_individual}.

Comparing Figure \ref{fig_len_scaling_pressure_constant} and Figure  \ref{fig_len_scaling_pressure_individual} it can be seen that as is expected, the mean of all pressure measurements are relatively consistent for both groups of VPs. The standard deviation of all pressure measurements is seen to be larger for group 2, relative to group 1. This increase in standard deviation is most noticeable when comparing the maximum and minimum pressure values. The standard deviation of the average pressure at each location is relatively low for both groups of VPs. While there is an increase in the standard deviation of pressure measurements when using a constant scaling term, relative to individual scaling terms, the extent of this increase is reasonable when weighed up against the reduction in dimensionality achieved by using a single scaling term.

Further analysis is carried out by comparing the mean and standard deviations of each flow rate measurement; and the correlation between all pressure and flow rate measurement for each of the two groups of VPs. Similar behaviour is seen to the comparison of pressure measurements highlighted above.

\clearpage

\clearpage

\section{MCMC trace plots}
\label{appendix_MCMC_trace_plots}

\begin{figure*}
\centering
\includegraphics[width=6in]{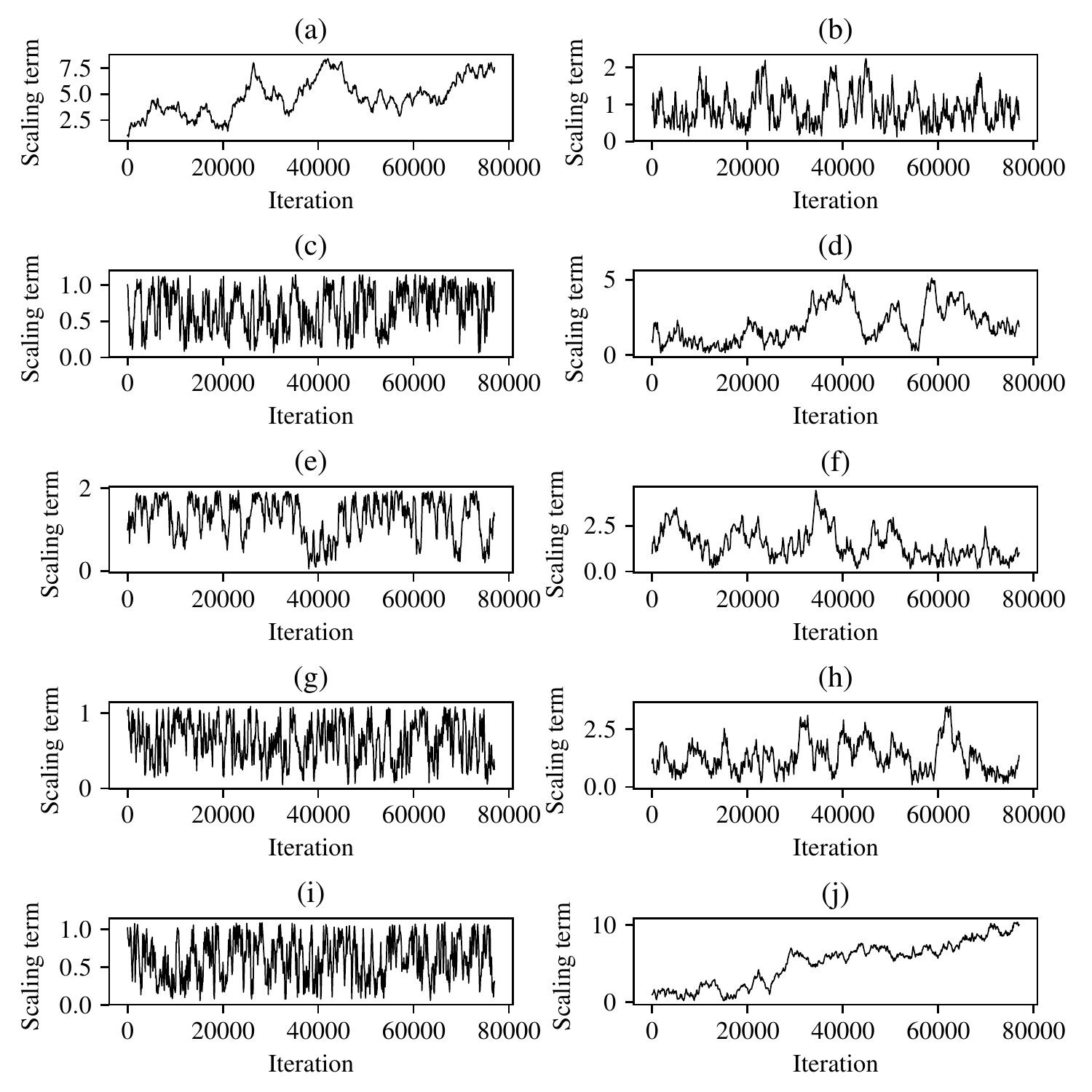}
\caption{The parameter scaling terms at each $100^{th}$ iteration of the Markov chain applied to the: aorta chain initialising value (a), aorta chain decay term (b), right arm chain daughter/parent ratio (c), right arm chain decay term (d), left arm chain daughter/parent ratio (e), left arm chain decay term (f), right leg chain daughter/parent ratio (g), right leg chain decay term (h), left leg chain daughter/parent ratio (i), and the left leg chain decay term (j) for the $\beta$ properties of the network.}
\label{fig_TP_beta_chain}
\end{figure*}

\begin{figure*}
\centering
\includegraphics[width=6in]{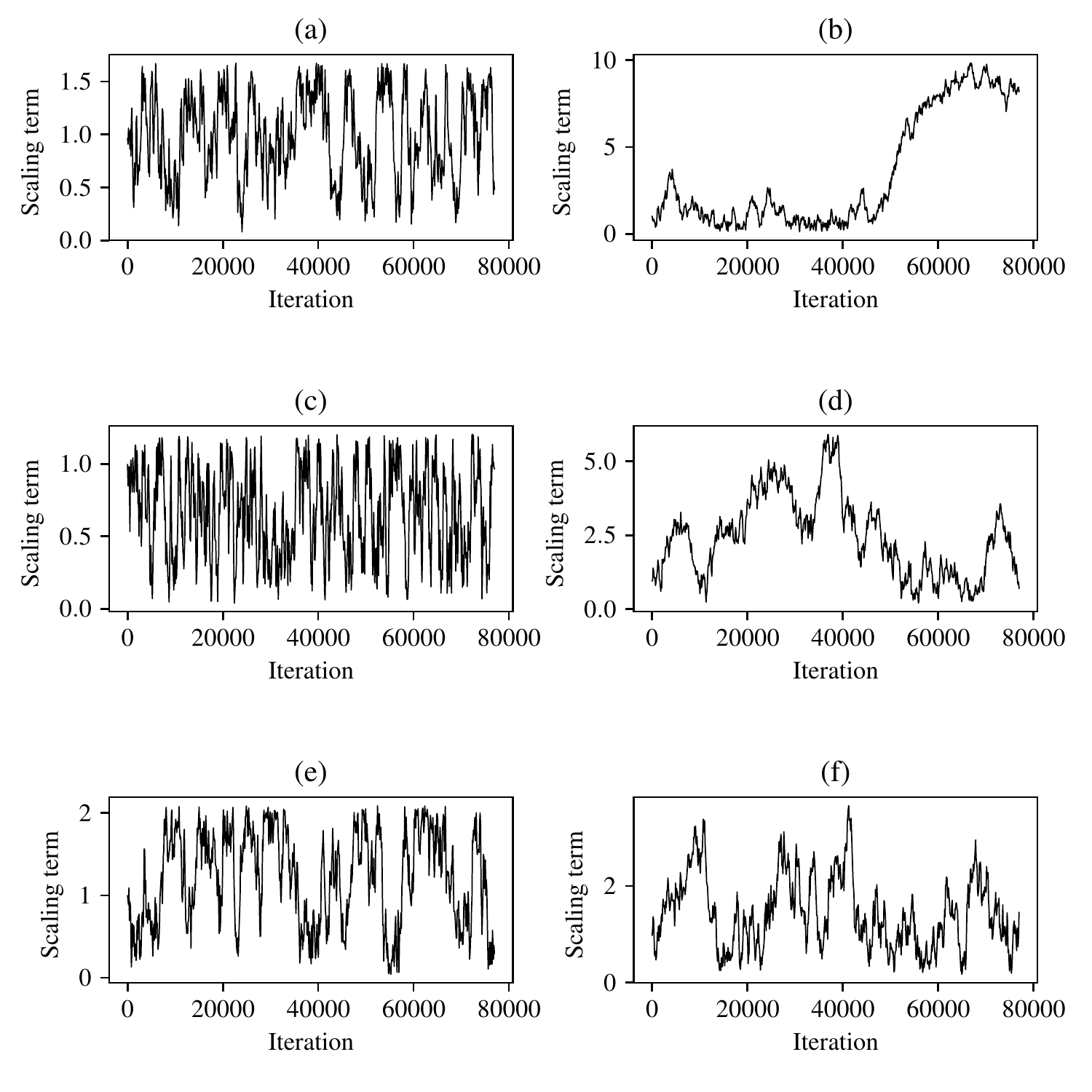}
\caption{The parameter scaling terms at each $100^{th}$ iteration of the Markov chain applied to the: brachiocephalic trunk daughter/parent ratio (a), brachiocephalic trunk decay term (b), right common carotid daughter/parent ratio (c), right common carotid decay term (d), left common carotid daughter/parent ratio (e), and the left common carotid decay term (f) for the $\beta$ properties of the network.}
\label{fig_TP_beta_ind_I}
\end{figure*}

\begin{figure*}
\centering
\includegraphics[width=6in]{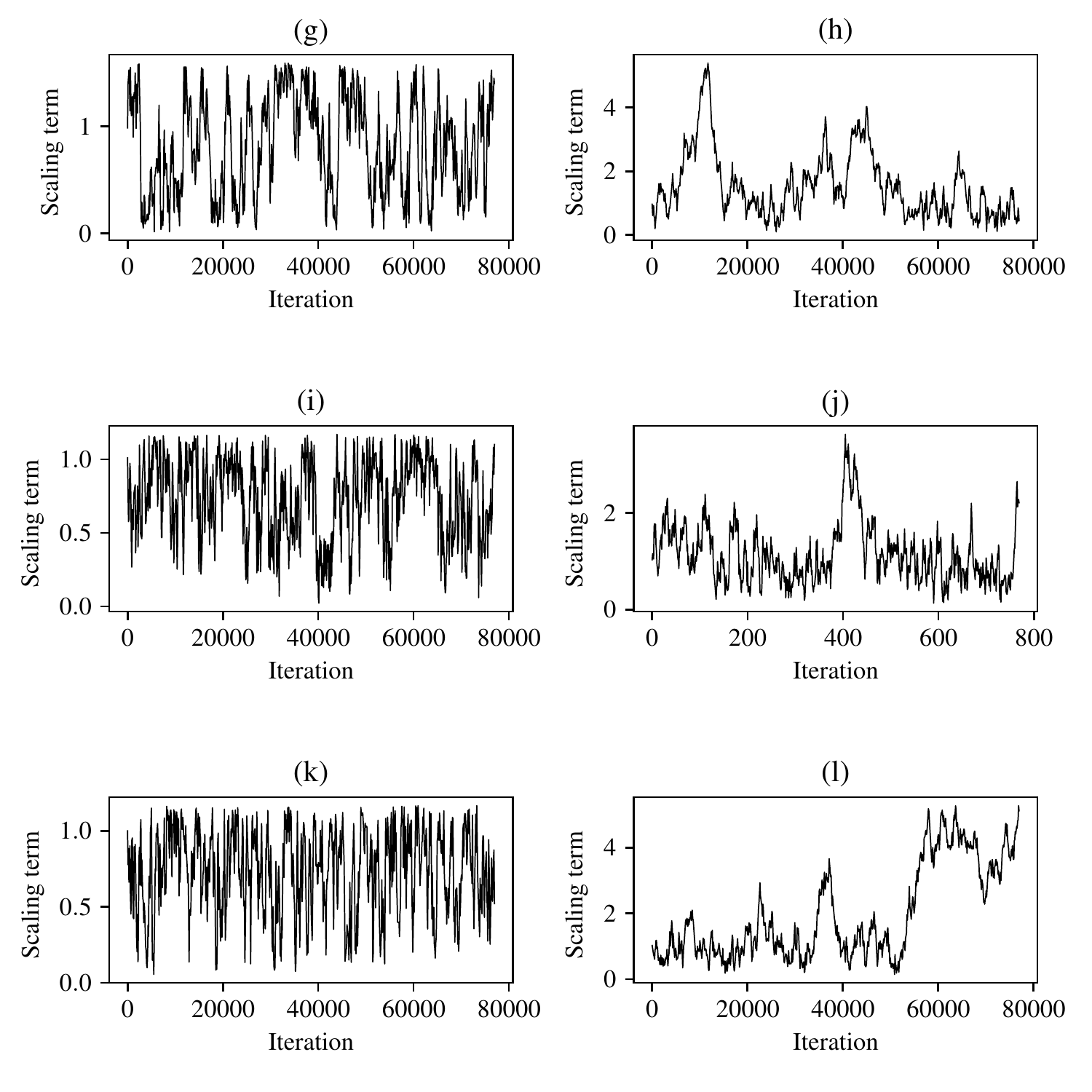}
\caption{The parameter scaling terms at each $100^{th}$ iteration of the Markov chain applied to the: celiac trunk daughter/parent ratio (g), celiac trunk decay term (h), right common iliac daughter/parent ratio (i), right common iliac decay term (j), left common iliac daughter/parent ratio (k), and the left common iliac decay term (l) for the $\beta$ properties of the network.}
\label{fig_TP_beta_ind_II}
\end{figure*}

\begin{figure*}
\centering
\includegraphics[width=6in]{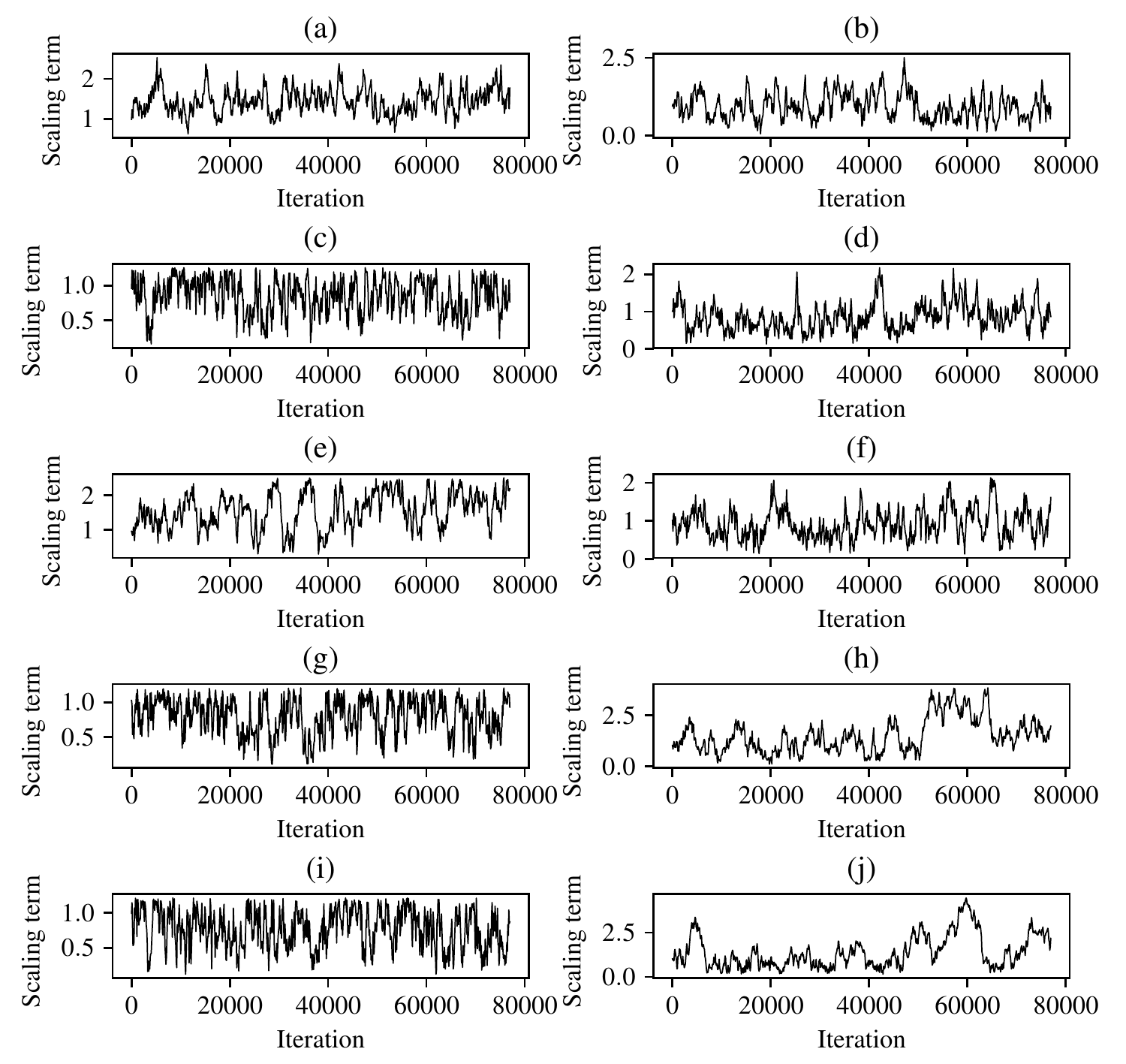}
\caption{The parameter scaling terms at each $100^{th}$ iteration of the Markov chain applied to the: aorta chain initialising value (a), aorta chain decay term (b), right arm chain daughter/parent ratio (c), right arm chain decay term (d), left arm chain daughter/parent ratio (e), left arm chain decay term (f), right leg chain daughter/parent ratio (g), right leg chain decay term (h), left leg chain daughter/parent ratio (i), and the left leg chain decay term (j) for the $r_0$ properties of the network.}
\label{fig_TP_r_chain}
\end{figure*}

\begin{figure*}
\centering
\includegraphics[width=6in]{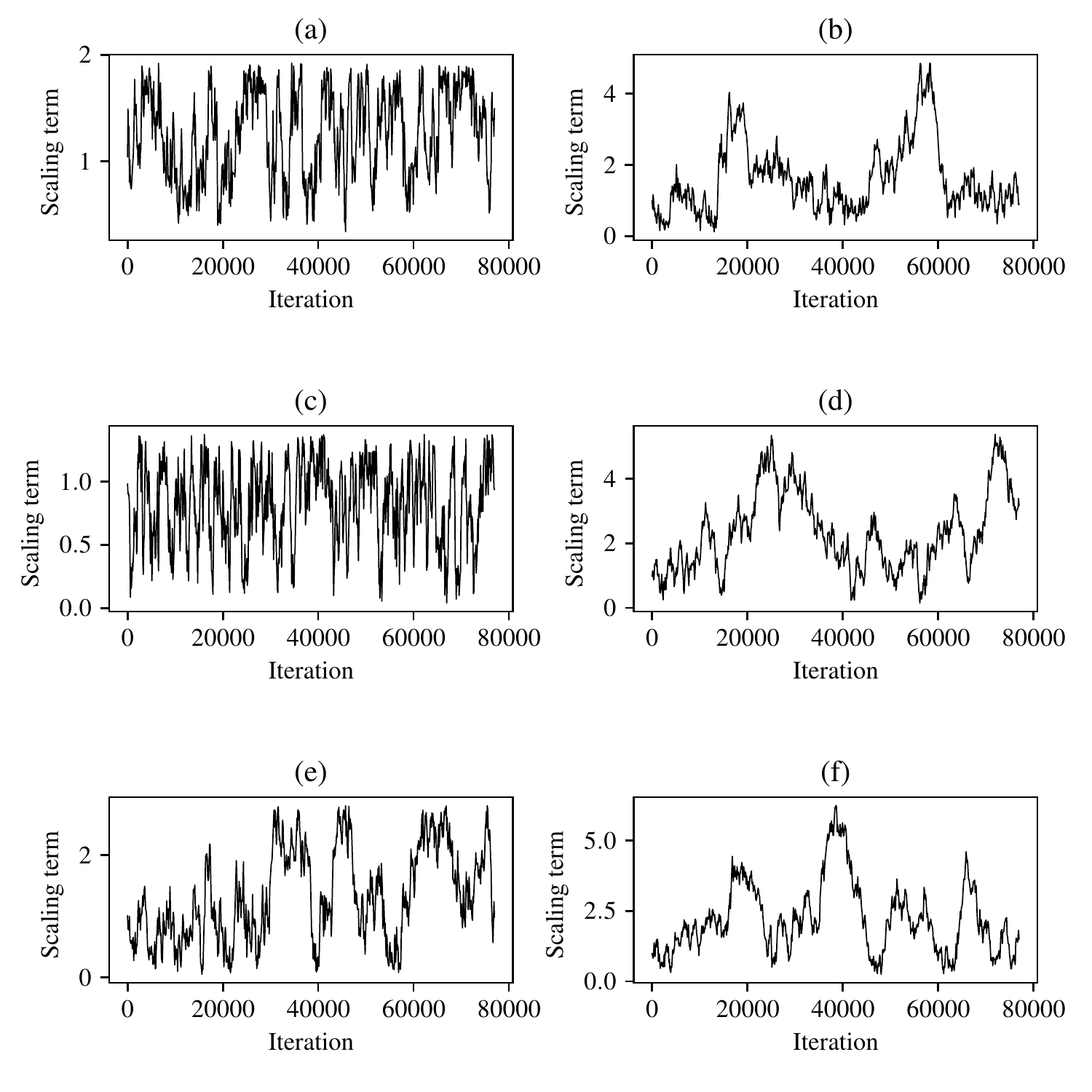}
\caption{The parameter scaling terms at each $100^{th}$ iteration of the Markov chain applied to the: brachiocephalic trunk daughter/parent ratio (a), brachiocephalic trunk decay term (b), right common carotid daughter/parent ratio (c), right common carotid decay term (d), left common carotid daughter_parent ratio (e), the left common carotid decay term (f) for the $r_0$ properties of the network.}
\label{fig_TP_r_ind_I}
\end{figure*}

\begin{figure*}
\centering
\includegraphics[width=6in]{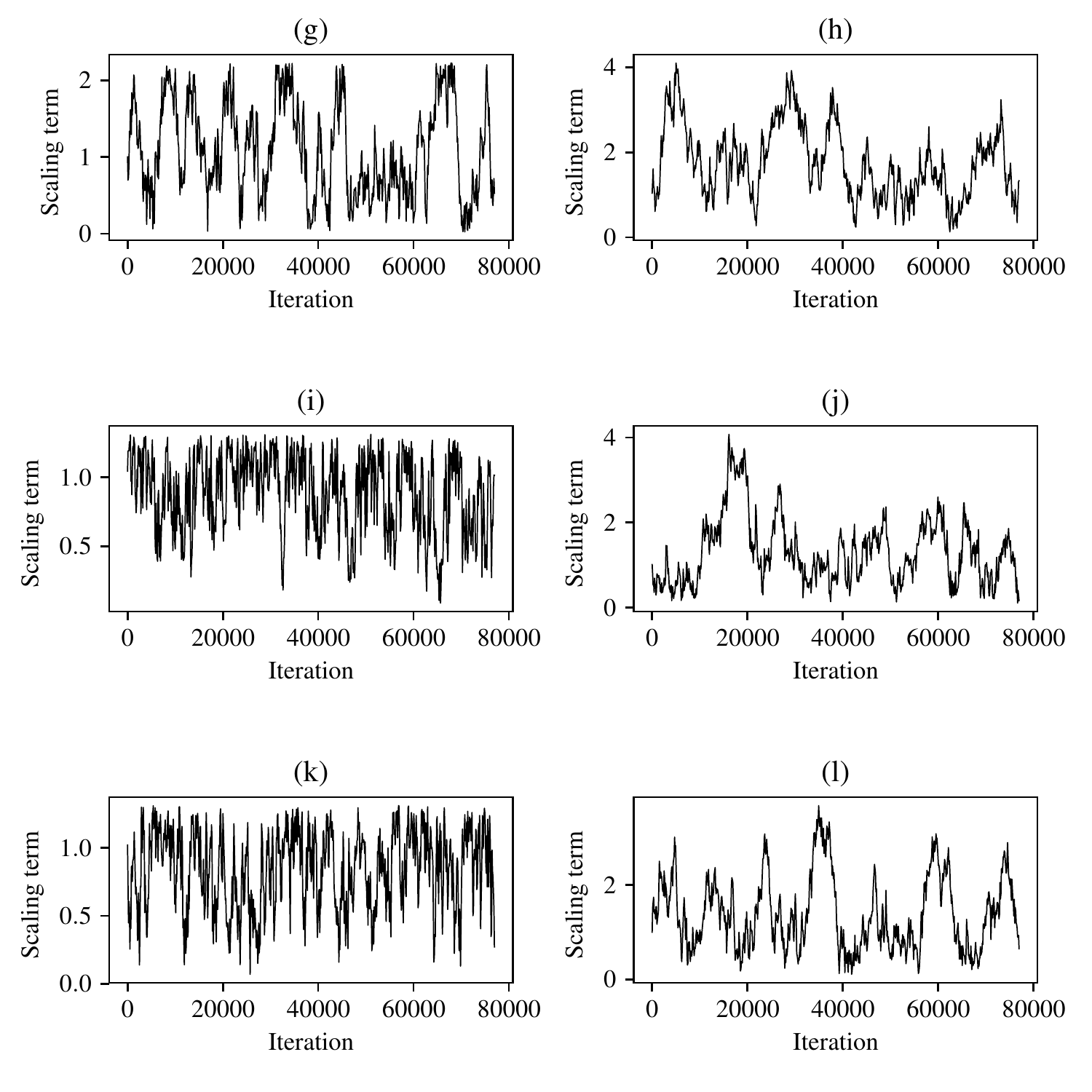}
\caption{The parameter scaling terms at each $100^{th}$ iteration of the Markov chain applied to the: celiac trunk daughter/parent ratio (g), celiac trunk decay term (h), right common iliac daughter/parent ratio (i), right common iliac decay term (j), left common iliac daughter/parent ratio (k), and the left common iliac decay term (l) for the $r_0$ properties of the network.}
\label{fig_TP_r_ind_II}
\end{figure*}

\begin{figure*}
\centering
\includegraphics[width=6in]{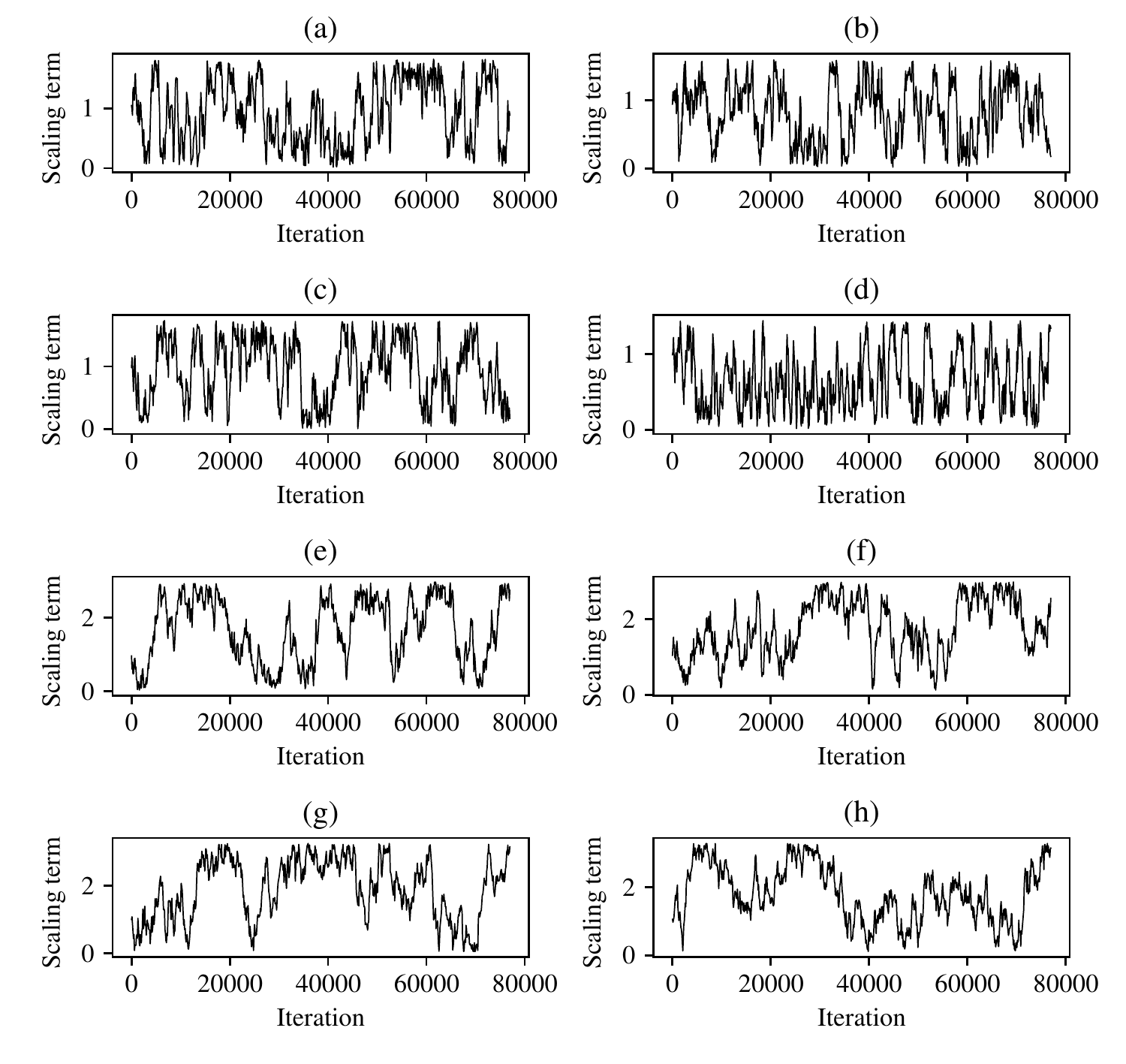}
\caption{The parameter scaling terms at each $100^{th}$ iteration of the Markov chain applied to the daughter/parent ratio of the: inferior mesenteric (a), right renal (b), left renal (c), superior mesenteric (d), second left posterior intercostal (e), second right posterior intercostal (f), first left posterior intercostal (g), first right posterior intercostal (h) for the $\beta$ properties of the network.}
\label{fig_TP_aorta_beta_constant}
\end{figure*}

\begin{figure*}
\centering
\includegraphics[width=6in]{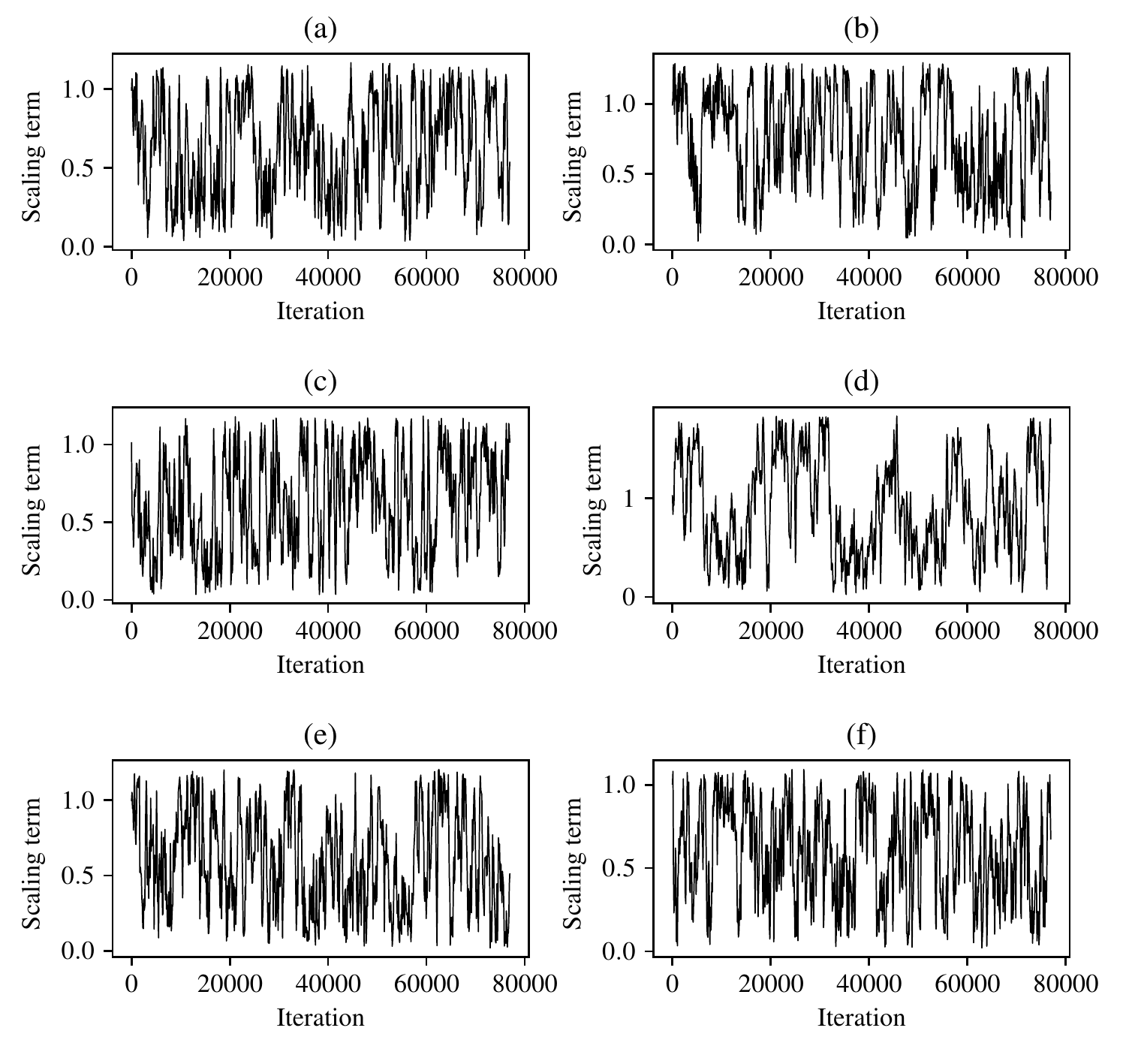}
\caption{The parameter scaling terms at each $100^{th}$ iteration of the Markov chain applied to the daughter/parent ratio of the: first right ulnar (a), right common interosseous (b), right radial (c), right vertebral (d), right external carotid (e), right internal carotid (f) for the $\beta$ properties of the network.}
\label{fig_TP_armR_beta_constant}
\end{figure*}

\begin{figure*}
\centering
\includegraphics[width=6in]{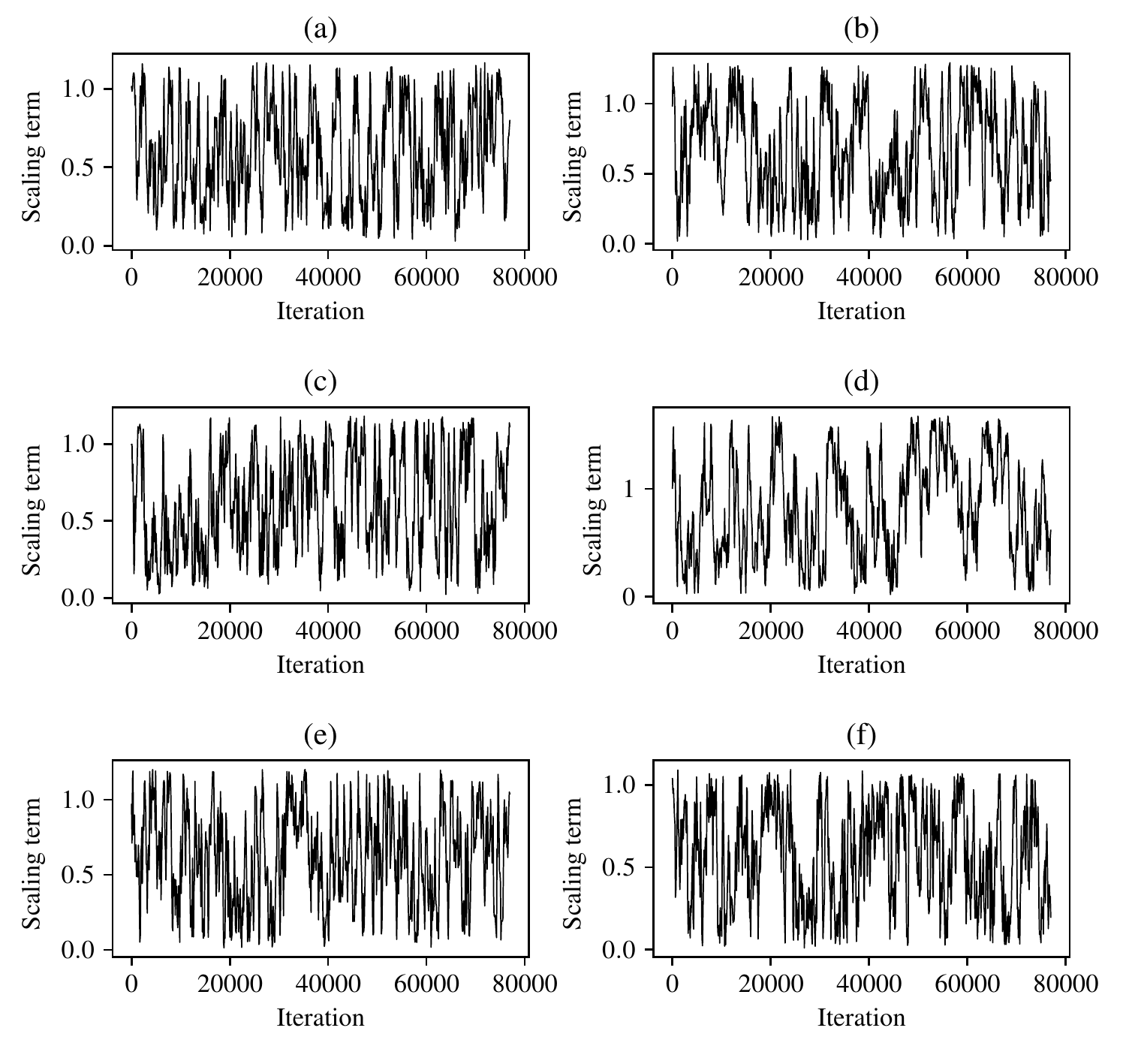}
\caption{The parameter scaling terms at each $100^{th}$ iteration of the Markov chain applied to the daughter/parent ratio of the: first left ulnar (a), left common interosseous (b), left radial (c), left vertebral (d), left external carotid (e), left internal carotid (f) for the $\beta$ properties of the network.}
\label{fig_TP_armL_beta_constant}
\end{figure*}

\begin{figure*}
\centering
\includegraphics[width=6in]{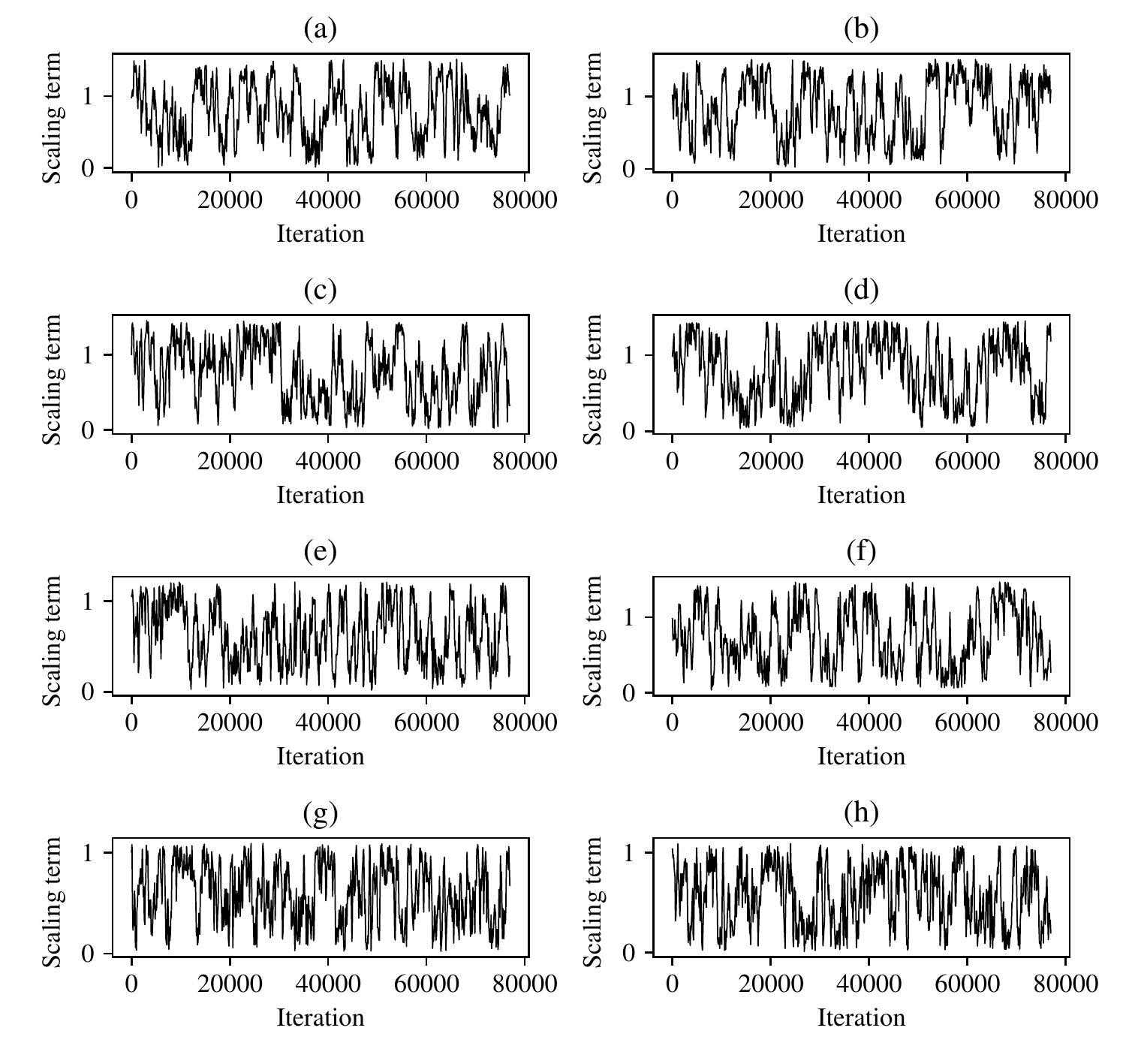}
\caption{The parameter scaling terms at each $100^{th}$ iteration of the Markov chain applied to the daughter/parent ratio of the: right anterior tibial (a), left anterior tibial (b), right posterior tibial (c), left posterior tibial (d), right profunda femoris (e), left profunda femoris (f), right internal carotid (g), left internal carotid (h) for the $\beta$ properties of the network.}
\label{fig_TP_leg_beta_constant}
\end{figure*}

\begin{figure*}
\centering
\includegraphics[width=6in]{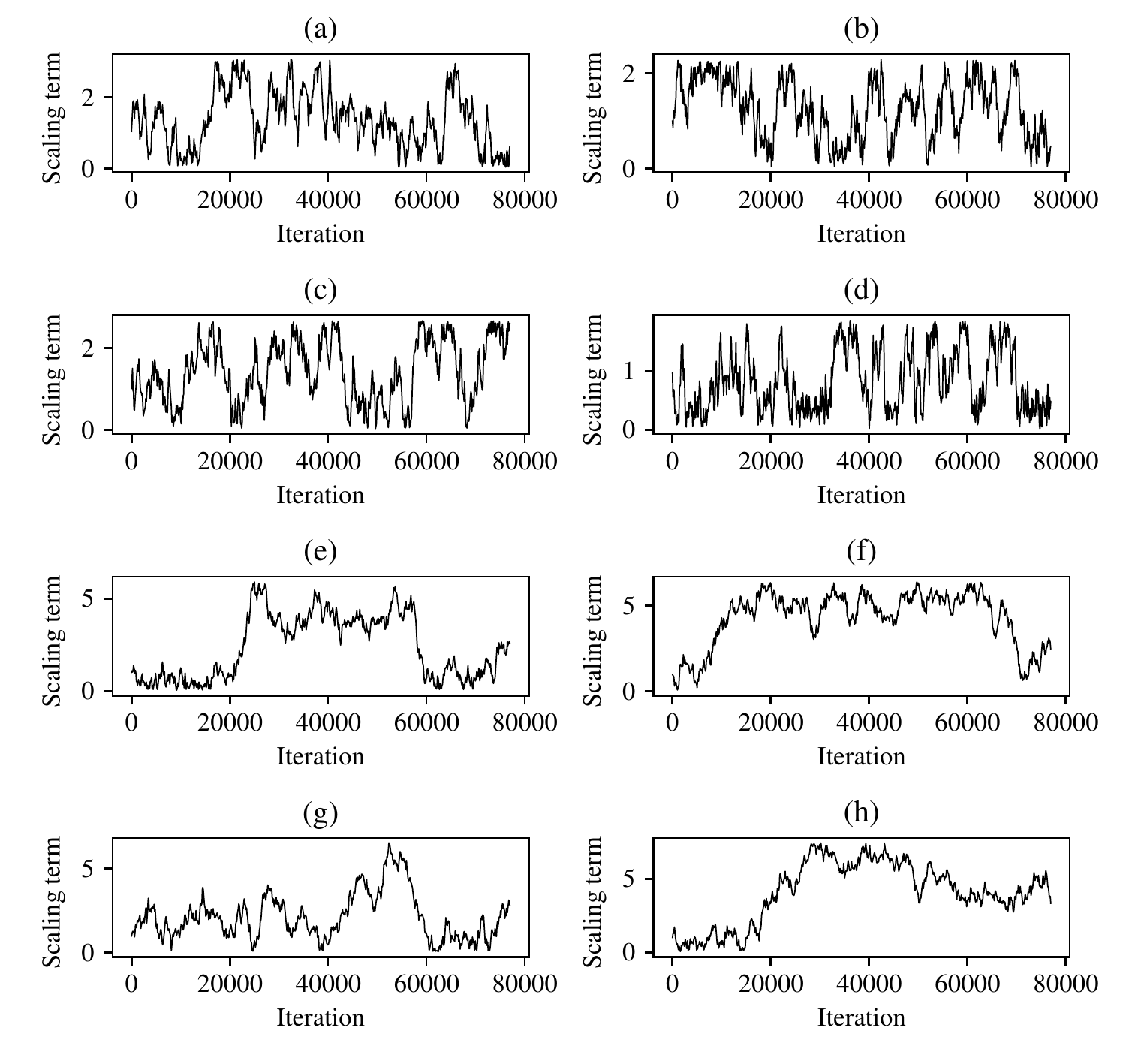}
\caption{The parameter scaling terms at each $100^{th}$ iteration of the Markov chain applied to the daughter/parent ratio of the: inferior mesenteric (a), right renal (b), left renal (c), superior mesenteric (d), second left posterior intercostal (e), second right posterior intercostal (f), first left posterior intercostal (g), first right posterior intercostal (h) for the $r_0$ properties of the network.}
\label{fig_TP_aorta_r_constant}
\end{figure*}

\begin{figure*}
\centering
\includegraphics[width=6in]{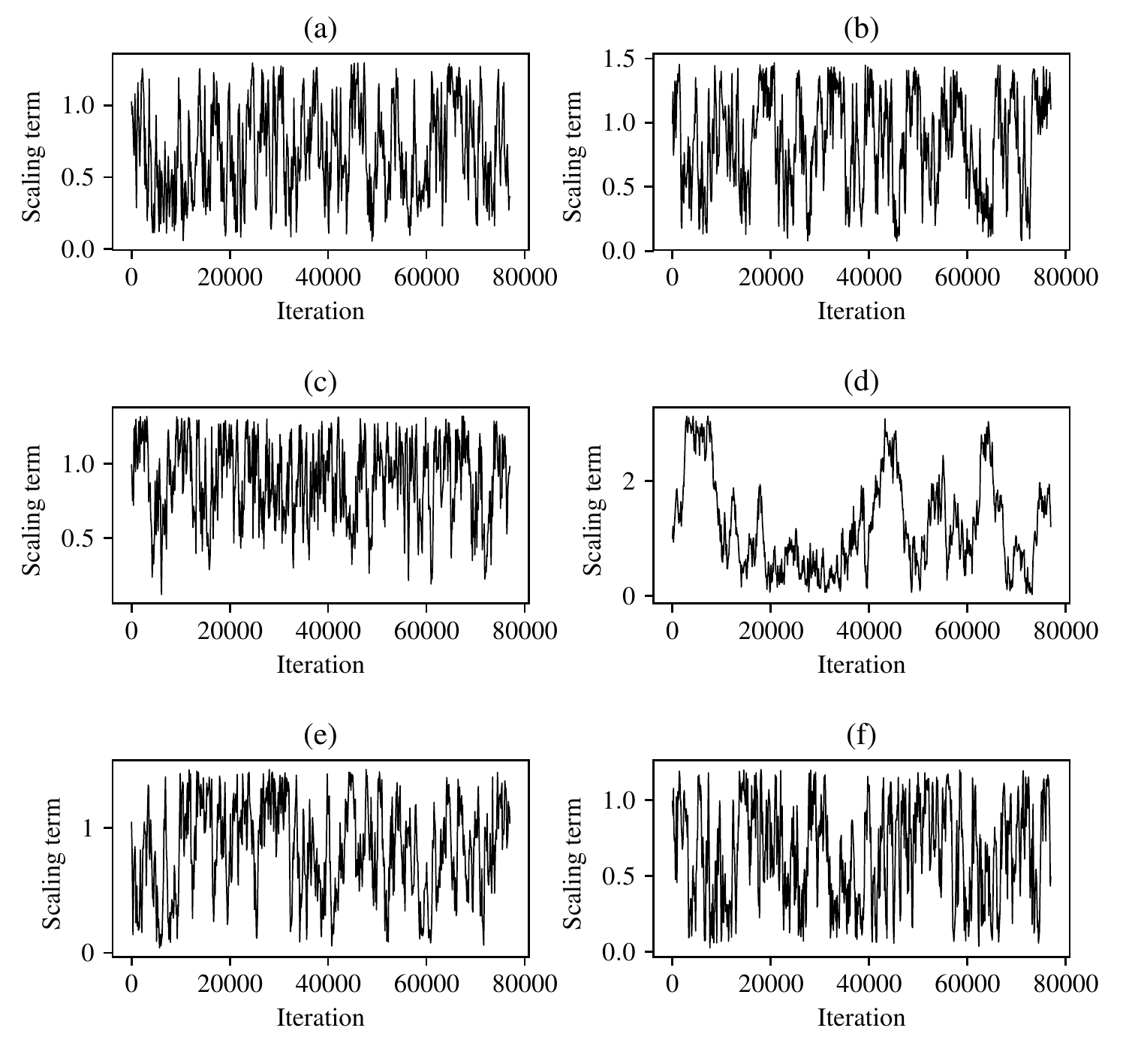}
\caption{The parameter scaling terms at each $100^{th}$ iteration of the Markov chain applied to the daughter/parent ratio of the: first right ulnar (a), right common interosseous (b), right radial (c), right vertebral (d), right external carotid (e), right internal carotid (f) for the $r_0$ properties of the network.}
\label{fig_TP_armR_r_constant}
\end{figure*}

\begin{figure*}
\centering
\includegraphics[width=6in]{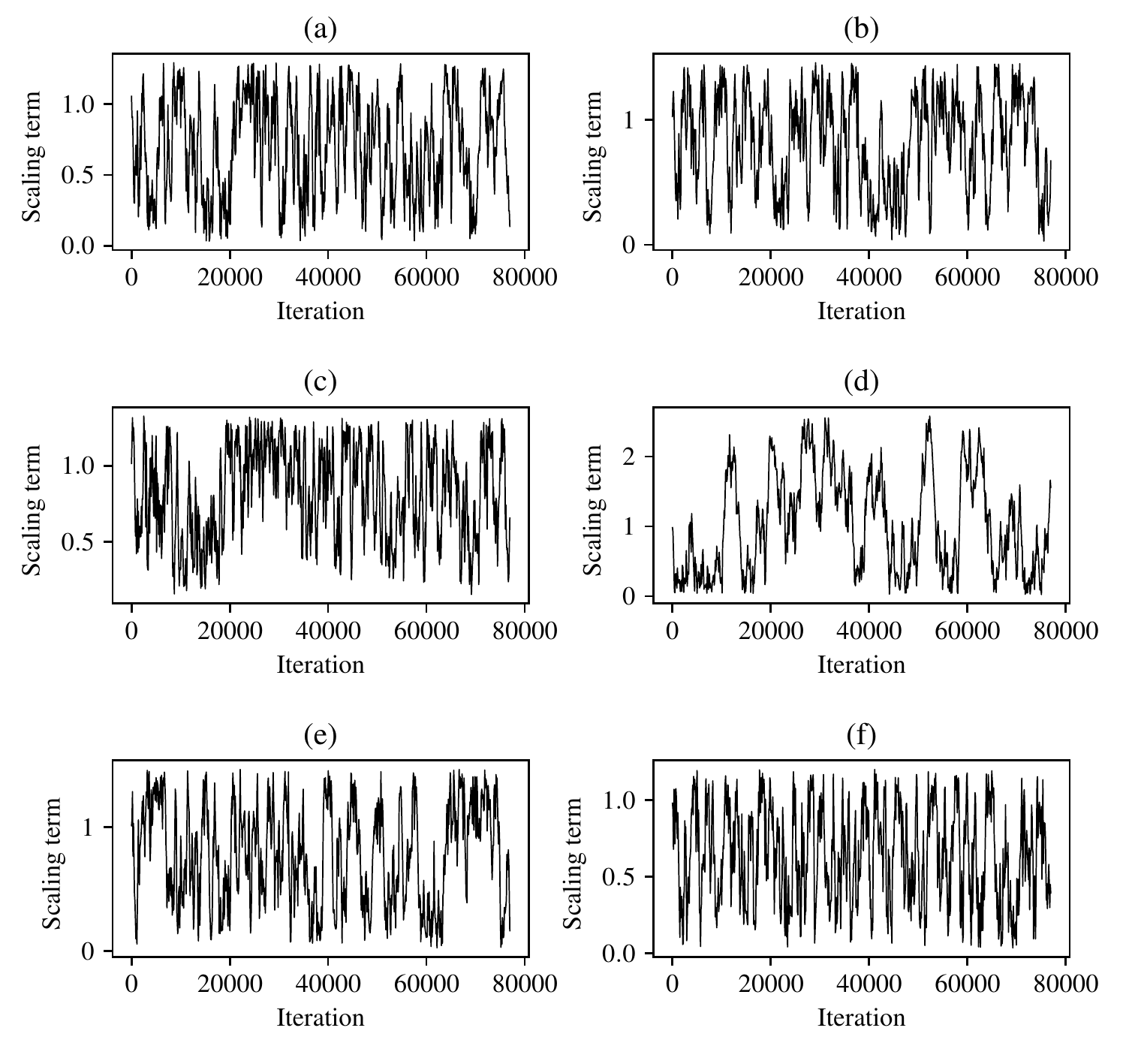}
\caption{The parameter scaling terms at each $100^{th}$ iteration of the Markov chain applied to the daughter/parent ratio of the: first left ulnar (a), left common interosseous (b), left radial (c), left vertebral (d), left external carotid (e), left internal carotid (f) for the $r_0$ properties of the network.}
\label{fig_TP_armL_r_constant}
\end{figure*}

\begin{figure*}
\centering
\includegraphics[width=6in]{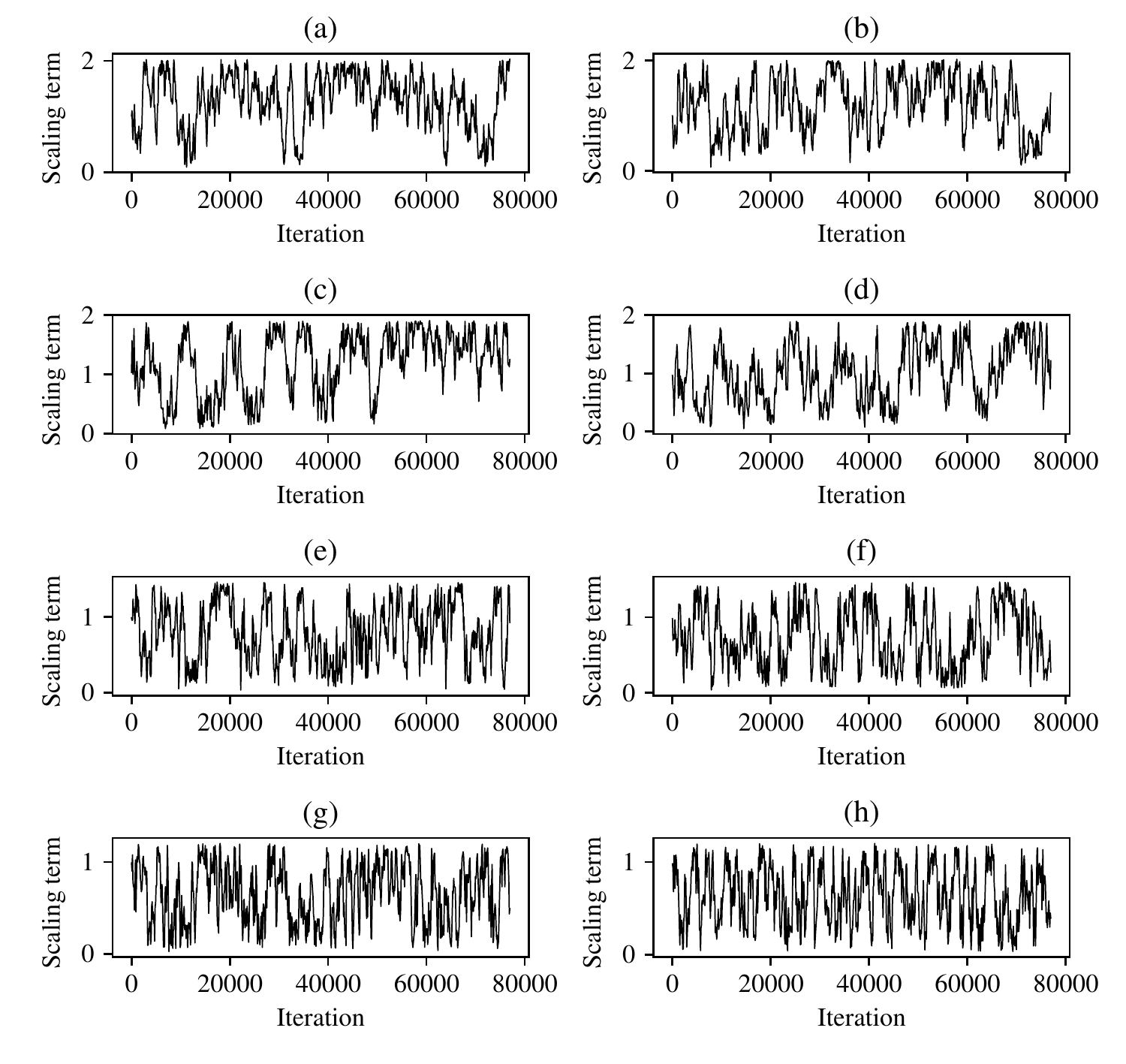}
\caption{The parameter scaling terms at each $100^{th}$ iteration of the Markov chain applied to the daughter/parent ratio of the: right anterior tibial (a), left anterior tibial (b), right posterior tibial (c), left posterior tibial (d), right profunda femoris (e), left profunda femoris (f), right internal carotid (g), left internal carotid (h) for the $r_0$ properties of the network.}
\label{fig_TP_leg_r_constant}
\end{figure*}

\begin{figure*}
\centering
\includegraphics[width=6in]{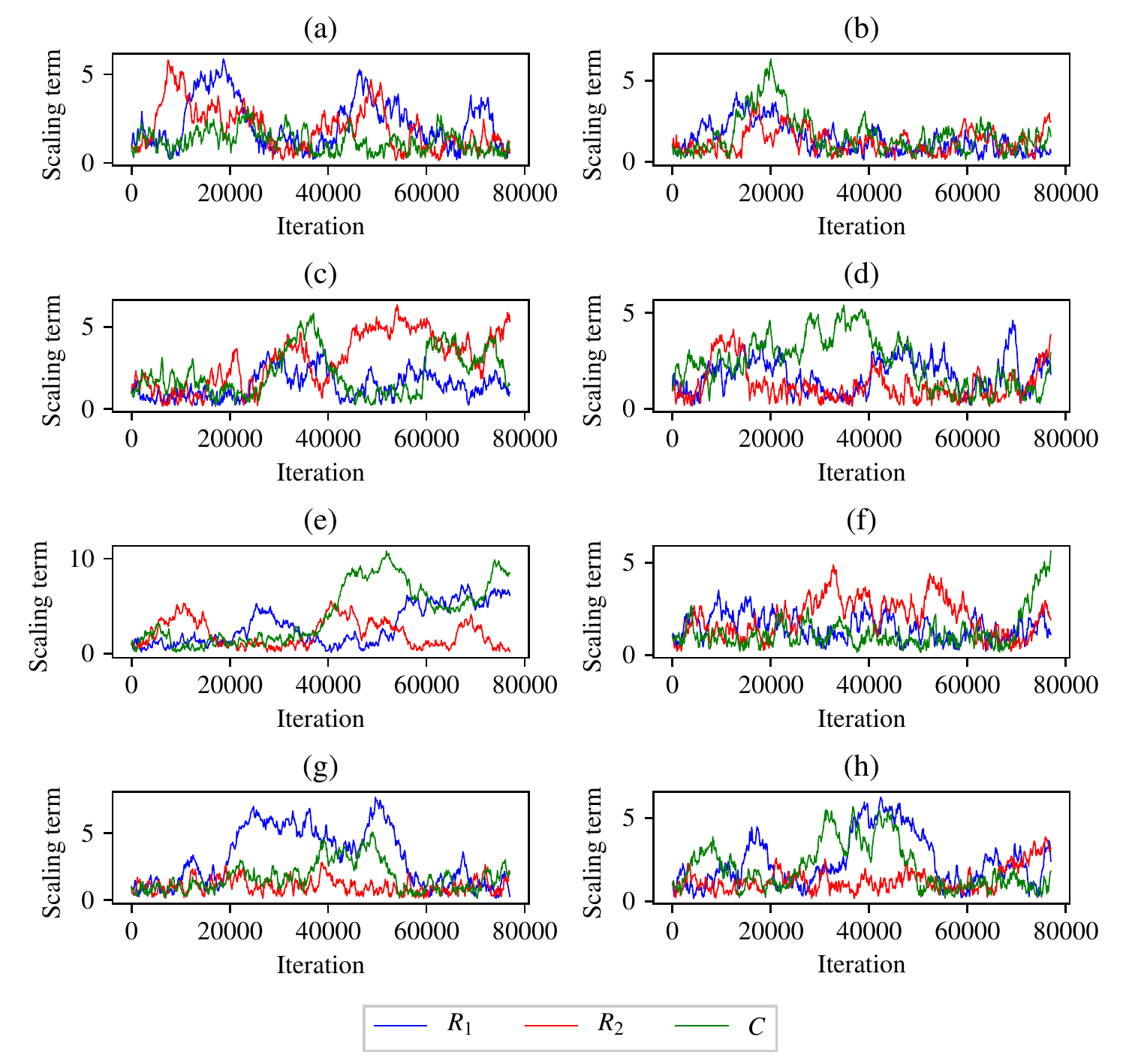}
\caption{The parameter scaling terms at each $100^{th}$ iteration of the Markov chain applied to the Windkessel model parameters at the terminal boundary of the: inferior mesenteric (a), right renal (b), left renal (c), superior mesenteric (d), second left posterior intercostal (e), second right posterior intercostal (f), first left posterior intercostal (g), and the first right posterior intercostal (h).}
\label{fig_TP_WK_aorta}
\end{figure*}

\begin{figure*}
\centering
\includegraphics[width=6in]{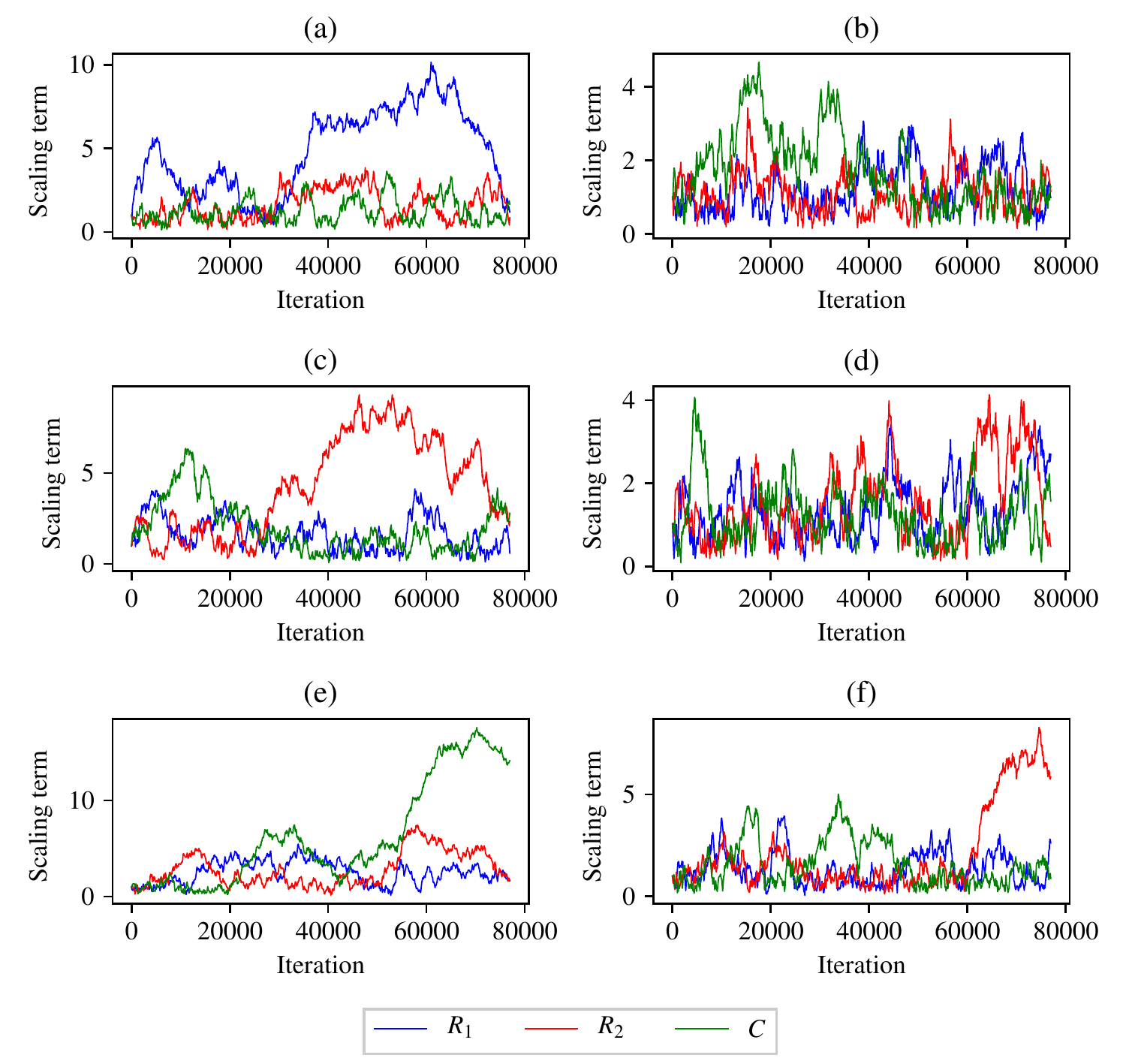}
\caption{The parameter scaling terms at each $100^{th}$ iteration of the Markov chain applied to the Windkessel model parameters at the terminal boundary of the: second right ulnar (a), right common interosseous (b), right radial (c), right vertebral (d), right  external carotid (e), and the right internal carotid (f).}
\label{fig_TP_WK_armR}
\end{figure*}

\begin{figure*}
\centering
\includegraphics[width=6in]{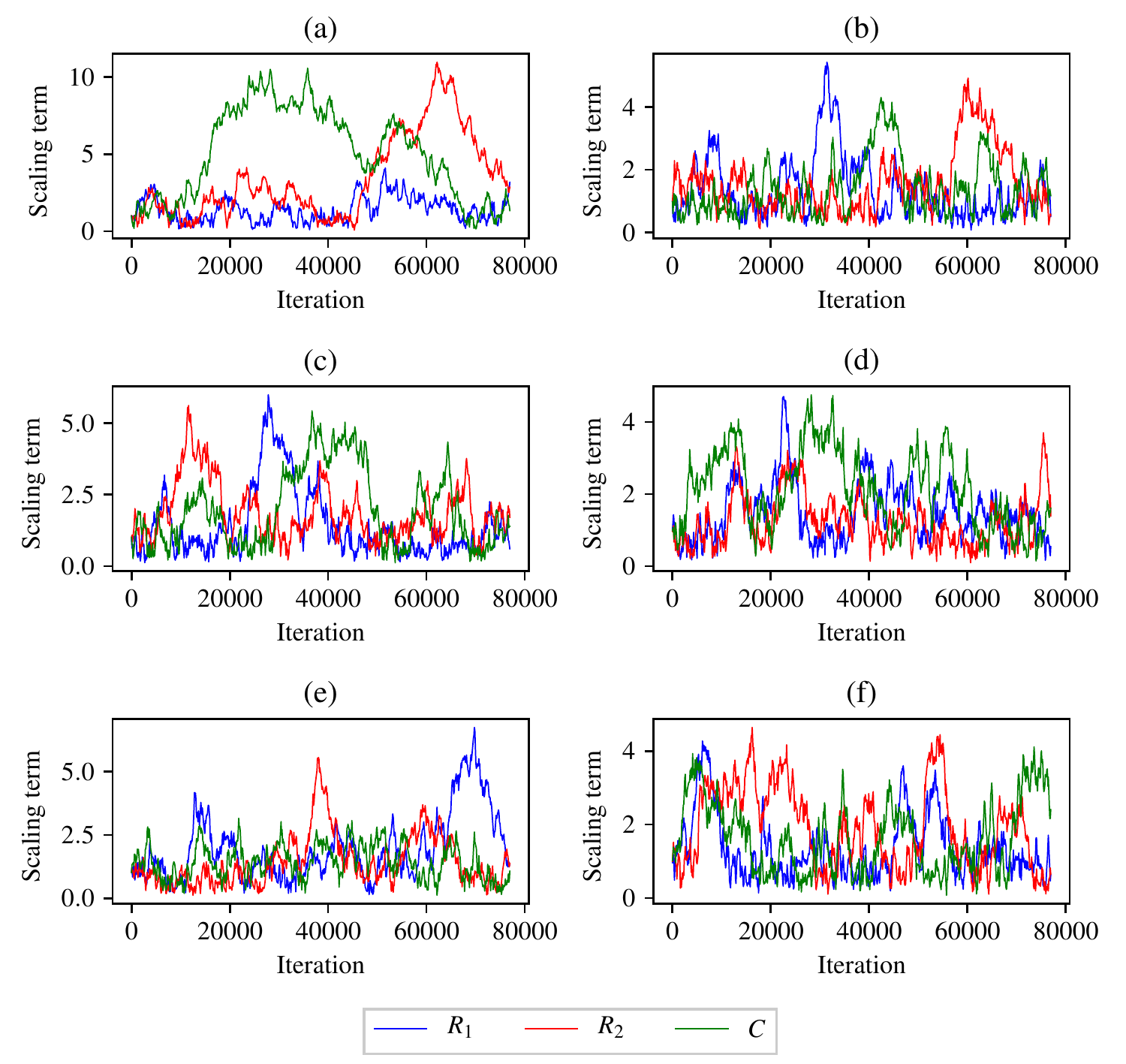}
\caption{The parameter scaling terms at each $100^{th}$ iteration of the Markov chain applied to the Windkessel model parameters at the terminal boundary of the: second left ulnar (a), left common interosseous (b), left radial (c), left vertebral (d), left external carotid (e), and the left internal carotid (f).}
\label{fig_TP_WK_armL}
\end{figure*}

\begin{figure*}
\centering
\includegraphics[width=6in]{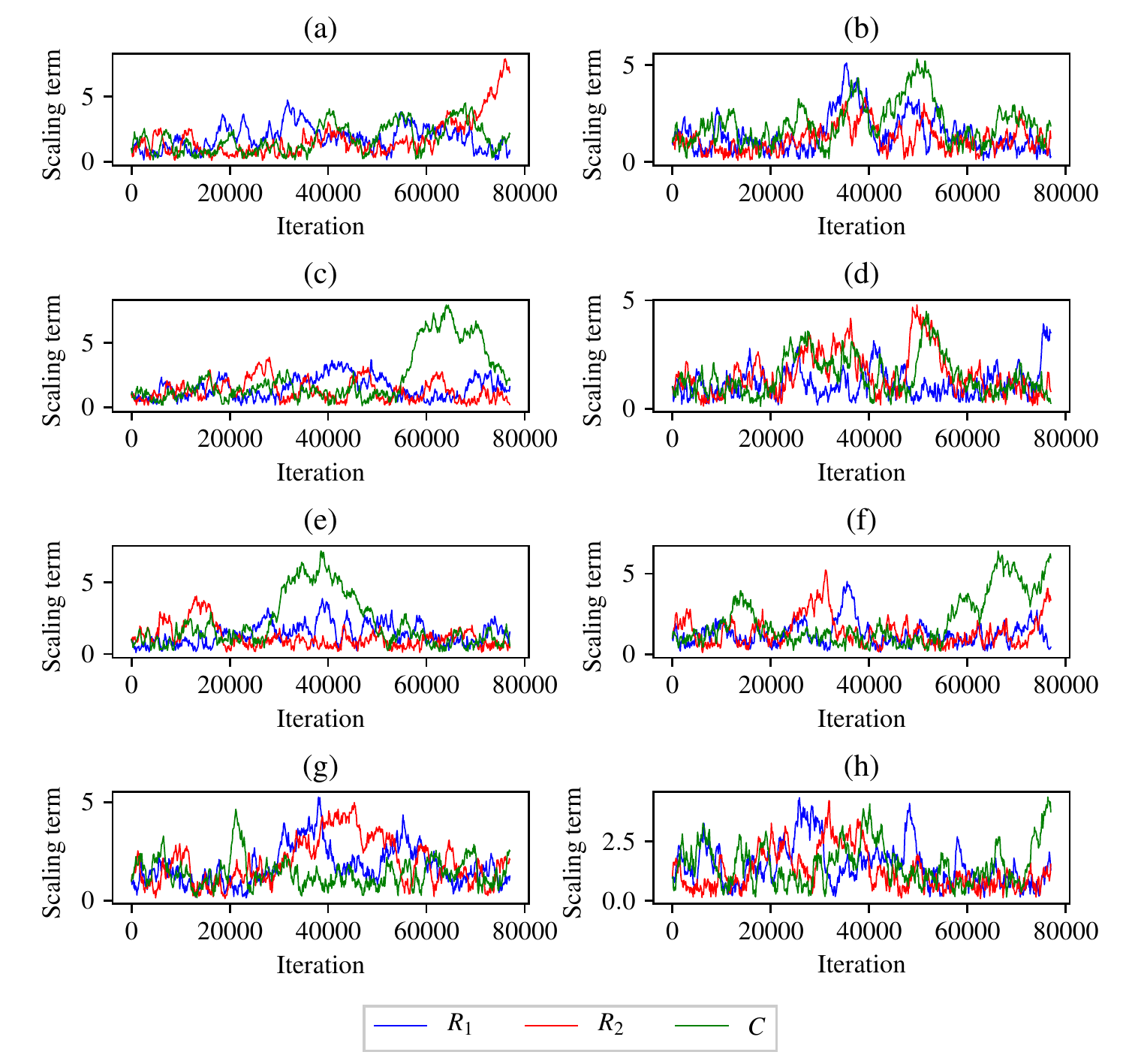}
\caption{The parameter scaling terms at each $100^{th}$ iteration of the Markov chain applied to the Windkessel model parameters at the terminal boundary of the: right anterior tibial (a), left anterior tibial (b), right posterior tibial (c), left posterior tibial (d), right profunda femoris (e), left profunda femoris (f), right internal iliac (g), left internal iliac (h).}
\label{fig_TP_WK_leg}
\end{figure*}

The trace plots of all parameters within each VPs arterial network are split into the following figures:
\begin{itemize}
\item The scaling terms applied to the reference parameters of the vessel wall mechanical property profiles of vessels with varying properties along their lengths are shown in Figures \ref{fig_TP_beta_chain}, \ref{fig_TP_beta_ind_I}, and \ref{fig_TP_beta_ind_II}.
\item The scaling terms applied to the reference parameters of the vessel radius property profiles of vessels with varying properties along their lengths are shown in Figures \ref{fig_TP_r_chain}, \ref{fig_TP_r_ind_I}, and \ref{fig_TP_r_ind_II}.
\item The scaling terms applied to the reference daughter-parent ratios of the vessel wall mechanical properties and the radii of all vessels bifurcating off of the aorta with constant properties are shown in Figures \ref{fig_TP_aorta_beta_constant} and \ref{fig_TP_aorta_r_constant} respectively.
\item The scaling terms applied to the reference daughter-parent ratios of the vessel wall mechanical properties and the radii are shown for vessels with constant properties in the right upper extremities in Figures \ref{fig_TP_armR_beta_constant} and \ref{fig_TP_armR_r_constant} respectively, and left upper extremities in Figures \ref{fig_TP_armL_beta_constant} and \ref{fig_TP_armL_r_constant} respectively.
\item The scaling terms applied to the reference daughter-parent ratios of the vessel wall mechanical properties and the radii of vessels with constant properties within the lower extremities are shown in Figures \ref{fig_TP_leg_beta_constant} and \ref{fig_TP_leg_r_constant} respectively.
\item The scaling terms applied to the reference parameters of the Windkessel models in the aortic region, right upper extremities, left upper extremities, and the legs are shown in Figures \ref{fig_TP_WK_aorta}, \ref{fig_TP_WK_armR}, \ref{fig_TP_WK_armL}, and \ref{fig_TP_WK_leg} respectively.
\end{itemize}
To aid in visualisation of the results shown in the above listed figures, all trace plots are thinned out by plotting only each $100^{\text{th}}$ iteration of the chain. This is done to filter out high frequency noise, clarifying the low frequency behaviour of the chain.\\

\section{Pressure and flow-rate profiles from random VPs}
\label{appendix_pressure_flow}

\begin{figure*}
\centering
\includegraphics[width=6in]{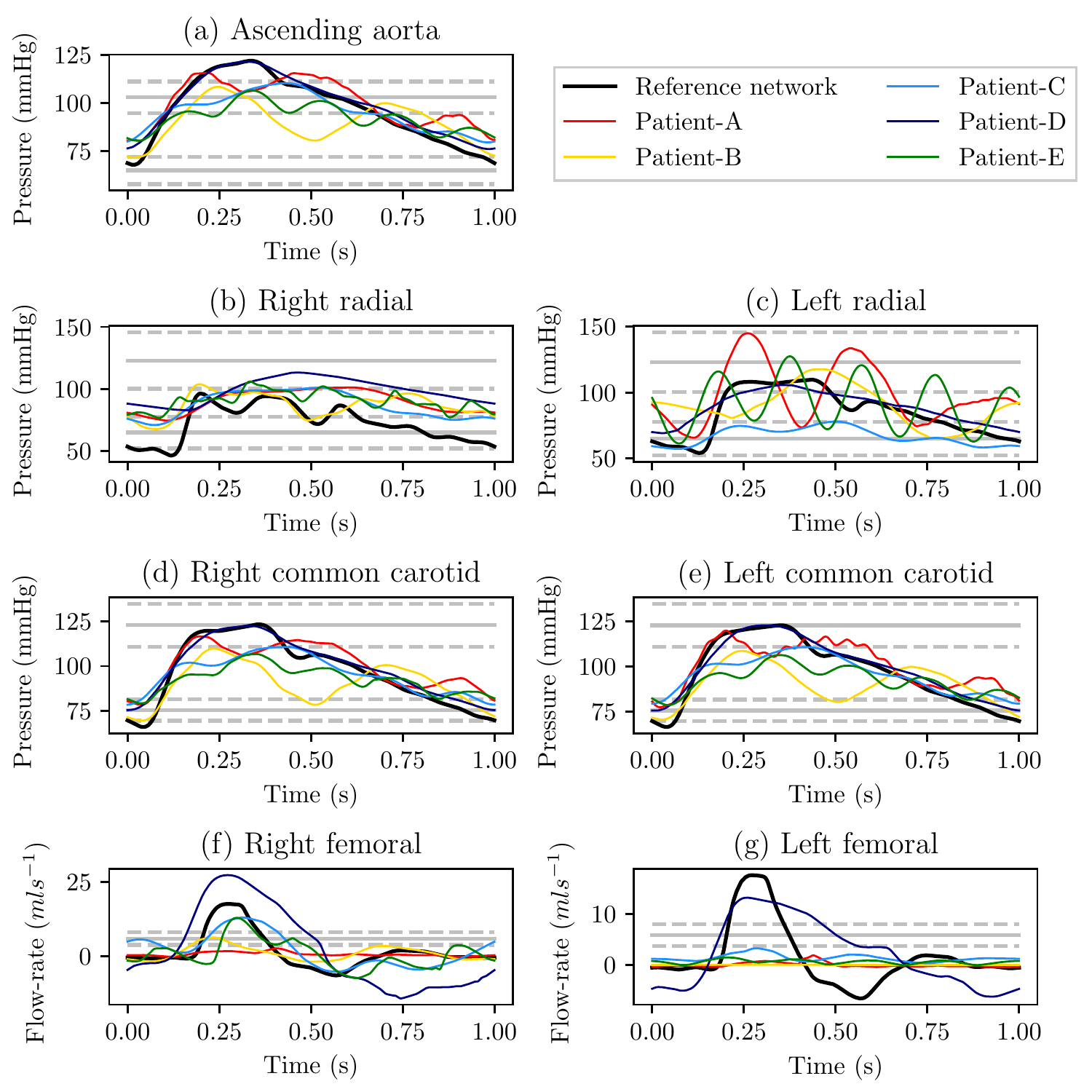}
\caption{Examples of pressure and flow-rate profiles taken from five VPs randomly drawn from the VPD. The subplots show the: pressure profiles in the ascending aorta (a), pressure profiles in the right radial artery (b), pressure profiles in the left radial artery (c), pressure profiles in the right common carotid artery (d), pressure profiles in the left common carotid artery (e), flow-rate profiles in the right second femoral artery (f), and flow-rate profiles in the left second femoral artery (g). In each figure the profiles taken from the reference network are shown in black; and the literature reported measurements and associated error are shown by the solid and dashed grey lines respectively.}
\label{fig_random_profiles_I}
\end{figure*}

\begin{figure*}
\centering
\includegraphics[width=6in]{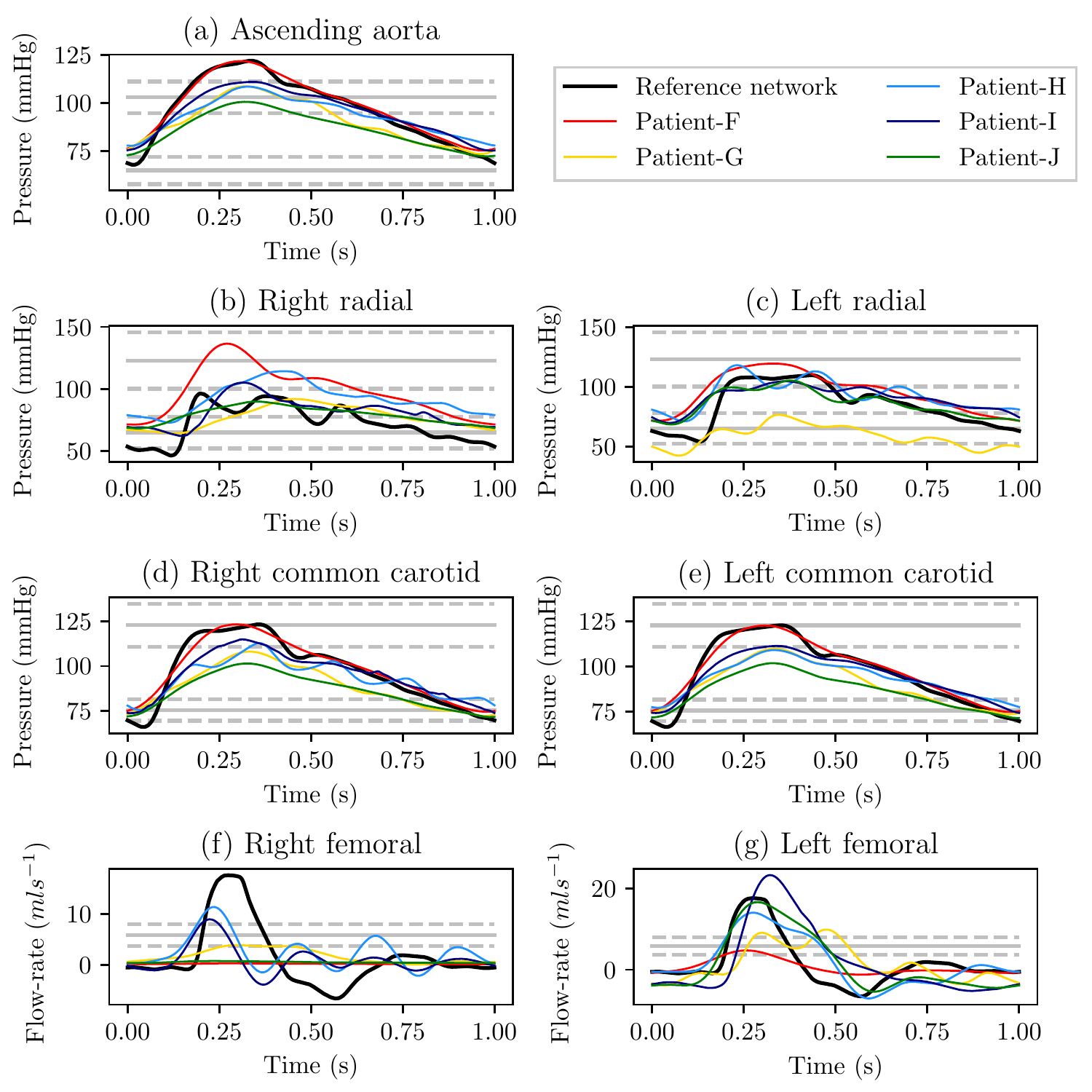}
\caption{Examples of pressure and flow-rate profiles taken from five VPs randomly drawn from the VPD. The subplots show the: pressure profiles in the ascending aorta (a), pressure profiles in the right radial artery (b), pressure profiles in the left radial artery (c), pressure profiles in the right common carotid artery (d), pressure profiles in the left common carotid artery (e), flow-rate profiles in the right second femoral artery (f), and flow-rate profiles in the left second femoral artery (g). In each figure the profiles taken from the reference network are shown in black; and the literature reported measurements and associated error are shown by the solid and dashed grey lines respectively.}
\label{fig_random_profiles_II}
\end{figure*}

\begin{figure*}
\centering
\includegraphics[width=6in]{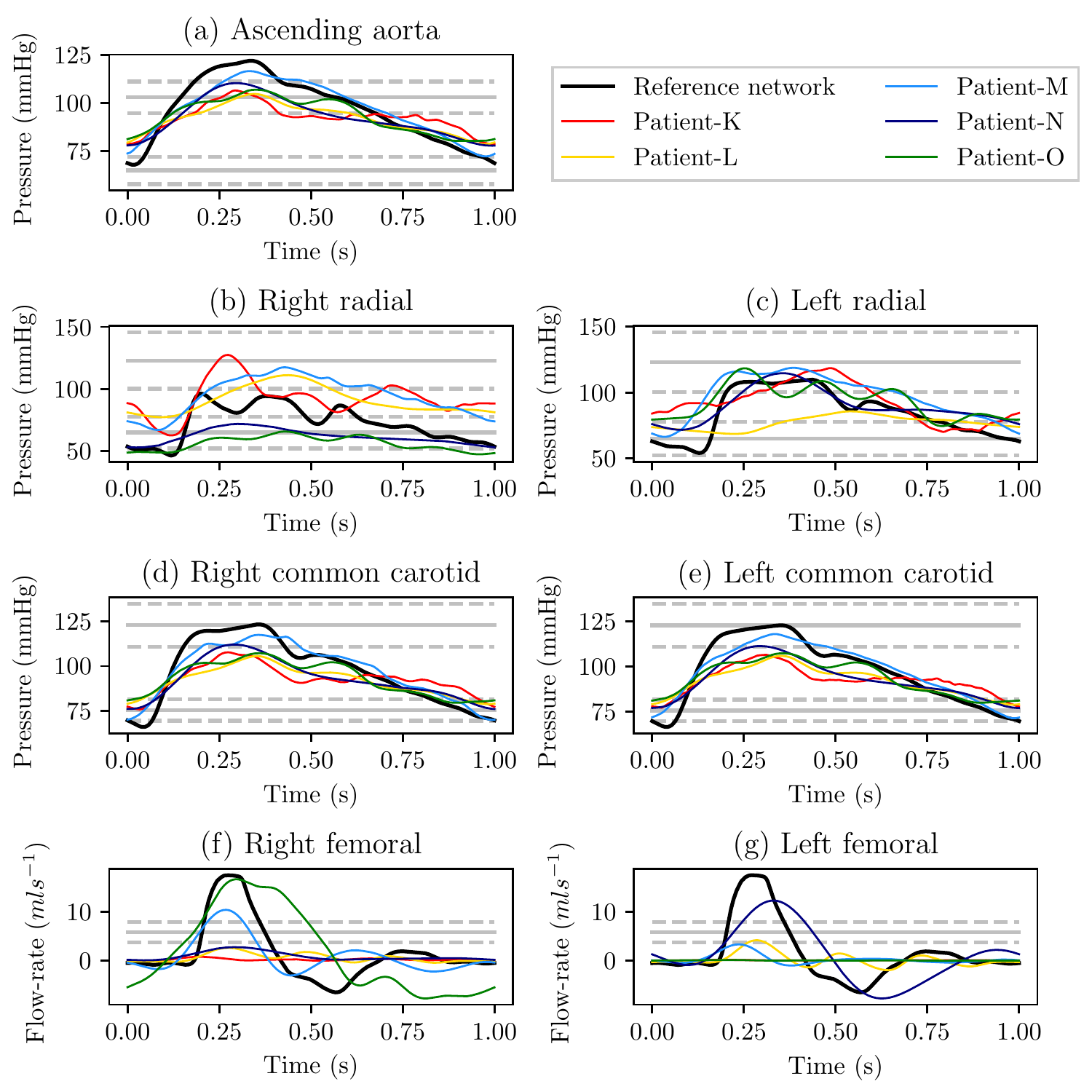}
\caption{Examples of pressure and flow-rate profiles taken from five VPs randomly drawn from the VPD. The subplots show the: pressure profiles in the ascending aorta (a), pressure profiles in the right radial artery (b), pressure profiles in the left radial artery (c), pressure profiles in the right common carotid artery (d), pressure profiles in the left carotid common artery (e), flow-rate profiles in the right second femoral artery (f), and flow-rate profiles in the left second femoral artery (g). In each figure the profiles taken from the reference network are shown in black; and the literature reported measurements and associated error are shown by the solid and dashed grey lines respectively.}
\label{fig_random_profiles_III}
\end{figure*}

Within the subsequent three figures (Figures \ref{fig_random_profiles_I}, \ref{fig_random_profiles_II}, and \ref{fig_random_profiles_III}) pressure and flow-rate profiles are shown from 15 VPs randomly sampled from the VPD. The pressure or flow-rate profile at each location measured within the reference network is also included, as well as the literature based measurements and associated error incorporated into the posterior distribution.

\end{document}